\documentclass[12pt]{article}
\usepackage{amsmath}
\usepackage{amssymb}
\usepackage{xspace}
\usepackage{epsfig}

\newcommand{\numub}{\ensuremath{\overline{\nu}_\mu}\xspace}
\newcommand{\nueb}{\ensuremath{\overline{\nu}_e}\xspace}
\newcommand{\numu}{\ensuremath{\nu_\mu}\xspace}
\newcommand{\nue}{\ensuremath{\nu_e}\xspace}
\newcommand{\nux}{\ensuremath{\nu_x}\xspace}

\newcommand{\nutau}{\ensuremath{\nu_\tau}\xspace}
\newcommand{\nuex}{\ensuremath{\overline{\nu}_x}\xspace}
\newcommand{\nubar}{\ensuremath{\overline{\nu}}\xspace}

\newcommand{\dmq}{\ensuremath{\Delta m^{2}}\xspace}
\newcommand{\mum}{\ensuremath{\mu^{-}}\xspace}
\newcommand{\mup}{\ensuremath{\mu^{+}}\xspace}

\newcommand{\cprA}{reproduced by permission of the AAS}
\newcommand{\cprB}{with kind permission of the European Physical Journal (EPJ)}
\newcommand{\cprC}[1]{copyright ({#1}), with permission from Elsevier}
\newcommand{\cprD}[1]{copyright ({#1}) by the American Physical Society}

\begin{document}

\title{Experimental results on neutrino oscillations}

\author{Ubaldo Dore$^1$ and Domizia Orestano$^2$ \\
1) \small Dipartimento di Fisica, Universit\`a di Roma ``La Sapienza" and I.N.F.N.,\\
\small Sezione di Roma, P. A. Moro 2, Roma, Italy\\ 
2) \small Dipartimento di Fisica, Universit\`a Roma Tre and I.N.F.N.,\\
\small Sezione di Roma Tre, Via della Vasca Navale 84, Roma,
Italy
}
\maketitle

E-mail:{\it ubaldo.dore@roma1.infn.it} and {\it orestano@fis.uniroma3.it}


\vskip 0.5cm
\noindent{\small\it This is an author-created, un-copyedited version of an article published in 
Reports on Progress in Physics. 
IOP Publishing Ltd is not responsible for any errors or omissions in this version of the manuscript or any version derived from it. The definitive publisher authenticated version is available online at 

http://www.iop.org/EJ/journal/RoPP.
}
\vskip 0.5cm
\begin{abstract}
The phenomenon of neutrino oscillation has been firmly established:
neutrinos change their flavour in their path from their source to
observers.
 This paper  is  dedicated  to the description of experimental
results in the oscillation field, of their present understanding and of
possible future developments in the experimental neutrino
oscillation physics.
\end{abstract}

\tableofcontents

\section{Introduction}
\par The interpretation of experimental results on solar and atmospheric
neutrinos in terms of neutrino oscillations had been put forward in the past, 
but it is only in the last years that this interpretation has been confirmed 
both for solar and atmospheric neutrinos.
\begin{itemize}
\item Solar neutrinos:
\par the  solar neutrino deficit
(measured flux of $\nue$ versus prediction of the  Solar Standard
Model, SSM) first observed  by the pioneering Ray
Davis chlorine experiment \cite{davis1} (final results in \cite{cleveland}),
and later by many other radiochemical experiments (SAGE \cite{sage}, 
GALLEX \cite{Gallex}, GNO \cite{gno})
and real time water Cherenkov detectors (Kamiokande \cite{fukuda96} and
Super-Kamiokande \cite{hosaka06}) had been interpreted  as 
due to 
oscillations.  The SNO \cite{SNO}
heavy water experiment has confirmed the oscillation hypothesis measuring a 
total flux of solar neutrinos well in agreement with the SSM
predictions, allowing to interpret the $\nue$ deficit as due to 
$\nue$ being transformed to
$\numu$ or $\nutau$, for which charged current interactions are
energetically impossible.
\par Oscillation parameters in agreement with the ones obtained in the 
solar 
experiments have been measured also by the  $\nueb$ reactor experiment
KamLAND~\cite{kamland}. 
\item Atmospheric Neutrinos:
\par the $\numu$ deficit observed in the flux of atmospheric  neutrinos 
coming 
from the
other side of the earth has been seen by Kamiokande 
\cite{sk1} and SuperKamiokande and has been 
 been interpreted  as due 
to muon neutrino oscillations \cite{Skat1}. This interpretation has been 
confirmed by other atmospheric neutrino experiments, MACRO \cite{Macro}
and Soudan-2 \cite{sou}, and 
 by long baseline accelerator experiments (K2K \cite{k2k} and later
MINOS\cite{minosa}). A review of the  discovery of neutrino oscillations 
can be found in reference \cite{kajita}.
\end{itemize}
\par
\begin{figure}[h]
\begin{center}
\epsfig{figure=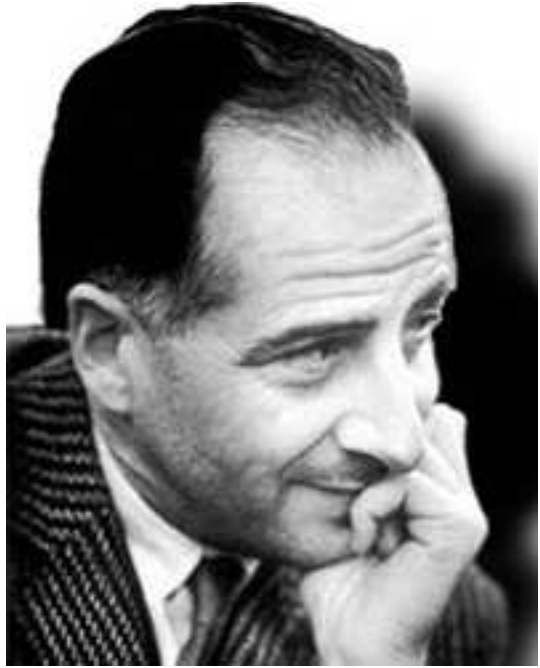}
\label{pontecorvo}
\caption {Bruno Pontecorvo }
\end{center}
\end{figure}

 The experimental establishment of neutrino oscillation
 can be considered
 a real triumph of Bruno
Pontecorvo \cite{Pontecorvo:1957cp,Pontecorvo:1957qd} (figure \ref{pontecorvo}), 
who first introduced this concept and
pursued this idea for many years when the general consensus did
support massless neutrinos, with no possibility of oscillations.

\bigskip
This paper is devoted to the review of experimental results in the 
oscillation field. 
\par 
In Section 2 we will give a brief presentation  of neutrino properties.
Section 3 will
contain  a brief presentation of  the theory of neutrino
oscillations. A  complete theoretical treatment
 can be found in many recent papers \cite{kaiser,bilenky,giunti,vissani,gonzalez,lipari}.
 Section 4  will describe
neutrino sources, solar, reactor, atmospheric and terrestrial.
Section 5 will be dedicated to the neutrino interactions that are
relevant for the study of oscillations. 
Section 6 will present the
results obtained   in  this field, up to year 2007, 
accompanied by a brief 
description of the  experimental apparatus used. Section 7 will 
illustrate  
the
present knowledge of the parameters describing the oscillations.
Section 8 will discuss the results achievable by currently in operation 
or approved experiments. Section 9 will discuss different scenarios
 for future neutrino experiments.

\section{Neutrino properties}
\par
\vskip 1cm
\par Neutrinos are fermions that undergo only weak interactions.
\begin{itemize}
\item
They can interact via the exchange of a W (charged currents) or via
the exchange of a $Z^{0}$ (neutral currents).
\item
The V-A theory requires, in the limit of zero mass,  that only left-handed 
 (right-handed) neutrinos (anti-neutrinos) are active.

\item In the Minimal Standard Model (MSM) there are 3 types of neutrinos
and the corresponding number of anti-neutrinos.
\item  Interactions have the same strength for the 3 species (Universality).
\item
Neutrinos are coupled to charged leptons, so we have 3 lepton doublets

 $\begin{pmatrix}e^{-}\cr\nue\end{pmatrix}$,
 $\begin{pmatrix}\mu^{-}\cr\numu\end{pmatrix}$,
 $\begin{pmatrix}\tau^{-}\cr\nutau\end{pmatrix}$.

Leptons  in each doublet carry an additive leptonic
number $L_e$, $L_\mu$, $L_\tau$, which has opposite
sign for antiparticles.
\par  The leptonic numbers are separately conserved.
\end{itemize}
\
\par One of the unsolved problem of neutrino physics is their
nature. Are they Majorana particles or Dirac particles? In the
Majorana scheme   there  is  only one neutrino with two helicity states.
 In the Dirac scheme
neutrinos can be left-handed or right-handed  and the same for
anti-neutrinos. For massless neutrinos in the V-A theory only
left-handed (right-handed) neutrinos (anti-neutrinos) can interact and
the two representations coincide.
\par The discovery of oscillations implies
that neutrinos have mass and consequently the
 helicity of the neutrino will not be Lorentz invariant,
as it happens for charged leptons. 
 A neutrino with nonzero mass is left-handed in one reference system and
it might be right-handed in another reference system.
\par For massive neutrinos  there are processes that are possible in the
Majorana scheme and forbidden in the Dirac one which  can 
be used to discriminate  between the two possibilities. One of these
processes is the neutrinoless double beta decay.
\par Double beta decay process is summarized by the reaction
$ A(Z,N) \rightarrow A(Z+2,N-2)+2 e^{-} +2\overline{\nue}$.
 This  process, second order in weak interactions, has been
observed for some nuclei for which it is energetically possible while the single 
beta decay into A(Z+1,N-1) is forbidden.
\par
\begin{figure}[htbp!]
\begin{center}
\epsfig{figure=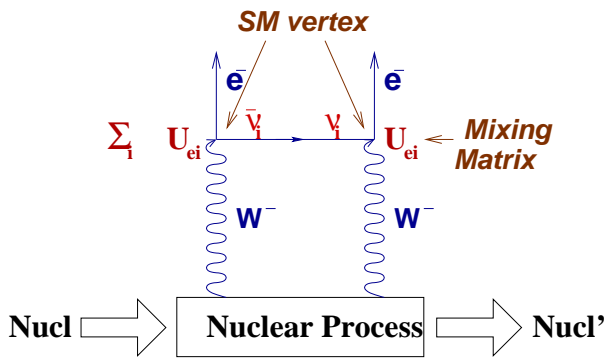,width=8.0cm}
\caption{Diagram describing the double beta decay}
\label{dbd}
\end{center}
\end{figure}

 In the
neutrinoless case $A(Z,N)\rightarrow A(Z+2,N-2)+2e^{-}$
only 2 electrons are present in the final state,
see Figure~\ref{dbd}).
This process is only possible for massive Majorana neutrinos, which are 
emitted 
with one helicity at one vertex and absorbed with opposite helicity at the other vertex. 
The amplitude associated with this helicity flip is proportional 
to the neutrino mass.
The rate of this process, possible only for
massive Majorana neutrinos,
would be given by
  $$R(0\nu)=(G \cdot M(nuclear)\cdot M_{eff})^2$$
  where:
\begin{itemize}
\item G is a phase space factor
\item M(nuclear) is the matrix element for the neutrinoless transition
between the two involved nuclei
\item $M_{eff}=\sum U_{ei}^2*m_i $, where
$U_{ei}$ are the elements of the mixing matrix, 
described in Section~\ref{oscil2},  giving the
mass of $\nue$ in terms of the mass eigenstates.
\end{itemize}
\par The theoretically appealing option  of  Majorana 
neutrinos has been the object of extensive experimental
programs and is  under investigation by
experiments both in run or in construction.
\par  No clear evidence for this process has been observed at the moment,
even though a claim of observation has been made with a mass
of $0.4~\mathrm{eV}/\mathrm{c}^2$ \cite{KlapdorKleingrothaus:2001ke}.
Current upper limit on M$_{eff}$, 
affected by large uncertainties due to the nuclear
matrix elements involved, is of the order of $1~\mathrm{eV}/\mathrm{c}^2$
\cite{Fiorini2007}.  \par For a review of
the present experimental  situation and future programs see reference
\cite{Fiorini2007}.

\par As it will be shown in Section~\ref{oscil2} 
oscillations can provide information only on the differences of
square masses of neutrinos and not on their masses. 
Attempts to measure directly neutrino masses have given up to now
only upper limits \cite{PDBook}:
\begin{itemize}
\item $m(\nue)$ limits are obtained from the end point of the
electron spectrum from Tritium decay, $m(\nu_e)~\le 2~\mathrm{eV}/\mathrm{c}^2$
\item $m(\numu)$ limits from the muon momentum end point in the $\pi$ decay,
\par $m(\numu)\le~190~\mathrm{KeV}/\mathrm{c}^2$
\item $m(\nutau)$ limits from the missing momentum of the 5 body  semileptonic
decay of $\tau$, $m(\nutau)\le~18~\mathrm{MeV}/\mathrm{c}^2$.
\end{itemize}
\par A limit on the neutrino mass can also be derived from cosmological
considerations, m$\le$0.13~eV~\cite{Seliok}.

\vskip 1 cm
\section{Neutrino oscillations}
\label{oscil2}
\par
\subsection{Vacuum oscillations}
\subsubsection{Three flavor mixing}\label{3flav}
\par In analogy to what happens in the quark sector the weak interaction
states, called flavor eigenstates, $\nue$, $\numu$, $\nutau$~~are a
linear
combination of the mass eigenstates ~$\nu_1,\nu_2,\nu_3$ ~that describe
the propagation of the neutrino field. These states are connected by an
unitary matrix U
$$ \nu_{\alpha}=\sum_j U_{\alpha j}\cdot\nu_j $$
with index $\alpha$ running over the three flavor eigenstates and index j
running over the three mass eigenstates.
The $3\times3$ matrix U is called the Pontecorvo--Maki--Nakagava--Sakata
and is analogous to the Cabibbo--Kobaiashi--Maskawa (CKM)  matrix in the quark sector.
\par In the general case a $3\times3$ matrix can be parametrized by 3 mixing
angles $\theta_1=\theta_{12}$, $\theta_2=\theta_{23}$, $\theta_3=\theta_{13}$
and a CP violating phase $\delta$

A frequently used parametrization of the U matrix is the following
\begin{eqnarray}
U & =&\begin{pmatrix}1&0&0\cr 0&c_{23}&s_{23}\cr 0&-s_{23}&c_{23}
\end{pmatrix}
\begin{pmatrix}c_{13}&0&s_{13}e^{-i\delta}&
 \cr
0&1&0\cr -s_{13}e^{+i\delta}
 &0&c_{13}
\end{pmatrix}
\begin{pmatrix}c_{12}&s_{12}&0\cr -s_{12}&c_{12}&0\cr 0&0&1
\end{pmatrix}
\end{eqnarray}
\par where $c_{jk}=\cos(\theta_{jk})$ ~~~and ~~~~  $s_{jk}=\sin(\theta_{jk})$.
\par The factorized form of the matrix turns out to be very useful in data
interpretation since the first matrix contains the parameters
relevant for atmospheric and accelerator neutrino oscillations,
the second one the parameters accessible to short distance reactor
experiments and the CP violating phase  $\delta$, while the third depends upon
the parameters involved in solar neutrino oscillations.
\par  Given three neutrino masses we can define two independent 
 square mass differences $\dmq_{12}$ and  $\dmq_{23}$.
\par As it will be shown in next sections 
$|\dmq_{12}|\ll |\dmq_{23}|$ and so $\dmq_{13}\simeq
\dmq_{23}$.
\par
\begin{figure}[hbtp!]
\begin{center}
\epsfig{figure=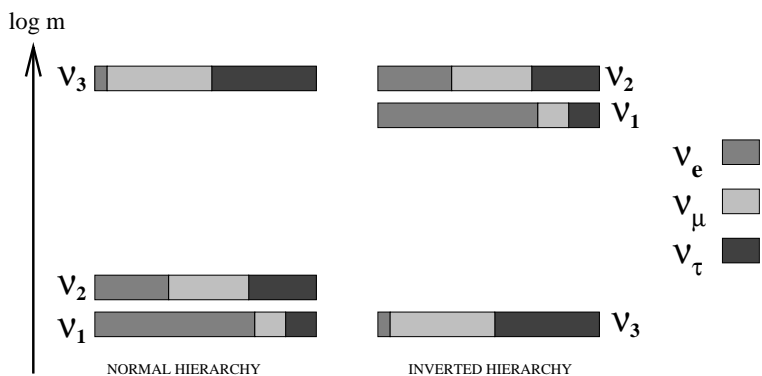,width=10.0cm}
\end{center}
\caption{The two possible mass spectra for normal and inverted hierarchies.}
\label{hiera}
\end{figure}

 The mass spectrum is formed by a doublet closely spaced $\nu_1$ and 
$\nu_2$,  
 and by a third state $\nu_3$ relatively distant.  
 This state can be heavier (normal hierarchy) or 
lighter (inverted  hierarchy) ( $\dmq_{23}$ positive or 
negative), the situation is depicted in Figure~\ref{hiera}.
Results discussed in Section~\ref{expres} 
indicate that $\dmq_{12}\approx 10^{-4}$~eV$^2$ and 
$\dmq_{23}\approx 10^{-3}$~eV$^2$. 
\vskip 0.6cm

\subsubsection{The two flavor mixing}
\label{twoflav}

The mechanism of oscillations can be explained easily by using as an 
example the mixing between two flavor
states and two mass
states $m_1$ and $m_2$.
The mixing matrix is reduced to $2\times2$ and is
characterized by a single parameter, omitting irrelevant phase factors:

\begin{gather*}
\begin{pmatrix} \cos\theta&\sin\theta\\-\sin\theta&\cos\theta
\end{pmatrix}
\quad
\end{gather*}

For the time evolution of a neutrino created for example 
as $\nue$ with a momentum p  
at time t=0 we can write (with the $\hbar$=c=1 choice of units)
\par\hskip5cm $|\nu(0)> = |\nue
>=\cos\theta|\nu_{1}>+\sin\theta |\nu_{2}>$
\par\hskip5cm 
   $|\nu(t)> =\cos\theta e^{-iE_1 t}|\nu_1>+
\sin\theta e^{-iE_2t}|\nu_{2}>$
\par where$ E_i= \sqrt{p^2+m_i^2}$.
\par At a distance L
$\approx$ t from the source the probability of detecting it in a different 
flavor, for example as a $\numu$, is
\par $$ P(\nue \rightarrow \numu)=|<\nu_\mu|\nu(t)>|^{2} 
=\sin^{2}(2\theta) \sin^{2}(\Delta
M^{2}L/4E)$$
\par where $\dmq= m_1^{2}-m_2^{2}$.
Choosing to express $\dmq$ in $eV^2$,
L in meters and E in MeV (or in km and GeV respectively)
\par $$ P(\nue \rightarrow \numu) =\sin^{2}(2\theta) \sin^{2}(1.27\Delta
M^{2}L/E)$$
\par In the two flavor scheme the survival probability for $\nue$ is given by
\par $$P(\nue \rightarrow \nue)~~=1- P(\nue \rightarrow \numu).$$

\begin{figure}[htb!]
\begin{center}
\epsfig{figure=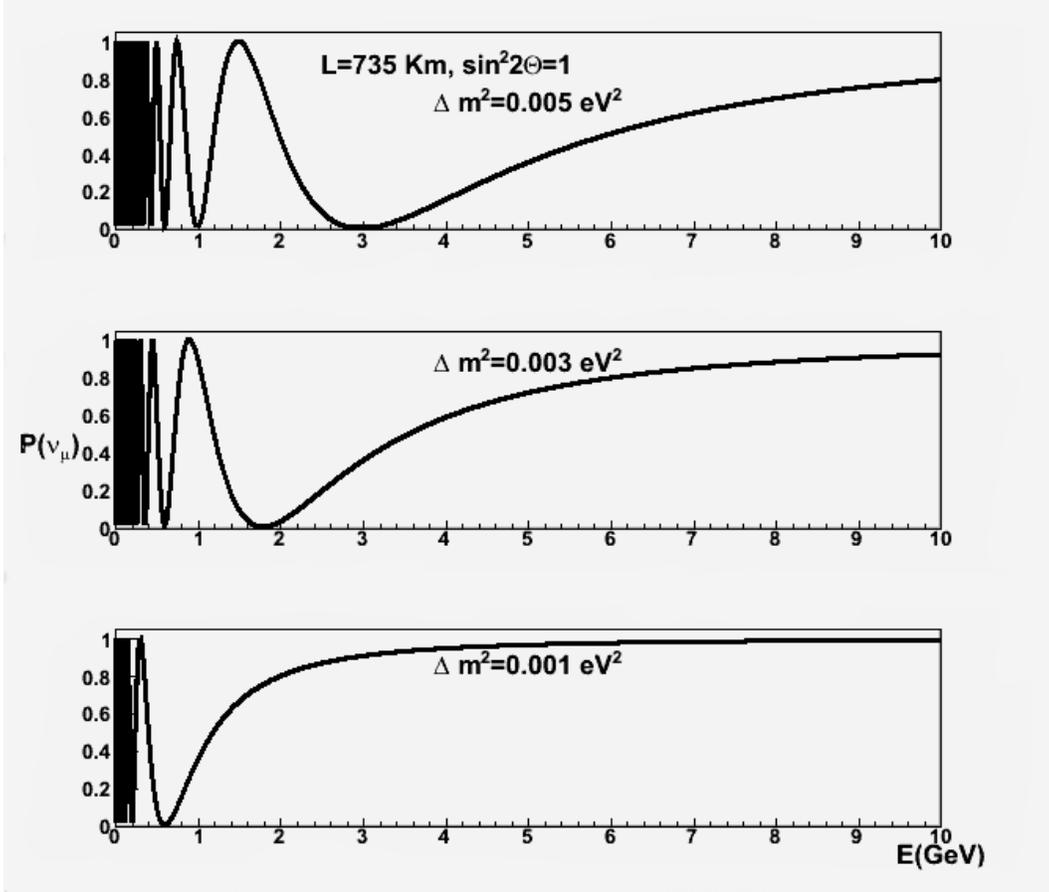,width=14.0cm}
\end{center}
    \caption {Oscillating behavior}
    \label{fig:oscill_behaviour}
\end{figure}

\par We can define the  oscillation length as ($\hbar$=c=1) 
 $$L_{osc}= 4\pi E/\dmq$$ that, adopting the units above,
can be rewritten as
   $$L_{osc}= 2.48E/\dmq$$
and so $$P=\sin^{2}(2\theta) \sin^{2}(\pi L/L_{osc})~~ $$
\par The oscillation probability has an oscillating behavior with the first
maximum
at~~$ L/L_{osc}=1/2$.
Figure~\ref{fig:oscill_behaviour} shows examples of oscillation
patterns as a function of the neutrino energy for fixed L and different 
values of $\dmq$.

It should be noted that in the two flavor approximation  CP and T 
violating
terms vanish and
$$P(\nu_{\alpha} \rightarrow \nu_{\beta})=P(\nu_{\beta} \rightarrow \nu_{\alpha})$$
$$P(\nu_{\alpha} \rightarrow \nu_{\beta})=P(\overline{\nu}_{\alpha} \rightarrow \overline{\nu}_{\beta})$$

\subsection{Matter oscillations}\label{msw}
\par In the previous section it has been assumed that neutrinos propagate
in vacuum. The presence of matter modifies the oscillation probability 
because one must include the amplitude for forward elastic scattering. 

 The  scattering processes can be expressed in terms of a refraction index 
different for
$\nue$ and $\numu$ or $\nutau$. The difference in refraction index
can then introduce additional phase shifts thus modifying the
oscillation probability via the Mikheyev-Smirnov-Wolfenstein (MSW) effect~\cite{MSW:MS,MSW:W}.
\par Let us define the effective potentials experienced by neutrinos in matter
$$V_{\mu,\tau}=\pm\sqrt{2}G_{F}(-N_e/2 +N_p/2 -N_n/2)$$
$$V_{e}=\pm\sqrt{2}G_{F}(-N_e/2 +N_p/2 -N_n/2+N_e)$$
where the sign plus 
is for neutrinos and the  sign minus for anti-neutrinos.
$G_{F}$ is the Fermi coupling constant, N$_{e}$, N$_{p}$ and N$_{n}$ are
the electron, the proton and the neutron number density. 
The additional term in the second equation arises from the W exchange
contribution to the  scattering process 
$\nue+e \rightarrow \nue+e$ (see Section~\ref{nuescatt}).
\par The relevant quantity for neutrino propagation is
$\Delta V_{e,\alpha}=V_{e}-V_{\alpha}$, the difference between
potentials for electron neutrinos and neutrinos of flavor
$\alpha$ ($\alpha=\mu$ or $\tau$).

$$\Delta V_{e,\alpha}=\pm\sqrt{2}G_{F}\cdot N_{e}={\pm}
7.6\cdot10^{-14}~eV\cdot\rho \cdot (Z/A)$$
where $\rho$ is the density of matter (in g/cm$^3$).
Defining $B=2E\Delta V$, $\epsilon=B/\dmq$,
assuming a constant density (a reasonable assumption for terrestrial long 
baseline
experiments), in the two flavor mixing treatment
we can replace vacuum parameters with matter parameters
$$ \sin 2\theta_{m}=\sin 2\theta/\sqrt{(\cos 2\theta- \epsilon)^2+\sin 
^2 2\theta}$$
$$ \dmq_{m}=\dmq\sqrt{(\cos 2\theta- \epsilon)^{2}+\sin ^2 2\theta}$$
$\epsilon$ sign is positive for neutrinos and a positive $\dmq$ value, 
and is reverted for anti-neutrinos or for negative $\dmq$.
\par The oscillation probability can be written as 
  $$P=\sin^{2}(2\theta_{m}) \sin^{2}(\dmq_{m} L/4E)$$
\par In the limit $\epsilon\ll \cos 2\theta$
matter effects become negligible.
\par The above formulas are valid in the case of propagation in a constant 
density medium, variable density becomes important in the propagation of 
neutrinos in the Sun, treatment of this situation can be found in~\cite{kuo}.
\par The treatment of  matter effects in the three
flavor case is complicated and  can be found in~\cite{matter3flav}.

\subsection{Approximations for the oscillation probabilities}
\label{appro}
\par The oscillation probability in the 3 flavor case contains  two mass 
differences,
three mixing angles, the phase $\delta$ and the matter effect contribution. 
Approximate formulas in terms of $\alpha=\dmq_{12}/\dmq_{23}$ and in 
terms of 
 the  mass effect term B, have been developed in the limit 
$\alpha \ll 1$ and $B/\dmq_{23}\ll 1$~\cite{arafune,Richter,freund,cerve}.

For example, for the $\numu\rightarrow\nue$ oscillation probability has been  
written in  ref \cite{freund}
as : 
\begin{eqnarray*}
P(\numu\rightarrow \nue)=sin^{2}\theta_{23}
sin^{2}2\theta_{13}{{sin^{2}[(1-A)\Delta]}\over{(1-A)^{2}}} \\
\pm J\alpha sin\delta_{CP }sin{\Delta}{{sin(A\Delta)}\over{A}}{{sin[(1-A)\Delta]}\over{(1-A)}} \\
+J\alpha cos\delta_{CP } cos{\Delta}{{sin(A\Delta)}\over{A}}{{sin[(1-A)\Delta]}\over{(1-A)}}\\
+\alpha^{2}cos^{2}\theta_{23}sin^{2}2\theta_{12}{{sin^{2}(A\Delta)}\over{A^{2}}}
\end{eqnarray*}
\begin{equation}\label{eq:approx}
\end{equation}

\par Where J=$
cos\theta_{13}sin2\theta_{12}sin2\theta_{13}sin2\theta_{23}$
\par $\alpha=\dmq_{12}/\dmq_{23}, \Delta=\dmq_{23}L/4E,A=B/\dmq_{23}$,
B as defined in Section~\ref{msw}.

The appearance probability, neglecting matter effects, for
accelerator   neutrinos in the 3 flavor mixing scheme
 using $\dmq_{12}=8\times10^{-5}~\mathrm{eV}^2$ ($\alpha\approx 0$), 
and $L/E\simeq$~ 1 and therefore $\sin^{2}(\dmq_{12}L/4E) \simeq 0 $ is:
$$ P(\numu \rightarrow \nue)=\sin^{2}(2 \theta_{13})
\sin^{2}(\theta_{23})\sin^{2}(\dmq_{23}L/4E) $$
 
It can be shown that in the same approximation:
$$ P(\numu \rightarrow \nutau)=
\cos^{4}(\theta_{13})\sin^{2}(2 \theta_{23})
\sin^{2}(\dmq_{23}L/4E) $$
 $$ P(\nue \rightarrow \nutau)=\sin^{2}(2 \theta_{13})
\cos^{2}(\theta_{23})\sin^{2}(\dmq_{23}L/4E) $$

\par For $\sin^{2}(2\theta_{13}) \simeq 0$ the only probability different
from 0
is $ P(\numu \rightarrow \nutau)$
that can be written as $$ P(\numu \rightarrow \nutau)=\sin^{2}(2 \theta_{23})
\sin^{2}(\dmq_{23}L/4E) $$
depending upon two parameters $ \theta_{23}$ and $\dmq_{23}$
that will coincide with the two parameters of the simplified
treatment.  \par In this approximation the $\numu$ survival probability in
atmospheric or accelerators  neutrino experiments will be given by
$1-P(\numu \rightarrow \nutau)$.

\begin{figure}[hbtp!]
\begin{center}
\epsfig{figure=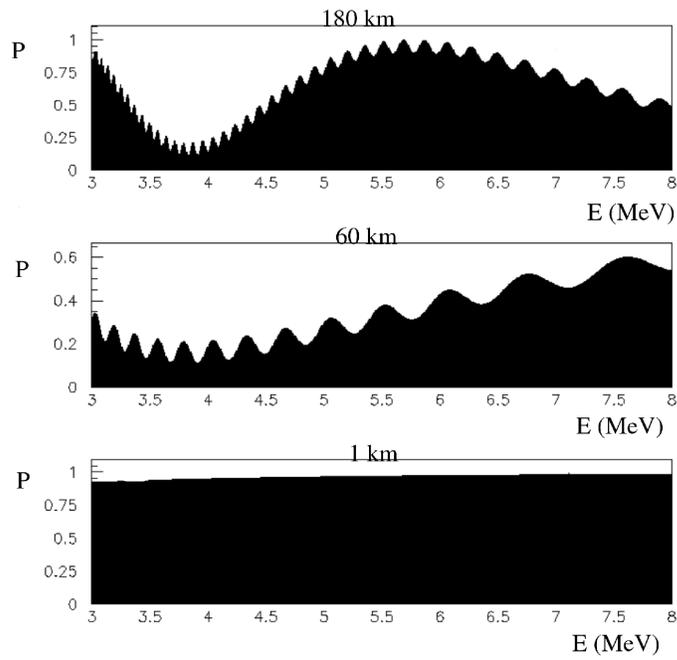,width=10.0cm}
\end{center}
    \caption {Probability for  $ P(\nueb \rightarrow \nueb)$ versus
neutrino energy (in MeV) at L = 180, 60, 1~km from the source.}
    \label{fig:dist3}
\end{figure}
\par In reactor experiments(see section \ref{reactor}), in which matter 
effect can be 
neglected 
because  the energy and matter density 
involved are small,  the 
survival probability of a 
$\nueb$ will be given by
$$ P(\nueb \rightarrow \nueb)=1-P1-P2$$
with
$$ P1= \cos^{4}(
\theta_{13}) \sin^{2}(2\theta_{12})\sin^{2}(\dmq_{12}L/4E)$$
$$ P2=\sin^{2}(2 \theta_{13})\sin^{2}(\dmq_{23} L/4E)$$
\par At short distances $\sin^{2}(\dmq_{12}L/4E) \simeq 0 $ and
the term P1 can be neglected.
\begin{eqnarray}
P(\nueb \rightarrow \nueb)=1- P2
=1-\sin^{2}(2 \theta_{13})sin^{2}(\dmq_{23}L/4E)
\label{eq:disapp}
\end{eqnarray}
\par will be sensitive to $\theta_{13}$~ and ~ $\dmq_{23}$.
\par At large distances the term P1 will be dominant  and  in the limit
of $ \sin^{2}(2 \theta_{13})\simeq 0$ we will have
 $$ P(\nueb \rightarrow \nueb)=1-P1=1-\cos^{4}(\theta_{13})
\sin^{2}(2\theta_{12})\sin^{2}(\dmq_{12}L/4E).$$
Figure \ref{fig:dist3} shows $P(\nueb \rightarrow \nueb)$ as a function of
the neutrino energy for
E($\nu$)=3-8 MeV (typical of reactor neutrinos), L=180, 60, 1~km
and with $\sin(2 \theta_{13})=0.05$, $\sin(2 \theta_{12})=0.314$,
$\dmq_{12}=7.9\times10^{-5} eV^{2}$,$\dmq_{23}=2.5\times10^{-3}
 eV^{2}$.

\subsection{Experimental determination of neutrino oscillation parameters}

\par Table~\ref{sensitivity} gives the $\dmq$ values
accessible to different
neutrino sources according to $\dmq\approx E/L$.
\begin{table}[bhtp]
\begin{tabular}{c|c|c|c}\hline
Neutrino source & Distance from source (Km) & Energy (GeV) &
$\dmq$~(eV$^2$) \\\hline
solar & $10^{8}$ & $10^{-3}$& $10^{-11}$\\
atmospheric from top & 20 & 1,10& $0.05,0.5$ \\
atmospheric from bottom  & $10^{4}$ &1,10& $10^{-4}$, $10^{-3}$  \\
reactors & 1& $10^{-3}$ & $10^{-3}$\\
reactors large distance& 100 & $10^{-3}$ & $10^{-5}$\\
accelerators           & 1 & 1,20 & 1,20\\
accelerators long distance&100,1000 & 1,20& $10^{-3},0.2$\\
\hline
\end{tabular}
\caption{Sensitivity to $\dmq$ of experiments studying different neutrino sources}
\label{sensitivity}
\end{table}
\par
\begin{figure}[htbp!]
\begin{center}
\epsfig{figure=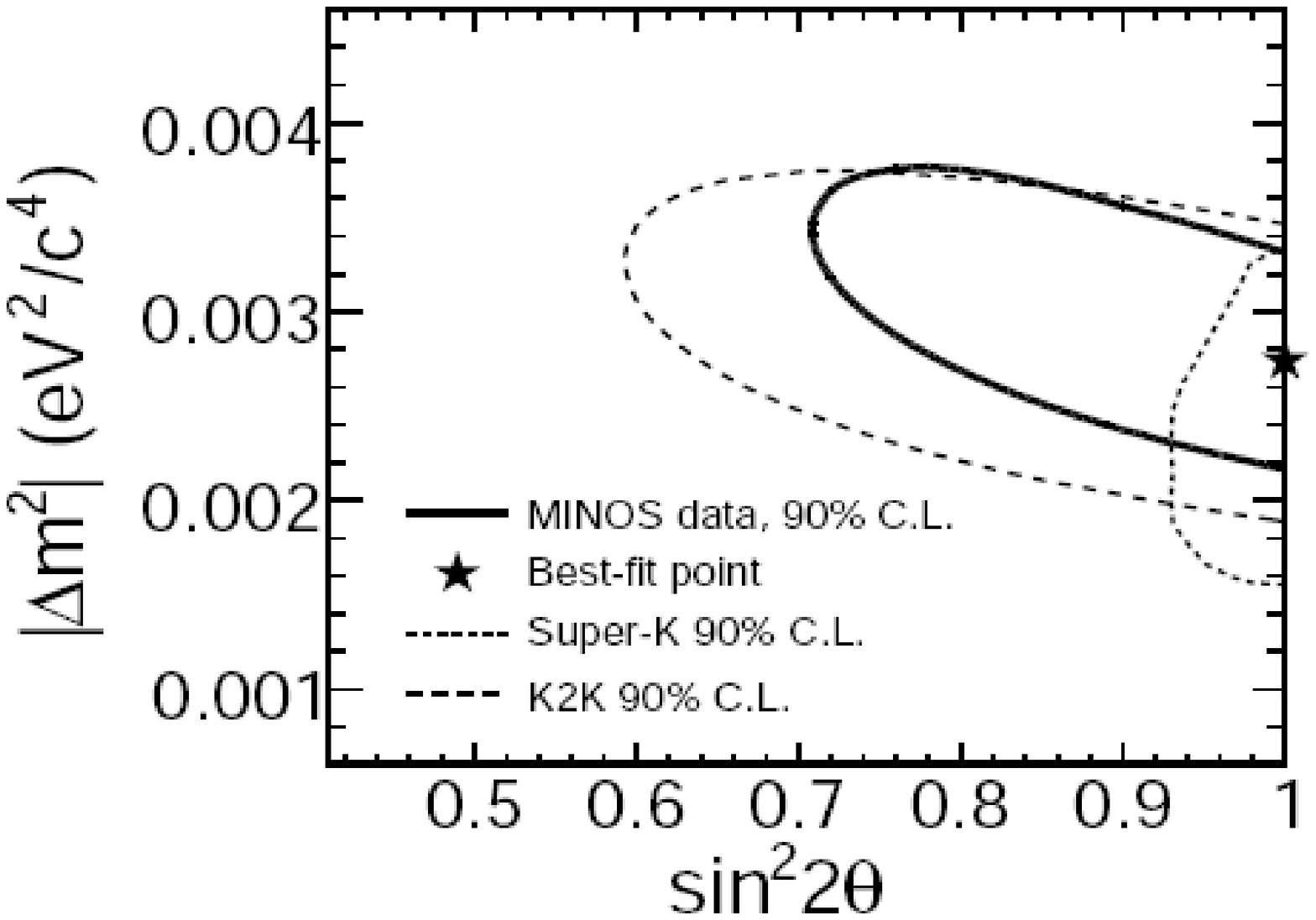,width=8.0cm}\epsfig{figure=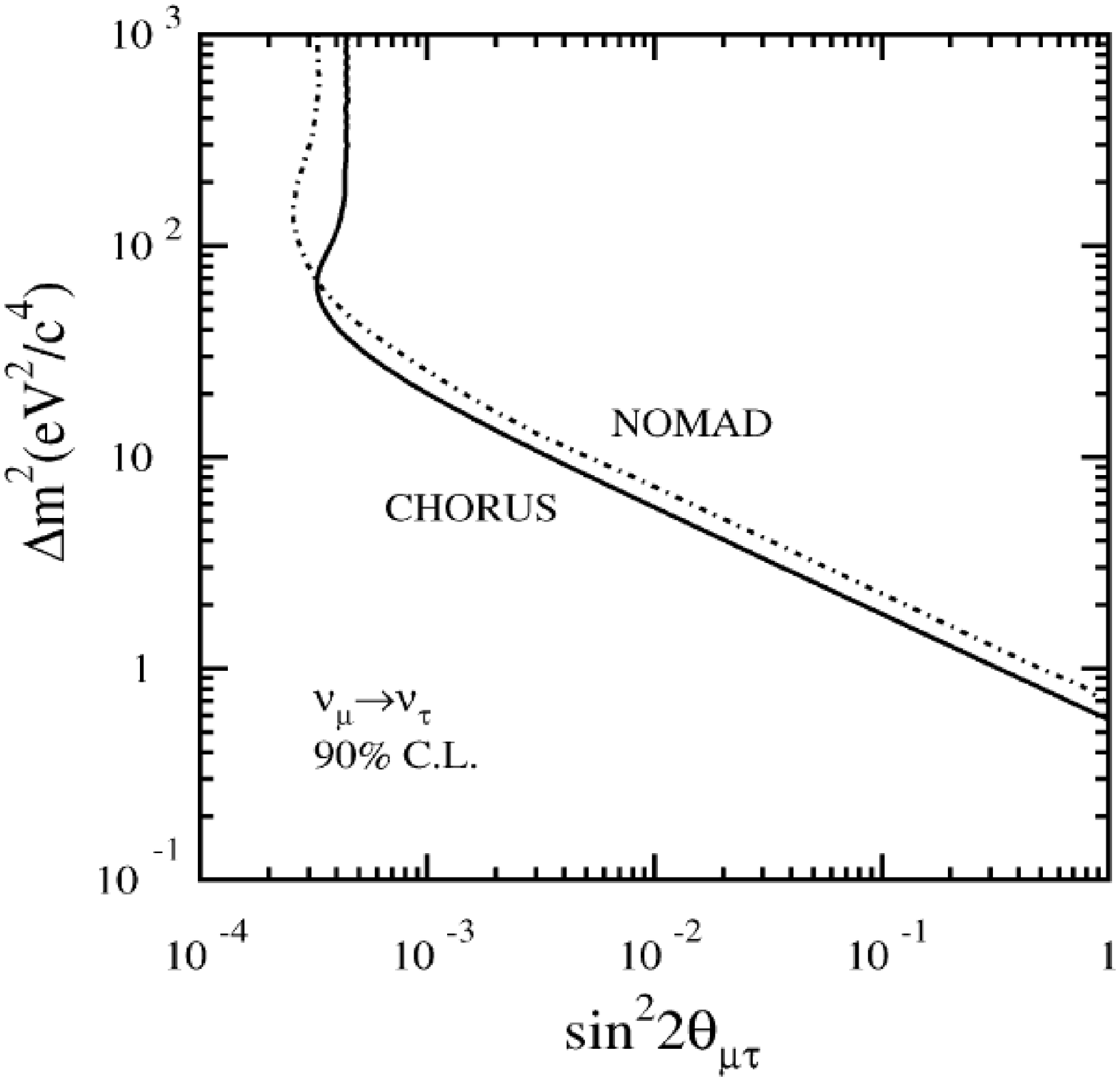,width=8.0cm} 
\caption
{Experimental results represented in the ($\dmq$,
$\sin^{2}(2\theta)$) plane: for positive (A), figure 
from ref.~\cite{minos}, 
and negative (B),  from ref.~\cite{choruso}, \cprC{2008} (results also from ref.~\cite{nomado}).
}
\end{center}
\label{fig:parplane}
\end{figure}

 In the two flavor scheme
the determination of the oscillation probability P gives a relation
between $\dmq$and $\sin^{2}(2\theta)$
 in the ($\dmq$,$\sin^{2}(2\theta)$) plane.
A measurement of P gives   a region in the parameter plane whose 
extension
depends on the resolution of the oscillation probability measurement.
In the case of a negative result an exclusion region
can be drawn. Examples of the two cases are shown in Figure~\ref{fig:parplane}.

\par Oscillations can be studied in two different approaches by the so-called
disappearance and appearance experiments.
\subsubsection{Disappearance experiments}
\par The flux of neutrinos of a given flavor $\nu_{\alpha}$ at a
distance L from the source, $\Phi(L) $, is compared to the
flux at the source, $\Phi(0)$.
The ratio $\Phi(L)/\Phi(0)$ will give the survival probability of
the neutrino, but no information on
the type of neutrino to which $\nu_{\alpha}$ has
oscillated. These   experiments crucially depend upon the knowledge  of
$\Phi(0)$. This approach is the only possible one for the low energy
$\nue$ or $\nueb$
(solar or reactor neutrinos)
since CC interactions of $\numu$ or $\nutau$ are kinematically  forbidden.
\par The uncertainties related to the knowledge of $\Phi(0)$ can be
canceled measuring the ratio of fluxes measured by two detectors positioned
at distances L ({\it far detector}) and $l\ll$L ({\it near detector}) from the source.

\subsubsection{Appearance experiments}
\par  Starting with a source of $\nu_{\alpha}$,  flavor $\nu_{\beta}$ 
neutrinos
will be searched for at a distance L.
In these experiments the main source of systematic errors
are the contamination
from  $\nu_{\beta}$ at the production point and background
mistaken as $\nu_{\beta}$ CC interactions.
Typical examples of this approach
are experiments with accelerators producing  $\numu$ neutrino beams.
These   $\numu$ beams have a small contamination of $\nue$. 
So in a  search
for $\numu \rightarrow \nue$  oscillation a possible signal must be
extracted from the contribution of beam $\nue$.

\par The following general considerations can be made

\begin{itemize}
\item
 the smallness of cross sections requires large mass targets in order to 
have
an appreciable number of interactions. In general, target and detector
coincide, both in appearance and disappearance experiments.
\item
appearance  experiments require the determination of
the flavor of the involved
neutrinos.  The detection of flavor does not give problems in the
case of $\numu$, while the detection of $\nue$ can give problems at high
energies, where electromagnetic showers from gamma
coming from  $\pi^{0}$ decays can mimic electrons.
The detection of $\tau$ is made
difficult by the short lifetime of this particles.
$\numu$ and $\nutau$ can of course be identified only above the energy
threshold for Charged Current  interactions.

\end{itemize}

\section{Neutrino sources}

\subsection{Solar neutrinos}\label{solarsources}
\par Neutrinos are produced in the thermonuclear reactions that take place
in the Sun core.  The process is initiated by the reactions
par$$ p+p \rightarrow H^{2}+e^{+}+\nue$$
\par$$ p+e^{-}+p\rightarrow H^{2}+\nue$$
\par followed by a chain of processes illustrated in Table~\ref{ppcycle}, 
whose 
net result is
$$ 4p +2 e^{-} \rightarrow He^{4}+2 \nue +\gamma.$$
\par Another source of neutrinos is the CNO Cycle,
whose  contribution to the solar neutrino flux is 
negligible~\cite{Clayton}.
\begin{table} [!ht]
\begin{center}
\begin{tabular} {|c|c|c|}\hline
Reaction & $\%$ of terminations & neutrino energy (MeV) \\\hline
$p+p\rightarrow H^2+e^++\nu$ & (99.75) & 0-0.420 \\
$p+e^-+p\rightarrow H^2+\nu$ & (0.25) & 1.44 \\
$H^2+p\rightarrow He^3+\gamma$& (100) &\\\hline\hline
$He^3+He^3\rightarrow He^4+2p$& (86) &\\\hline
OR & & \\\hline
$He^3+He^4\rightarrow Be^7+\gamma$ & & \\
$Be^7+e^-\rightarrow Li^7+\nu $ & (14) & 0.861 (90$\%$), 0.383 (10$\%$)\\
$Li^7+p\rightarrow2He^4$ & & \\\hline
OR & & \\\hline
$Be^7+p\rightarrow B^8+\gamma$ & & \\
$B^8\rightarrow (Be^8)^*+e^++\nu$ & (0.015) & 14.06 \\
$(Be^8)^*\rightarrow 2He^4$ & & \\ \hline
\end{tabular}

\caption {pp  chain  in Sun, from reference~\cite{bahcall_rmp54}}
\label{ppcycle}
\end{center}
\end{table}

\par The Q value of the reaction is 26 MeV and the corresponding energy
is released mainly in the form of electromagnetic radiation.
The average energy of the emitted neutrinos is $\approx$ 0.5~MeV.

\par The computation of the rate of these processes has been initiated by
Bahcall in the sixties  and his more recent evaluation,  using
different models for Sun parameters, has been published
in reference~\cite{bahcall:2004pz}.
\par
\begin{figure}[htbp!]
\begin{center}
\epsfig{figure=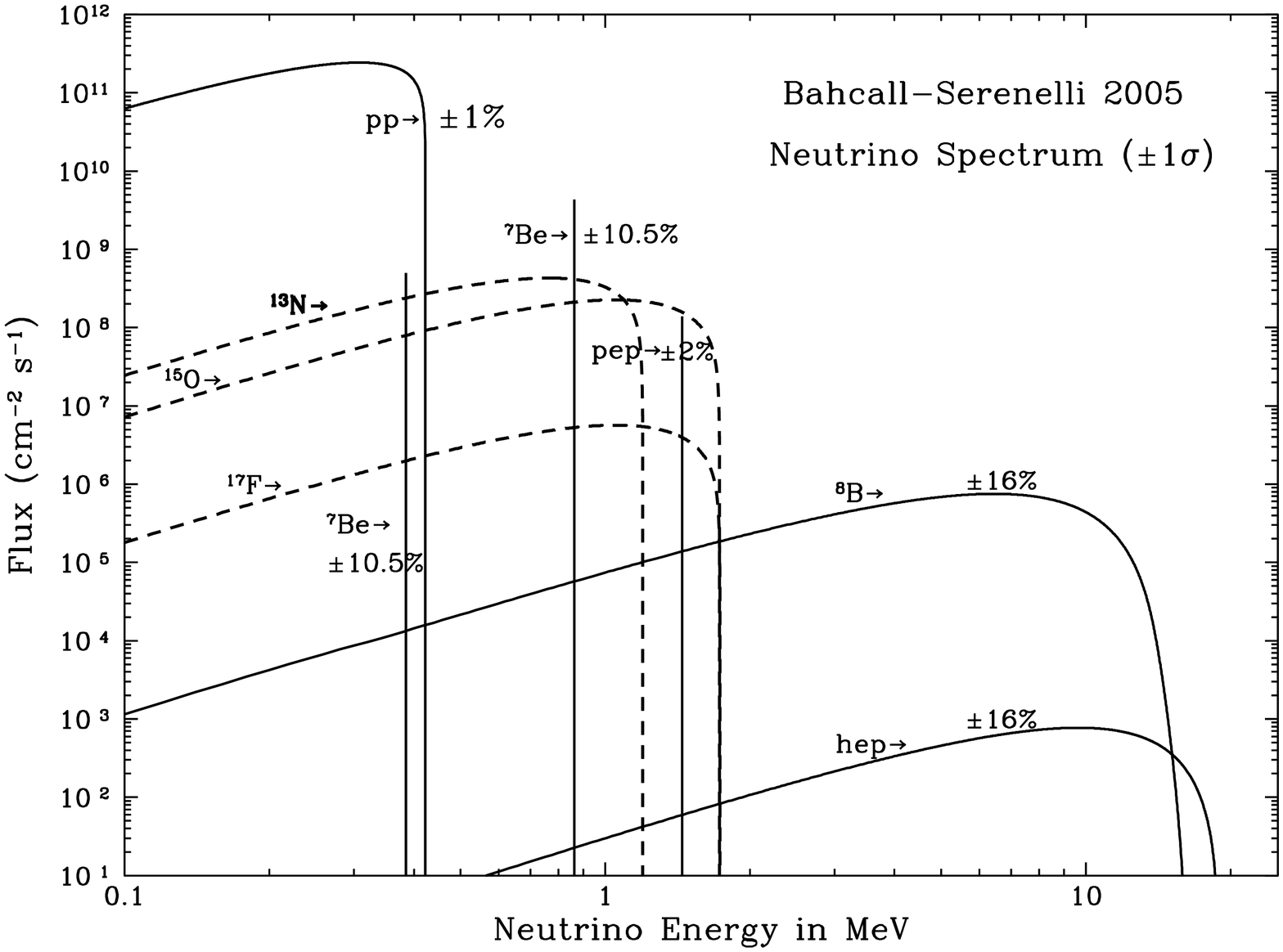,angle=-90,width=8.0cm}
\caption {Spectra of neutrinos from different processes in the
Sun~\cite{bahcall:2004pz}. \cprA.} \label{solspec}
\end{center}
\end{figure}

 The contribution from the pp cycle is  very well determined
and constitutes $\approx 99\%$ of the solar neutrino flux on earth.
Figure~\ref{solspec} shows the energy distribution of the different
sources of solar neutrinos. The information is summarized in
Table~\ref{solfluxes}. 

\begin{table}[!ht]
\begin{center}
\begin{tabular} {|c|c|c|c|c|}\hline
process & flux &error&mean energy&energy max\\
&    $10^{10}~cm^{-2}s^{-1}$ &$\%$ &MeV &MeV\\\hline
 pp    &   6.0  &1.  & 0.267 &0.42\\
pep  &  1.4 x$10^{-2}$&1.5& 1.44&1.44\\
hep  &7.6x$10^{-7}$&15&9.68&18.8 \\
Be$^{7}$&4.7x$10^{-1}$ &10&0.81&0.87\\
B$^{8}$&5.8x$10^{-4}$ &16.&6.73&14.0\\
N$^{13}$&6.1x$10^{-2}$ &30. &0.70&1.2\\
O$^{15}$&5.2x$10^{-2}$ &30 &0.99&1.73\\\hline
\end{tabular}
\caption{Rates of neutrino fluxes from  the Sun
~\cite{bahcall:2004pz} with error estimates from~\cite{bahcallunc}.}
\label{solfluxes}
\end{center}
\end{table}

\par
The error column in
Table~\ref{solfluxes} 
shows that
the Standard Solar Model (SSM) predicts with high precision the rate of the pp
fusion, which also produces most of the neutrino flux on earth.
The flux for pp neutrinos is predicted with a small error and 
so deviation from these predictions are a strong indication of 
oscillations.  
The final confirmation of the SSM
has been given by the SNO experiment, which found the total all neutrino flavours flux
(above 5 MeV) in agreement with the model prediction (see section \ref{snoexp}).

\par The different techniques used in solar neutrino detectors have 
different energy thresholds, so they are sensitive to different components 
of the solar neutrino spectrum.
The threshold for Chlorine detectors \cite{davis1} is
 0.814~MeV, well above the end point of the neutrino energy in the pp
process, while Gallium detectors \cite{gavrin} have a threshold at 
0.233~MeV which makes them
sensitive to pp neutrinos. For water counters \cite{hosaka06} the energy 
threshold is
fixed by the minimum  electron energy that can be detected above background (few MeV).
\vskip 3cm

\subsection{Reactor  neutrinos}
\par Nuclear reactors are an intense source of $\nueb$, generated in the
 beta decay  of fission fragments produced 
in the fission.
Each fission releases about 200~MeV and 6~$\nueb$. The average energy
of $\nueb$ is of the order of few MeV, well below the $\mu$ and $\tau$
production thresholds in CC interactions, therefore only disappearance
experiments are possible. These experiments require the flux and
the energy spectrum of neutrinos to be known with great precision.
\par Neutrinos are detected through the reaction  $ \nueb+p =e^{+}+n$
which has a threshold at 1.8 MeV.

\begin{figure}[ht]

\centerline{\psfig{figure=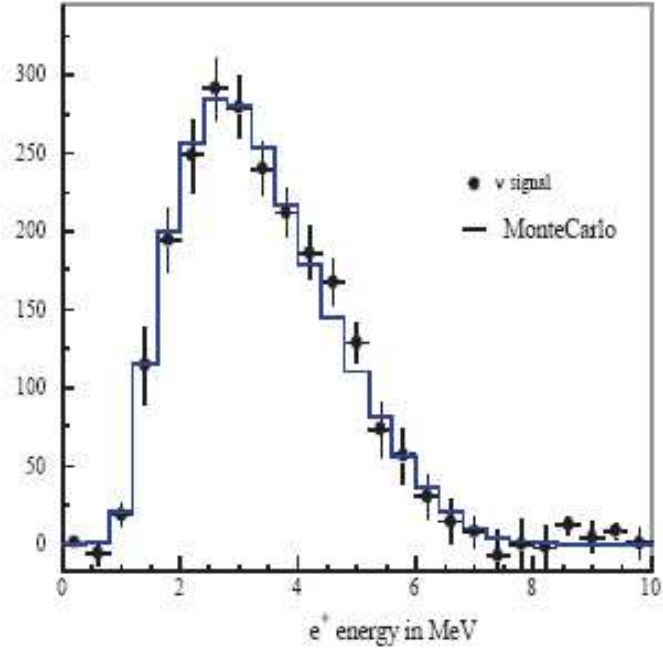,width=10.0cm}}
\caption {Example of energy spectrum of positrons produced in reactor 
neutrinos interactions,
showing the level of accuracy of flux predictions. 
Figure from~\cite{apollonio}, \cprB.}
\label{SpecChooz}
\end{figure}

\par  The determination of the neutrino flux is based upon the knowledge of
the thermal power of the reactor core and of the fission rate of the
relevant isotopes $U^{235},~U^{238},~Pu^{239},~Pu^{241}$. The
$\beta$ spectrum of the fission fragments is then 
converted in the \nueb
spectrum, which can be predicted at the $10^{-2}$ level.
The agreement  of predictions and data is demonstrated in 
Figure~\ref{SpecChooz}
where  the measured positron spectrum in the CHOOZ detector
is compared to Monte Carlo
prediction~\cite{apollonio}.

\subsection{Atmospheric   neutrinos}
\par Atmospheric neutrinos are generated by the interaction of 
primary cosmic ray
radiation (mainly protons) in the upper part of the atmosphere.
The average distance  traveled by pions and kaons before 
decay $\gamma c\tau$ (with $c\tau=7.8~m$ for pions and $c\tau=3.7~m$ 
for kaons) is such that they decay in flight, while some of the muons
produced in their decay 
($c\tau=658~m$) reach earth undecayed.
Neutrinos and anti-neutrinos are produced in the processes
$$\pi^{+}\rightarrow \mu^{+}+\numu $$
$$\pi^{-}\rightarrow \mu^{-}+\numub$$
$$K^{+}\rightarrow \mu^{+}+\numu+X $$
$$K^{-}\rightarrow \mu^{-}+\numub+X$$
$$\mu^{+}\rightarrow e^{+}+\numub+\nue$$
$$\mu^{-}\rightarrow e^{-}+\numu+\nueb$$
\par One of the most recent flux computations
has been made by Honda and collaborators
~\cite{Honda:2006qj},
who also provide references to previous computations.

\par
\begin{figure}[htpb!]
\begin{center}
\epsfig{figure=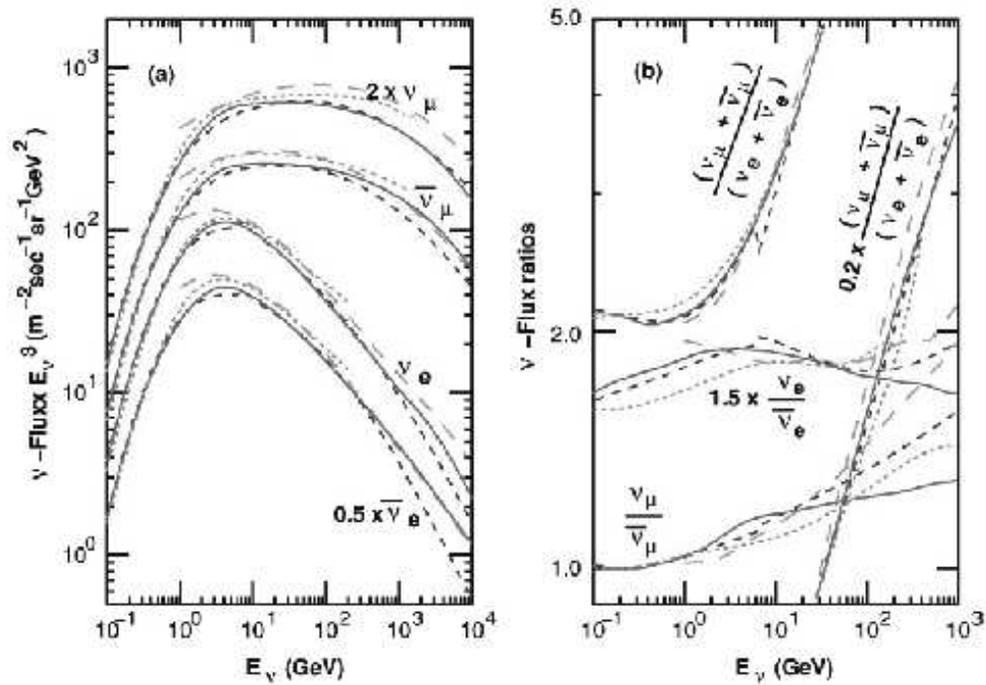,width=14.0cm} 
\end{center}
\caption
{Atmospheric neutrino fluxes and the ratio $\numu/\nue$. Figure from 
reference ~\cite{Honda:2004yz}, \cprD{2004}.  Solid line
Honda et al.~\cite{Honda:2004yz}, dotted line Honda et
al.~\cite{Honda:1995hz}, dashed line Fluka
group~\cite{Battistoni:2002ew,ATL97:113} long dashed Agrawal et
al.~\cite{PhysRevD.53.1314}.} \label{honda}
\end{figure}

 If all the muons could decay the ratio $\numu / \nue$ would be 2.
This ratio is larger
at high energies as shown by the energy spectra of atmospheric
neutrinos in Figure~\ref{honda}, in which results of Honda's computation are
compared to those of other models.
\par The neutrino flux for $E\approx 1$~GeV is $\approx 0.1  m^{-2}s^{-1}$
and is up-down symmetric.
\vskip 3cm
\subsection{Accelerator    neutrinos}
\label{beams}
\par
\begin{figure}[!ht]
\centerline{\psfig{figure=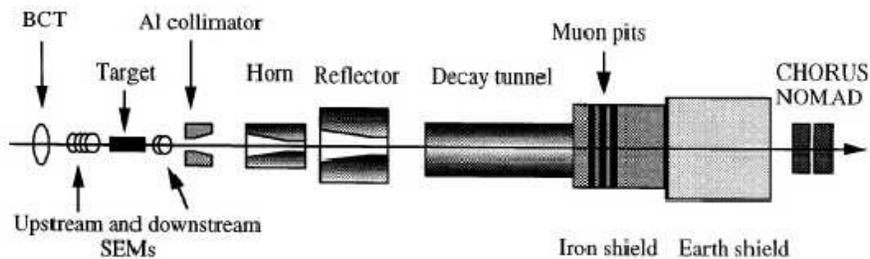,width=14.0cm}}
\caption {schematic view of a neutrino beam. Figure from ~\cite{Yellow}}
\label{nubeam}
\end{figure}

\begin{figure}[!ht]
\begin{center}
\epsfig{figure=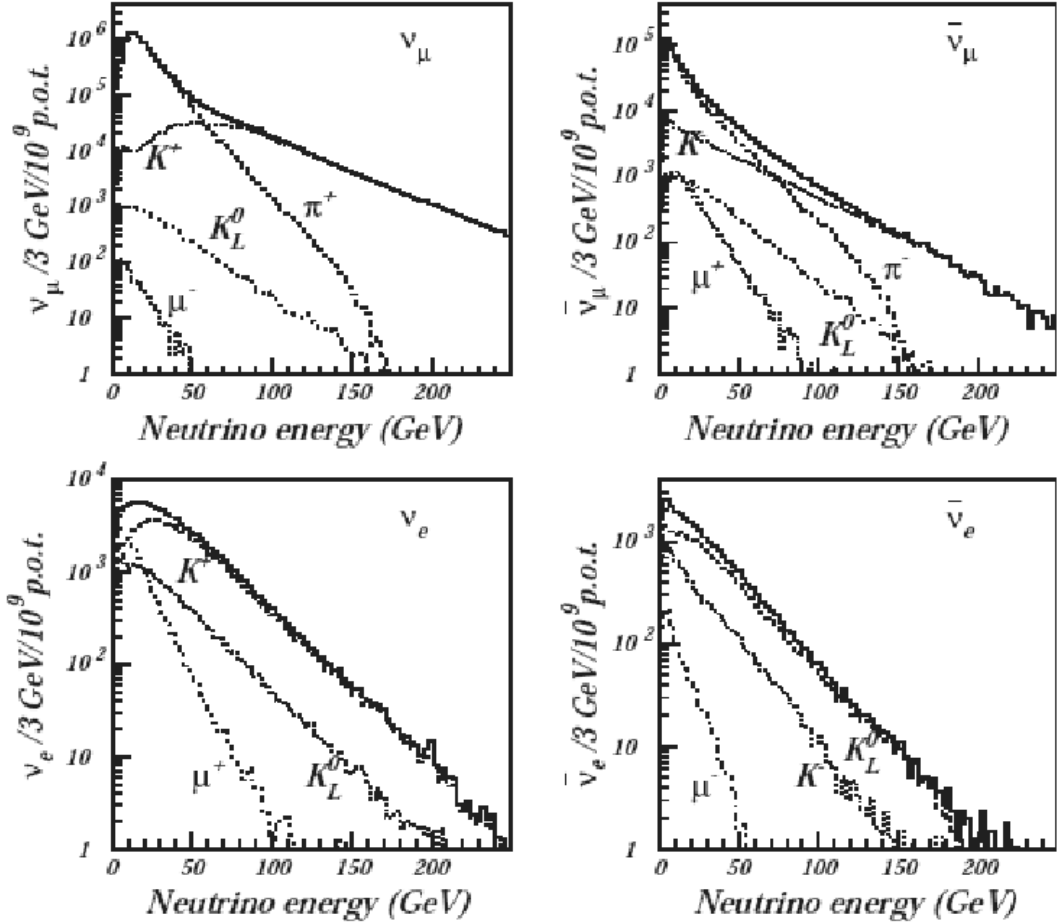,width=16.0cm} 
\end{center} 
\caption{Neutrino fluxes in the CERN WBB beam from reference \cite{WBBNomad}.} \label{beamene}
\end{figure}

 Neutrino beams are produced by  proton accelerators.
The extracted proton beam interacts on a target and the produced particles
are focused by a magnetic system (horn) whose polarity selects the
desired charge of the particles.
Pions (and kaons) are allowed to decay in an evacuated  tunnel followed
by an absorber stopping all particles except  neutrinos
and anti-neutrinos. The resulting beam contains mainly $\numu$
($\numub$) when positive (negative) particles are focused. A 
small contamination of $\numub$ ($\numu$) and $\nue$ ($\nueb$) is due 
at high energy to the kaon
semileptonic decay $K^{+}\rightarrow \pi^0+ e^{+}+\nue$, while at low energy 
there is a contamination from muon decay.
A schematic drawing of the CERN Wide Band Neutrino Beam (WBB) from the SPS 
is shown in Figure~\ref{nubeam}.
It is a typical high energy $\numu$ beam 
(a similar beam has been built at FNAL) whose composition
is given in Table~\ref{beamcomp}.
The momentum distribution of the neutrino produced is shown in Figure~\ref{beamene}.

This beam has been used for several neutrino experiments 
CDHS~\cite{cdhsdet}, CHARM~\cite{charmdet},
CHARM2~\cite{charm2det} and in the oscillation search for $\numu 
\rightarrow \nutau $
 by the CHORUS~\cite{chorusdet} and NOMAD~\cite{nomaddet} experiments.

\begin{table}[!ht]
\begin{center}
\begin{tabular}{|c|c|c|}\hline
 neutrinos & relative abundance & average energy (GeV)\\\hline
$\numu$&      1.      &              24.3             \\
$\numub$&  0.0678        &        17.2                    \\
$\nue$ &   0.0102      &         36.4                   \\
$\nueb$  &   0.0027     &         27.6                   \\\hline
\end{tabular}
\caption{Composition of the WBB beam at the CERN SPS from
 reference \cite{WBBNomad}.}
\label{beamcomp}
\end{center}
\end{table}
\par The relative  abundance of $\nutau$ has been extimated of 
the order 
of~$10^{-6}$ (see for example ref \cite{chorusdet})
\par
\begin{figure}
\begin{center}
\centerline{\psfig{figure=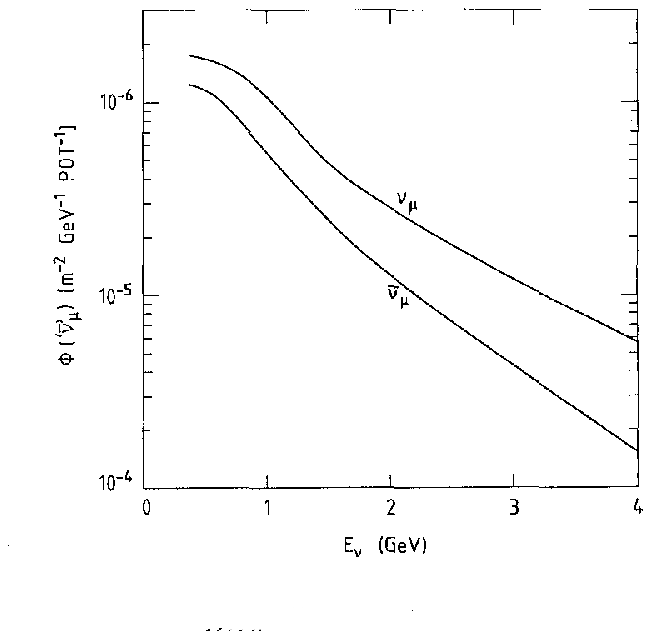,width=14.0cm}}
\caption {Neutrino fluxes in the medium energy CERN PS 
beam. From reference~\cite{charm}, \cprC{1984}.}
\label{beamene2}
\end{center}
\end{figure}

 The neutrino energy is correlated to the momentum of the protons. Figure
~\ref{beamene2}
shows the momentum spectrum of neutrino produced by 19 GeV protons
extracted from the CERN PS.
This beam has been used from the CDHS~\cite{cdhs}, CHARM~\cite{charm} and 
BEBC  (The Big European Bubble Chamber)~\cite{angelini}
for neutrino oscillation searches.
\par
\begin{figure}[hbtp!]
\begin{center}
\epsfig{figure=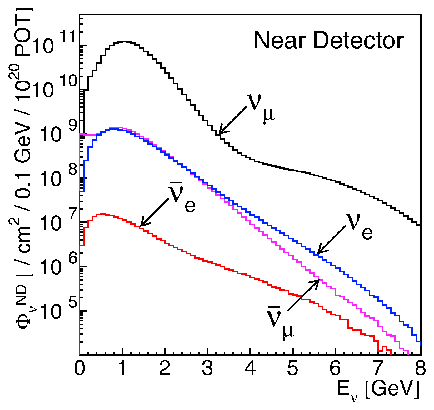,width=8.0cm}
\caption { The K2K beam, from  from reference \cite{k2k}, \cprD{2006}.} \label{K2Kbeam}
\end{center}
\end{figure}

\begin{figure}[hbtp!]
\begin{center}
\epsfig{figure=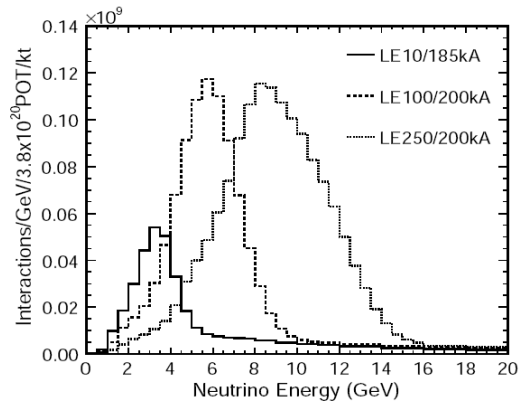,width=8.0cm} \caption {
The NUMI  beam: spectra in the low, medium and high energy beam
configurations, from reference~\cite{minos}.  } \label{NUMI}
\end{center}
\end{figure}

 The discovery of $\numu$ oscillation in the $\dmq\approx
10^{-3}~\mathrm{eV}^{2}$ region has pushed for low energy beams and long distance
experiments ($\dmq\approx E/L$).
The energy spectrum of the neutrino beam from the 12~GeV
protons of the KEK proton synchrotron at the
K2K\cite{k2k} near detector is
shown in Figure~\ref{K2Kbeam}. Figure~\ref{NUMI} shows the spectrum for
the NUMI beam
from the 120 GeV main injector at Fermilab, used for MINOS~\cite{michae},
in three different possible configurations.

\par
\begin{figure}[hbtp!]
\begin{center}
\epsfig{figure=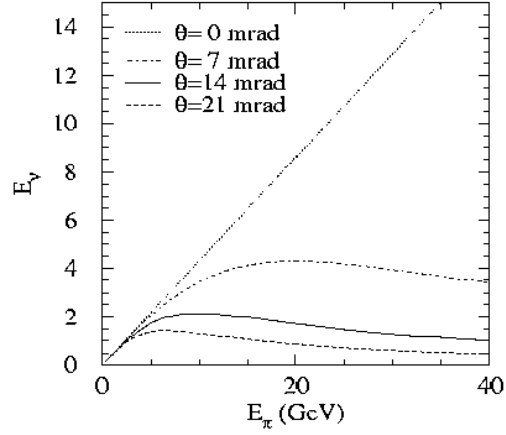,width=8.0cm}
\caption { The off axis beam: neutrino energy as a function
of the pion energy for neutrinos produced at an angle $\theta$
relative to the pion beam direction,from reference~\cite{novaprop}. } 
\label{OffAxis}
\end{center}
\end{figure}

 Off axis beams have also been designed to meet the need of low energy
beams of well defined energy.
They were first proposed by the E889~\cite{bnllbo} collaboration in 1995.
If neutrinos are observed at an angle with respect to
the incoming proton beam,
thanks to the kinematical characteristics
of the two body decay, the neutrino energy
becomes almost  independent from the pion
energy
$$ E_{\nu}=0.43 \cdot E_{\pi}/(1+\gamma_{\pi}^{2}\cdot\theta_{\pi\nu}^{2})$$
with $\gamma_{\pi}=E_{\pi}/m_{\pi}$.
Figure~\ref{OffAxis}
shows the neutrino energy as a function of the pion
energy for different angles.
Detecting the neutrinos off axis has the  advantage of giving a relatively well defined
momentum and of 
cutting the high energy part
of the spectrum, see for example Figure~\ref{t2kbeam}. 
 \par  A completely different approach has been used in the production
of anti-neutrinos for the LSND experiment~\cite{lsnd}.
Low energy protons (0.8~GeV) interacting in 
an absorber produce
low energy pions. The decay $\pi^{+} \rightarrow
\mu^{+}+\numu$ is followed by the~
$\mu^{+} \rightarrow e^{+}+\numub+\nue$ one.
$\pi^{-}$  are absorbed when they stop, a small fraction can decay 
in flight, in this case  their decay muons
come at rest and then can  absorbed or decay. An isotropic source of neutrinos
 is produced, 
mainly $\numu$, $\numub$, $\nue$ and a small fraction of
$\nueb$ coming from the decay $\mu^{-}\rightarrow
e^{-}+\numu+\nueb$.  
These $\nueb$ will be the main source of background in the 
search of the oscillation $\numub \rightarrow \nueb$.   

\section{Neutrino interactions}
\setlength{\parindent}{0.0cm}

\par This section will be devoted to the neutrino interactions
 that play a key role in
the oscillation experiments. When the oscillations are revealed by
the presence of a certain flavor in the final state, only charged
current interactions (CC) are relevant. Neutral current interactions (NC) are
used when the total flux of neutrinos, regardless of their flavor,
is measured.
 \bigskip
\subsection{Neutrino-nucleon scattering}
\par \hskip 0.5cm \underline{a)   Energies
E($\nu$)$\approx$ 1-10 MeV }
(solar and reactor neutrinos)

 At these  energies 
$\nue$ and $\nueb$ can experience charged current
 reactions only by scattering on free, (quasi)-free nucleons
\par $ \nueb+p \rightarrow e^{+}+n$
\par  $\nue+n\rightarrow e^{-}+p$\\
The cross section of the first process has a threshold at 1.8 MeV, in fact
the positron kinetic energy is given by
  $$ T(e^{+})=E(\nueb) +M(p)-M(n)-M(e)=E(\nueb)-1.8~\mathrm{MeV} $$

\par \hskip 0.5cm \underline{b) Scattering at medium
energy   E($\nu$) $\simeq$ 1 GeV}  (atmospheric and accelerator neutrinos)
\par  Above the threshold for muon production the quasi elastic CC processes
start
\par $\overline{\numu}+p =\mu^{+} +n$
\par $ \numu +n =\mu^{-} +p $
\par  followed by    $\pi^0$ or charged $\pi$ production via resonances
 and by deep inelastic processes at higher thresholds.

\begin{figure}[htbp!]
\centerline{\epsfig{figure=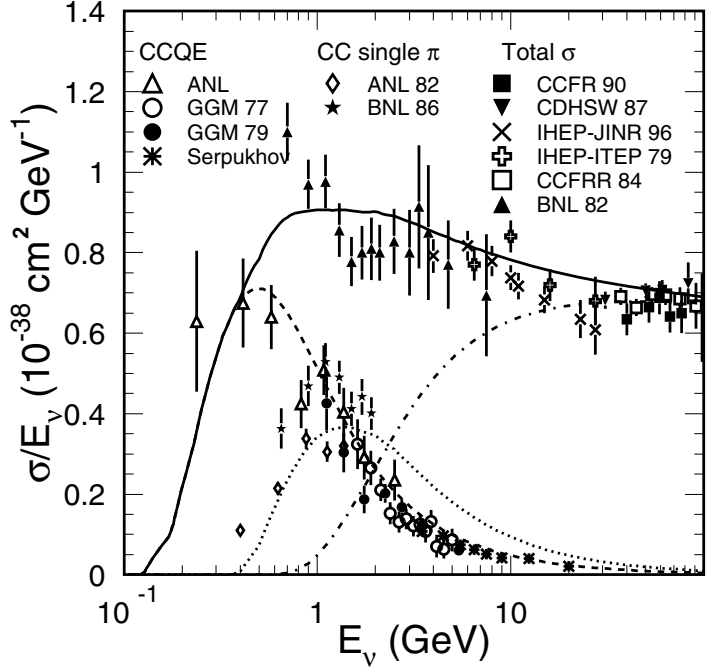,width=10.0cm}}
\caption{\textit{ Charged current total cross section from 
reference ~\cite{k2k}, \cprD{2006},  divided 
by $E_\nu$ for
  neutrino nucleon charged current interactions. The solid line shows
  the calculated total cross section. The dashed, dot and dash-dotted
  lines show the calculated quasi-elastic, single-meson and
  deep-inelastic scattering, respectively. The data points are
  taken from the following experiments:
  \mbox{({$\triangle$})ANL\protect\cite{Barish:1977qk}},
  \mbox{({$\bigcirc$})GGM77\protect\cite{Bonetti:1977cs}},
  \mbox{({$\bullet$})GGM79(a)\protect\cite{Ciampolillo:1979wp},(b)\protect\cite{Armenise:1979zg}},
  \mbox{({$\ast$})Serpukhov\protect\cite{Belikov:1985kg}},
  \mbox{({$\Diamond$})ANL82\protect\cite{Radecky:1982fn}},
  \mbox{({$\star$})BNL86\protect\cite{Kitagaki:1986ct}},
  \mbox{({$\blacksquare$})CCFR90\protect\cite{Auchincloss:1990tu}},
  \mbox{({$\blacktriangledown$})CDHSW87\protect\cite{Berge:1987zw}},
  \mbox{({$\times $})IHEP-JINR96\protect\cite{Anikeev:1996dj}},
  \mbox{({$+$})IHEP-ITEP79\protect\cite{Mukhin:1979bd}},
  \mbox{({$\Box$})CCFRR84\protect\cite{MacFarlane:1984ax}}, and
  \mbox{({$\blacktriangle$})BNL82\protect\cite{Baker:1982ty}}.  }}
\label{1gevcross}
\end{figure}
\par Figure~\ref{1gevcross}, from reference~\cite{k2k}, shows experimental 
measurements
of neutrino cross section together with the calculated value as a function of
the neutrino energy.

\par \hskip 0.5cm \underline{ c) High energy E($\nu$) $\gg$ 1~GeV}
(accelerator neutrinos)
 \par The deep inelastic scattering on quarks dominates at high energy. Cross
sections for $\nu_e$ and $\nu_{\mu}$ are
 
\par  $\sigma(\nu)=0.67~10^{-38}cm^2E_\nu /\mathrm{GeV}$, for neutrinos
\par  $\sigma(\overline\nu)=0.34~10^{-38}cm^2E_\nu /\mathrm{GeV}$, for anti-neutrinos.
\par For $\nu_{\tau}$ the high mass of the $\tau$ lepton
modifies the threshold of the various processes and changes
also the cross section at higher energies.
The linear growth with E$_\nu$ of the cross section continues until the
 effect of the 
propagator becomes important. 
\bigskip
\subsection{Neutrino-electron  scattering}\label{nuescatt}

\par
\begin{figure}
\centerline{\psfig{figure=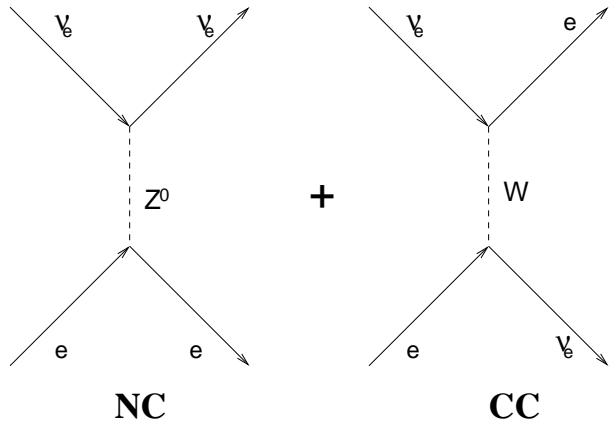,width=8.0cm}}
\caption {neutrino electron scattering} 
\label{feyn1}
\end{figure}

 The scattering of neutrinos on electrons is a 
purely weak process which is different for
\nue and other neutrinos. In fact both CC and NC
contribute to the \nue cross section while for \numu and \nutau only 
NC processes are possible (see Figure \ref{feyn1}). Again the cross sections
depend linearly upon E$_\nu$:
\par $$\sigma(\nue) =0.93~10^{-41}cm^2 E_\nu /\mathrm{GeV}$$
\par $$\sigma(\numu,\nutau) =0.16~10^{-41}cm^2 E_\nu /\mathrm{GeV}$$
\par and the ratio of the cross sections is 
$$\frac{\sigma(\nue)}{\sigma(\numu ~or~
\nutau)}\approx ~6$$
\par The following characteristics of $\nu$ scattering on electrons must 
be
noted: due to the small mass of the electron
\par a) cross sections of $\nu$ on electron at high energies
are smaller by a factor $\simeq10^{-3}$  compared to
 cross sections on nucleons. In fact cross sections are proportional to the 
mass of 
the scattering particle.
\par b) the electron will be emitted in the forward direction. The scattering
angle  
$\theta$ of the electron in the laboratory system is such that
$\theta^{2}\le 2m_{e}/E$, where 
E is the energy of the electron.  
\par
\begin{figure}[hbtp!]
\epsfig{figure=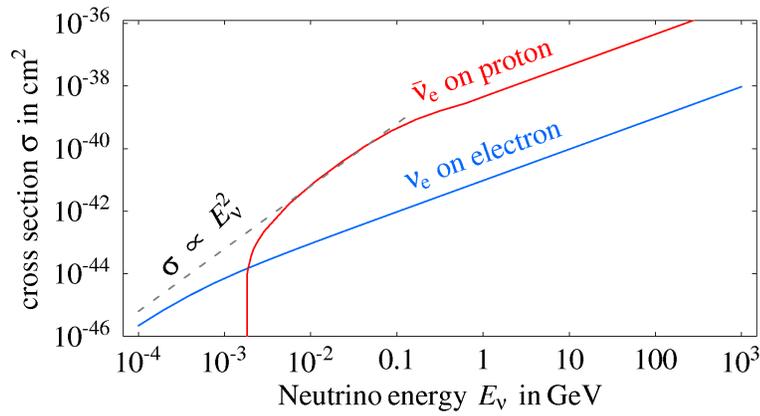,width=12.0cm} 
\caption
{Neutrino cross sections vs energy for scattering on nucleon or
electron, from reference~\cite{vissani}.}
\label{xs2} 
\end{figure}

 Figure \ref{xs2} compares the $\nueb$  charged current cross section for 
 scattering on protons with the  $\nue$ total cross section
on electrons. For  E$_\nu$ smaller than the mass m of
the target the cross section is proportional to  E$_\nu^{2}$, while for  
E$_\nu$
larger than m the   cross section is proportional to the product
$ m E_\nu$.

\subsection{Neutrino-nucleus  scattering}

\par At low energies the relevant reactions, exploited by radiochemical 
experiments, are

  $$ \nue + A(Z,N) \rightarrow e^{-}+A(Z+1,N) $$
  $$ \nueb + A(Z,N) \rightarrow e^{+}+A(Z-1,N) $$
\par  The  final nucleus is unstable and decays by electron capture.
In the rearrangement of atomic electrons that follows electron capture,
a photon or Auger electron is emitted.
\par Cross sections for these processes can be found 
in reference~\cite{bahcall:1996qv}.

\section{Experimental results}\label{expres}
This section summarizes the results obtained in the neutrino oscillation field
using neutrinos both from natural and artificial sources.

\subsection{Solar neutrinos}
\vskip 1.cm

\par The pioneering experiment of R. Davis did start the 'solar neutrino
puzzle': solar neutrinos observed on the earth are a fraction of those predicted
by the Solar Standard Model (SSM). The first results were published by Davis
in 1968~\cite{davis1} but only in 2002 the dilemma "problem with the 
neutrino" or "problem with
SSM" was solved by the SNO results. Indeed Davis had observed neutrino
oscillations. In the following, we will give a brief account of the different
experiments dedicated to the detection of solar neutrinos, which are divided into
two categories: radiochemical experiments and real time experiments.

\subsubsection{Radiochemical experiments:}
\par
In radiochemical experiment the \nue from the sun interact with a nucleus
via the reaction
 $$ \nue + A1(Z,A) \rightarrow e^{-}+A2(Z+1,A) $$
where the transition A1 to A2 leads to an unstable nucleus. The rate of
the
reaction is measured by counting the number of A2 nuclei, detected
via their decay.
\par The threshold of the previous reaction fixes
the minimum energy of the solar neutrinos that can be detected.

\paragraph{The Chlorine experiment:}
\par  Following a suggestion of B.Pontecorvo, R. Davis started a
neutrino experiment
in the Homestake Gold mine  in South Dakota, at a depth of 4800 meter 
water equivalent (MWE).
 After a test experiment
performed in 1964~\cite{davis}
showing that large underground experiment were feasible, Davis and
collaborators proceeded to build a large container filled with
100000 gallons of tetrachloroetilene. The observed  reaction was
\par $$  \nue +Cl^{37} \rightarrow Ar^{37}+e^{-} $$
\par The cross section, integrated on the B$^8$ spectrum, of this process  
has been computed to be   (1.14$\pm$0.037)$\times 10^{-42}$
cm$^{2}$~\cite{bahcall:1996qv}. In Davis's 
experiment
the rate of interactions is not measured directly.
 Using physical and 
chemical methods the amount of $Ar^{37}$ was extracted from the target 
material. The  $Ar^{37}$  is unstable,
 the counting was performed by observing the Auger electron or photon 
emitted in the decay. Since the $Ar^{37}$ decay half-time is 35 days,
the extraction had to be performed periodically,  1  run
each 2 months.
\par The  first indication of a neutrino deficit was given in 
1968~\cite{davis1}. Bahcall in the same year~\cite{baha68} did  show that 
these results 
were incompatible with his calculations  on the solar model.
 \par The 
results,  for runs 
taken from 1970 to 1995, give 
 for  the solar neutrino capture rate the value  
2.56$\pm$ 0.16 stat $\pm$ 0.16 syst SNU~\cite {cleveland}
(1 SNU=$10^{-36}$neutrino captures/(atom sec)).
 This result corresponds to a reduction
by a factor $\simeq$3 (Table~\ref{solarrates}) with respect to
the prediction
of the SSM, and represents the first evidence for
neutrino oscillation.
\par Note that since the threshold of the reaction on $Cl^{37}$
is 0.813 MeV, this experiment is marginally
sensitive to the  $Be^{7}$  solar neutrinos, and mainly to the $B^{8}$
neutrinos.

\paragraph{Gallium experiments:}
\par Three   radiochemical experiments have studied solar
neutrinos using the scattering on Gallium:
$$ \nue +Ga^{71} \rightarrow Ge^{71}+e^{-} $$
\par Since the threshold of this reaction is 0.233 MeV Gallium   experiments
are sensitive to the neutrinos from the primary pp reaction in the sun.
The three Gallium experiments are GALLEX~\cite{Gallex} and its continuation 
GNO~\cite{gno}, at the
Gran Sasso Laboratory in Italy, and SAGE~\cite{sage}, at the Baksan 
Neutrino Observatory in Russia. To ensure the correctness of the 
results 
all these
detectors have been calibrated with strong neutrino sources. 
\par
\paragraph{GALLEX and GNO:}

\par
\par The GALLEX experiment~\cite{Gallex}
 started data taking in 1991 at the Gran Sasso Laboratory,
at a depth of 3500 MWE. It used a large tank to contain 30 tons of
gallium
dissolved in 100 tons aqueous gallium chlorine solution. The target material was
periodically extracted  to count the $Ge^{71}$ produced in the neutrino
interaction.
The amount of $Ge^{71}$ was measured  by detecting its decay products, X
rays or Auger electrons following electron capture,
with proportional counters. Data were taken from 1991 to 1997.
\par
\begin{figure} [h]
\begin{center}
\epsfig{figure=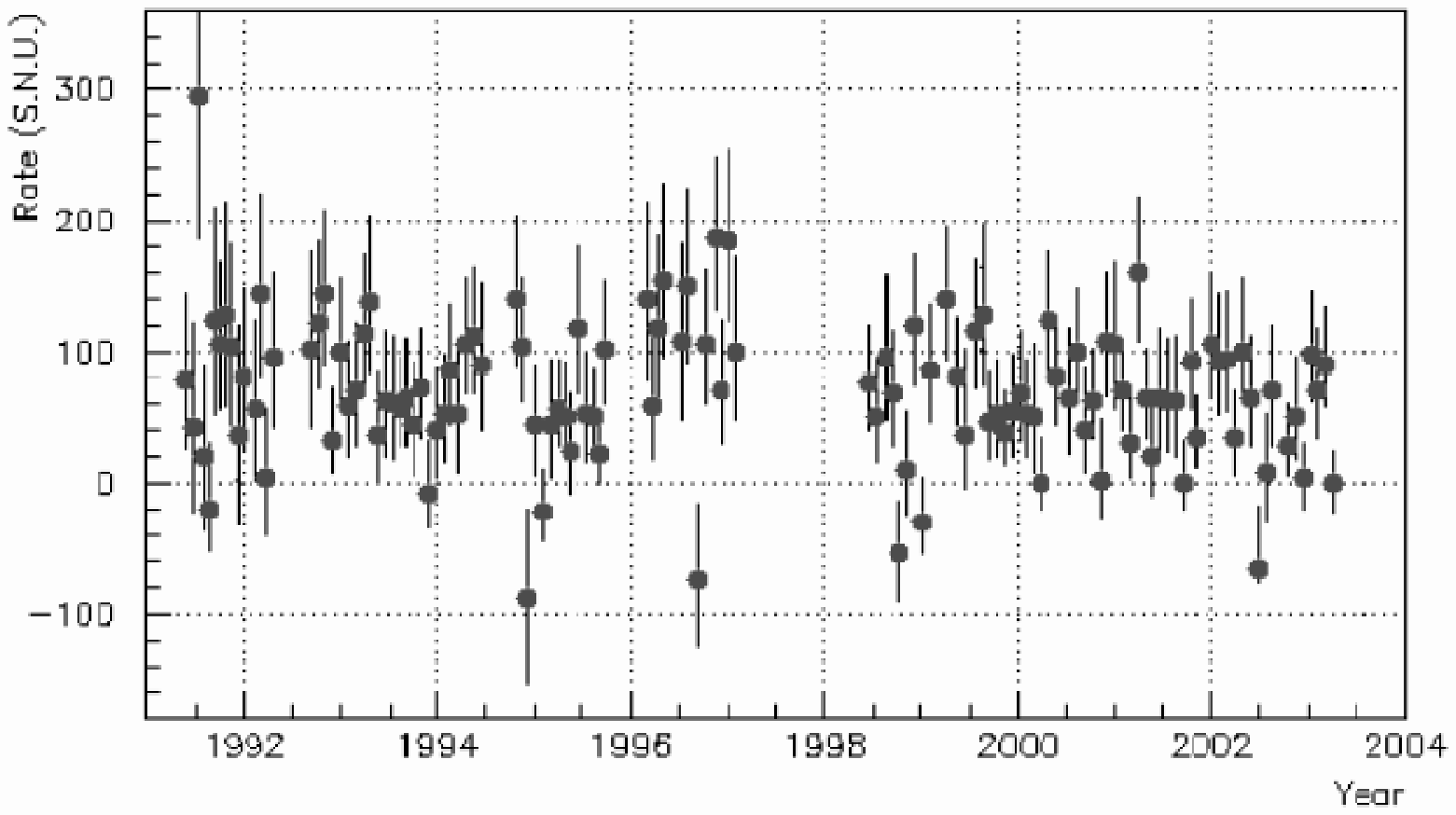,width=15.0cm}
\caption {Results from GALLEX and GNO data 
taking,from reference ~\cite{gno}, \cprC{2005}.}\label{galfig}
\end{center}
\end{figure}

  GALLEX  was followed by  the
GNO experiment~\cite{gno}, which took data from 1998 to  2003.
GALLEX+GNO performed separate measurements of the solar neutrino flux for the 123 runs taken
between 1991 and 2003. The time behavior of these  measurements is  shown
in Figure~\ref{galfig}, results are summarized in  Table~\ref{radio}.

\vskip 1cm
\paragraph{SAGE Experiment:}
\par The SAGE experiment~\cite{sage} is located in the Baksan neutrino 
observatory
4700~MEW under the sea level. An average mass of 45.6     tons of metallic
Gallium $Ga^{71}$ was used.

   \par In the period 1990-2003 107 neutrino runs were taken 
and the result of their
analysis  is shown in Table~\ref{radio}. 
\begin{table}[htbp!]
\centering
\begin{tabular}{|c|c|}
\hline
GALLEX~\cite{Gallex} &77.5 $\pm$ 6.2 $^{+4.3}_{-4.7}$ \\
GNO~\cite{gno}&62.9 $^{+5.5}_{-5.3}$$ \pm$2.5 \\
GNO+GALLEX~\cite{gno} &69.3 $\pm$4.1 \\
SAGE~\cite{sage} &70.8 $^{+5.3}_{-5.2}$ $^{+3.7}_{-3.2}$\\
\hline
\end{tabular}
\caption{Gallium  experiments results, capture rates expressed in SNU }
\label{radio}
\end{table}

\par The weighted average of all Gallium results is~\cite{gavrin}
$$ Capture~ Rate =67.6\pm 3.71~~ SNU$$
which compared with a prediction of 128 SNU gives a Data/SSM ration of 0.53 (see Table~\ref{solarrates}).
\vskip 1.cm

\subsubsection{Real time experiments:}
\par Solar neutrinos have been studied in real time using huge water
 (Kamiokande~\cite{kamio1} and Super-Kamiokande (SK)~\cite{fukuda96})
 or heavy water (SNO~\cite{SNO})  containers  
surrounded by
 a very large number of photomultipliers used to detect the Cherenkov
light emitted by fast particles produced in neutrino interactions.
The Cherenkov threshold in water is
$\beta$=0.75. 
The use of this technique to detect solar neutrinos as
 been pioneered by the Kamiokande
experiment and by its follow-up Super-Kamiokande where the only reaction 
allowed
for neutrinos of $E\simeq$ MeV   is the scattering on electrons.
\par Two relevant characteristics of this process are 
\par a)  in the scattering on electrons take part not only  $\nue$ but
with a smaller cross section ( $\simeq$ 1/6) also $\numu$ and 
$\nutau$;
\par b) the direction of scattered electrons is tightly connected with the 
direction of the incoming neutrino.
\par
\begin{figure} [h]
\centerline{\psfig{figure=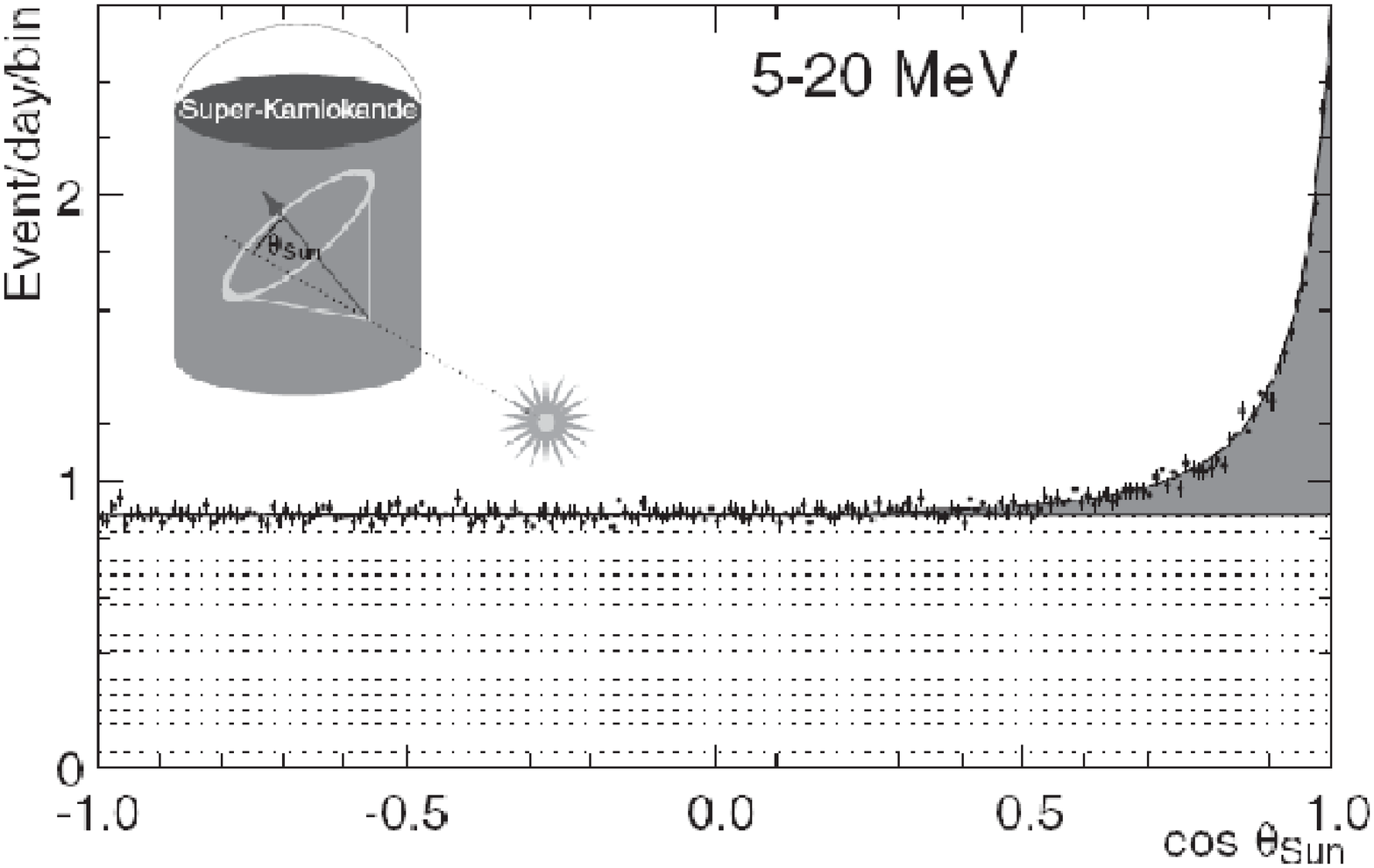,width=10.cm}}
\caption {Angular distribution of observed electrons in 
Super-Kamiokande, from reference \cite{hosaka06}, \cprD{2006}.  }
\label{angular}
\end{figure}

 Figure \ref{angular}  shows the angular distribution of observed electrons.

 The  SNO experiment (Sudbury Neutrino Observatory)~\cite{SNO} 
 has allowed also the study
of neutral current interactions   and charged current interactions
 using the quasi-free neutrons of deuterium.
\paragraph{Kamiokande and Super-Kamiokande:}
\label{SKC}
Kamiokande and Super-Kamiokande base their study
 on the detection of the neutrino scattering on electrons
$\nu + e \rightarrow \nu +e$.
\par The Kamiokande~\cite{kamio1} detector was originally built mainly 
 to search for proton decay, it did
start operation in 1983.
\par  The detector consisted of  a cylinder 16 m high, with  16.5 m
diameter  containing 3000 tons
of
pure water. The  surface was equipped with 1000 photomultipliers of 50 cm
diameter.
\par
\begin{figure}[hbtp!]
\begin{center}
\epsfig{figure=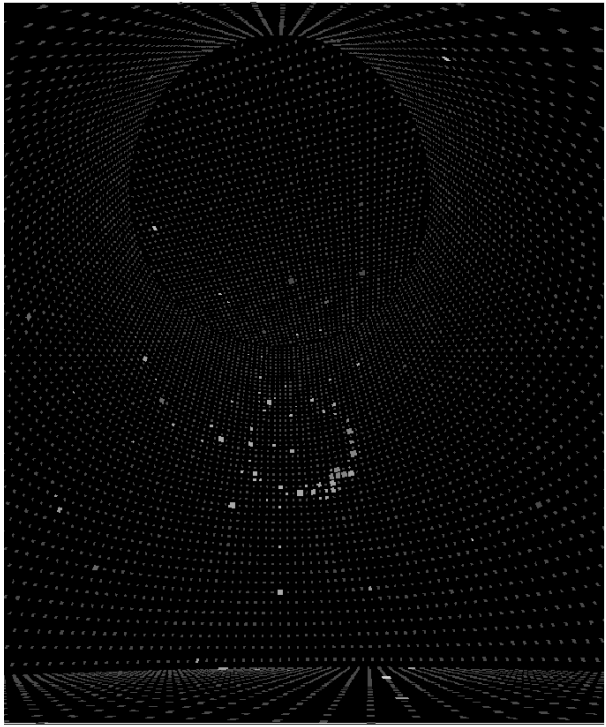,width=8.cm} \caption
{Cherenkov ring of a  few  MeV electron in the Super-Kamiokande  detector, from
http://www.ps.uci.edu/$\sim$tomba/sk/tscan/pictures.html}
\label{ring}
\end{center}
\end{figure}

 In water counters electrons are recognized by the characteristic
Cherenkov ring. Figure \ref{ring}
shows a few  MeV electron ring.

\par The energy threshold  to reject background was fixed to 9.3 MeV
and then lowered to 7 MeV  during  data taking. This threshold
made the experiment 
sensitive only to $B^8$ neutrinos and 800 events were collected.
\par The result of the experiment was~\cite{fukuda96}
 $$ \Phi(\nu_e)=(2.80 \pm 0.19 \pm0.33)\times 10^6 cm^{-2} sec^{-1}.$$
\par The ratio Data/SSM=0.55$\pm 0.04 \pm 0.07 $   confirmed the solar
neutrino deficit.
\par The Kamiokande detector  was followed by the  Super-Kamiokande one.
\par
\begin{figure}[hbtp!]
\begin{center}
\epsfig{figure=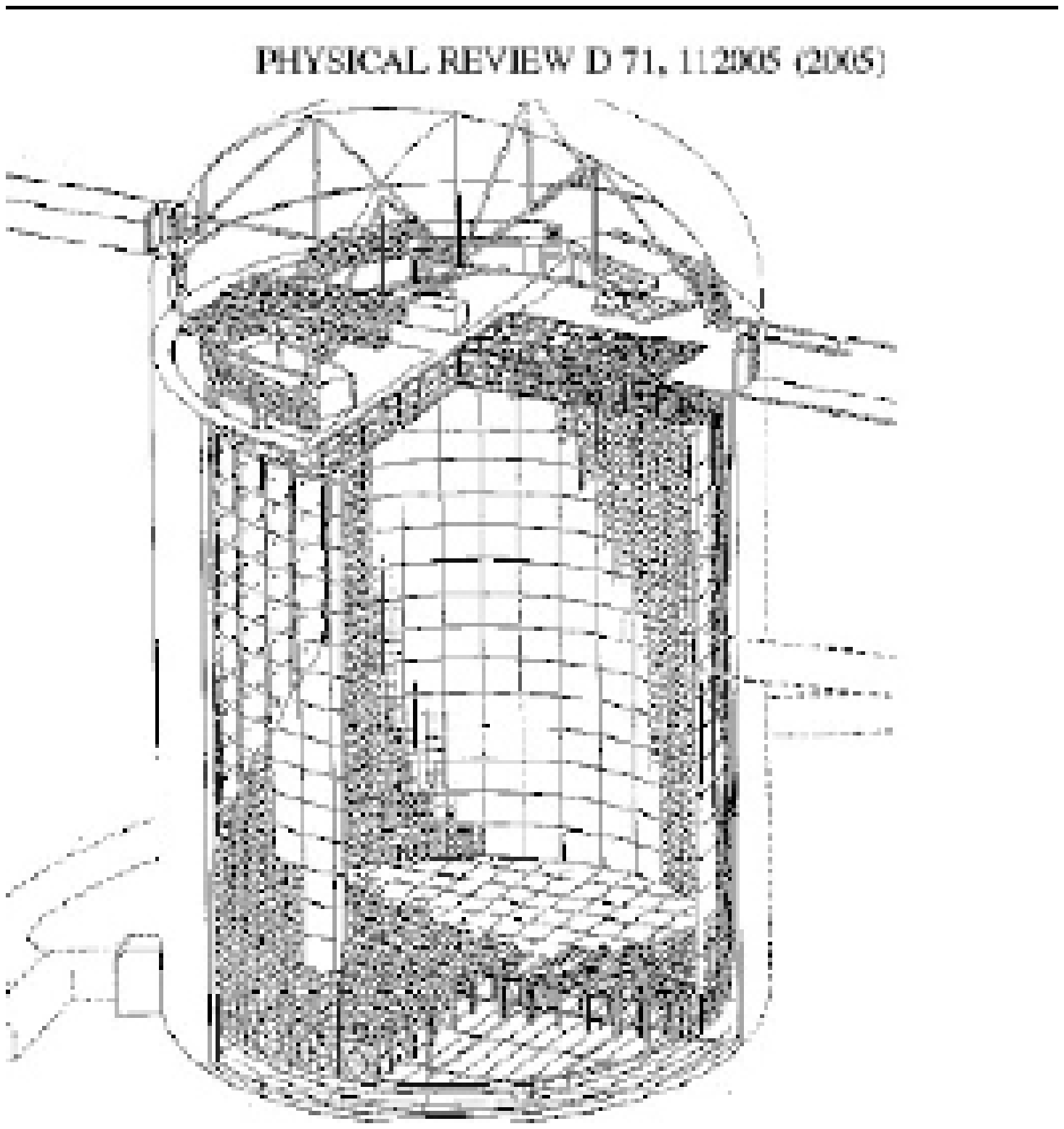,width=8.cm} \caption
{Layout of the Super-Kamiokande detector, from reference ~\cite{SK:atmos}, \cprD{2005}.}
\label{SKscheme}
\end{center}
\end{figure}

 A schematic drawing of the detector is shown in Figure~\ref{SKscheme}. 

\par Construction started in 1991 and was completed in 1995. Data taking
did start in 1996.\par The dimensions of the tank are 39.3~m diameter,
41.4~m height. The water mass is 50 kton, and the fiducial one is 22 kton. 
The surface of the inner part is covered by
11000 photomultipliers (PMs) covering 40$\%$ of his surface. 
The outer part is equipped with 1800 PMs and 
was used to veto entering charged particles.
\par
In   November 2001 an accident destroyed a large part of the PMs. The
detector was reconstructed  and at the end of 2002 the second phase of the
experiment, SK-2, started although with  smaller coverage (19\%) and was
concluded in 2005. Then the reconstruction of the detector was initiated and
concluded in 2006, SK-3. 
 
Data
taken from 1996 to 2001 constitute
phase 1  of the experiment.
22400 solar events have been  collected in this phase  
in 1496 days~\cite{hosaka06}, with a threshold 
of 5 MeV   (6 MeV in the first 280 days); the
corresponding interaction rate was
   $$ \Phi(\nu)=(2.35 \pm 0.02
\pm 0.08)\times 10^6 cm^{-2} sec^{-1}$$

\par The measured ratio  Data/SSM is $0.47 \pm 0.04 \pm 0014$.
\vskip 0.5cm

 \par Results of the analysis of phase 2
 will be
given in reference~\cite{Cravens}. With  the full PM coverage 
 restored (SK-3) data are being collected starting in January 2007.
Preliminary  results are presented in reference~\cite{phase3sk}

\vskip 0.5cm

\paragraph{SNO experiment:}
\label{snoexp}
\par The SNO~\cite{SNO}, Sudbury 
 Neutrino Observatory, is a 1000 tons heavy water
Cherenkov detector located 
2 km underground in INCO's Creighton mine near Sudbury, Ontario, 
Canada.
Three reactions can be observed in deuterium

\par 1)  $\nue+d\rightarrow p+p+e^-$ charged current interaction accessible only
to $\nue$
\par 2)$\nu_{x}+d\rightarrow p+n+\nu_{x}$   neutral current interaction accessible
to all neutrinos
\par 3) $\nu_{x}+e\rightarrow \nu_{x}+e$ accessible to $\nue$ and, with
smaller cross section, to $\numu$ and $\nutau$.

\par Reactions 1 and 3 are observed via the detection of the Cherenkov
light emitted by the electrons.
Reaction 2 is detected via the observation of the neutron in the
final state. This feature of SNO is extremely relevant since it allows flavor 
independent measurement of neutrino fluxes from the Sun, thus measuring the total neutrino flux independently from their oscillations.
This has been accomplished in two  concluded phases
\begin{description}
\item{Phase 1: 1999-2001}
\par The neutron has been detected via the 
observation of the Cherenkov light produced by the electron following the
reaction n+d$\rightarrow$T+$\gamma$(6.5 MeV). 
The observed events in phase 1 are~\cite{SNO}: 
\par \hspace{1.cm}   $1833 \pm 174$ ~~$\nue$ charged current events
\par \hspace{1.cm}~   $~273\pm 27$ ~~ electron scattering events
\par \hspace{1.cm}   $717\pm 177$~  neutral current events
\par Taking into account cross sections and efficiencies one obtains for 
the B$^8$ neutrino fluxes in 
units of $10^{6}cm^{-2}sec^{-1}$
\par \hspace{1cm} $\phi(CC)=1.76^{+0.06}_{-0.05} stat ^{+0.09}_{-0.09} 
syst$
\par \hspace{1cm} $\phi(ES)=2.39^{+0.24}_{-0.23} stat 
^{+0.12}_{-0.12}syst$
\par \hspace{1cm} $\phi(NC)=5.09^{+0.44}_{-0.43} stat ^{+0.46}_{-0.43}syst$
\par \hspace{1cm} 

Given that:

$\phi(\nue)=\phi(CC)$

$\phi(\numu,\nutau)=\phi(NC)-\phi(\nue)=\phi(NC)-\phi(CC)$

the following neutrino fluxes are obtained:

$\phi(\nue)=1.76^{+0.05}_{-0.05}stat^{+0.09}_{-0.09}syst$

$\phi(\numu,\nutau)=3.41^{+0.45}_{-0.45}stat^{+0.48}_{-0.45}syst$

\begin{figure}[hbtp!]
\begin{center}
	\epsfig{figure=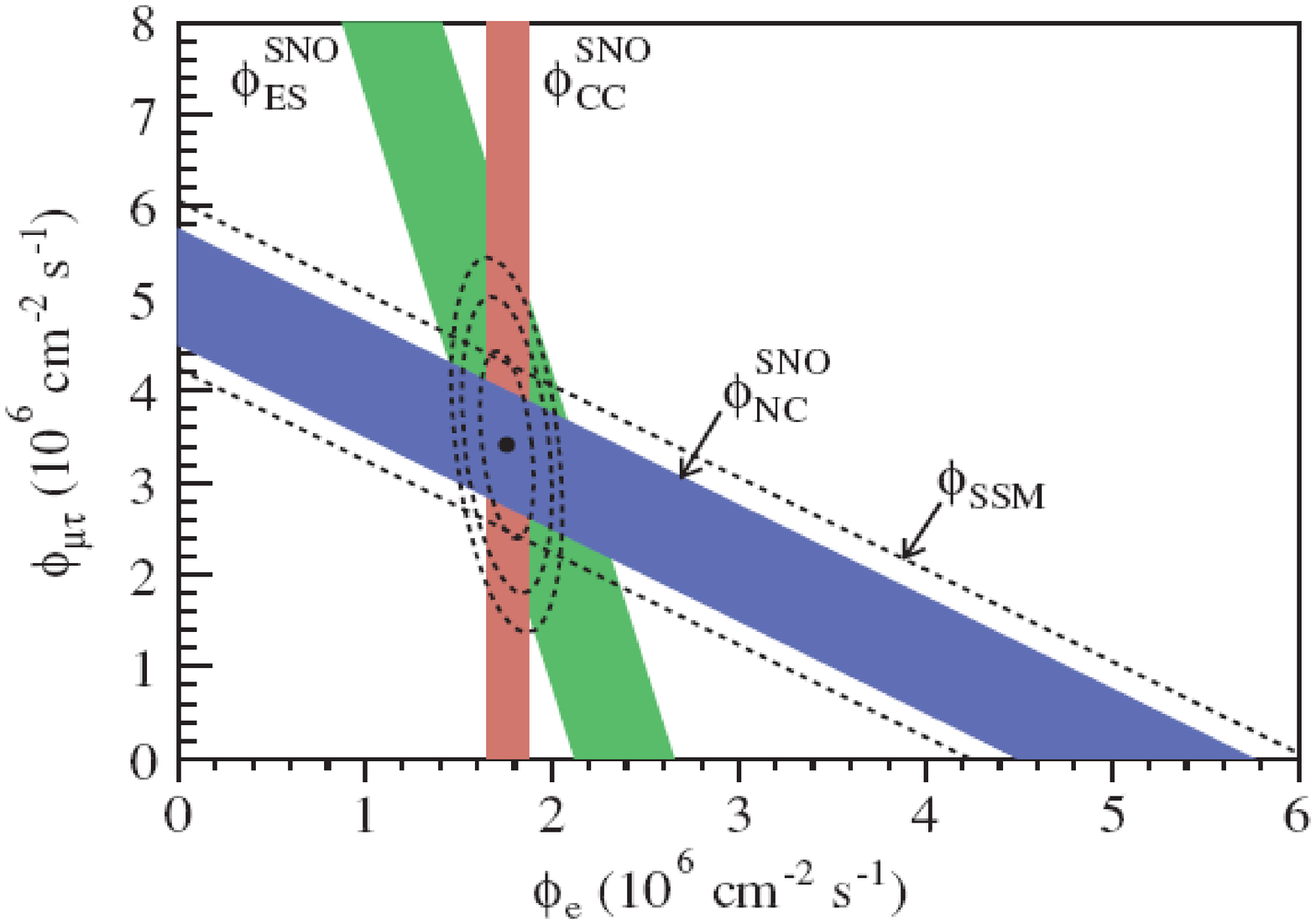,width=12.cm}
	\caption{SNO results for the various channels,from 
reference~\cite{SNOs}, \cprD{2005}.}
	\label{fig:snofig}
\end{center}
\end{figure}

\par These results are  graphically presented in Figure~\ref{fig:snofig}.

 From the above results  we can conclude that
\begin{itemize}
\item $R_{ee}=\Phi(CC)/\Phi(NC)=0.34 \pm0.023^{+0.029}_{-0.031}$.
2/3 of $\nu_e$
neutrinos have changed their flavor and
arrived on earth as $\nu_{\mu}$ and/or $\nu_\tau$

\item the flux of neutrinos of all flavors  (NC flux) is
in good
agreement
with the SSM predictions of (5.05$\pm$0.5)$\times 10^{6}~cm^{-2}sec^{-1}$~\cite{bahcall:2001}

\end{itemize}
\item{Phase 2: 2001-2002}
Two tons of NaCl have been added to the heavy water
increasing the efficiency of the neutron capture cross section.
In the Cl capture process multiple gamma rays are produced,
thus neutral current events can be statistically separated from
processes 1 and 3, where single electrons are produced.
Results for this phase are given in references~\cite{saltpa} and~\cite{SNOs}.
\item{Phase 3: 2003-2006}
 \par Neutron  detectors have been added, and the analysis is in 
progress.
\end{description}
\par The collaboration  
has decided to stop the experiment at the end of 2006, since the statistical
accuracy has reached the systematic one.
\par A new international laboratory is being constructed, SNOLAB, as an
extension of SNO and already a variety of experiments has been
proposed~\cite{snolab}.

\paragraph{Borexino experiment:}
\par In  2007 the Borexino experiment has published the first result,
a direct measurement of Be$^7$ solar neutrinos~\cite{bore}.
The  low  threshold of the experiment, 250 keV,
has allowed  to measure the Be$^7$ flux for the first time in
real time. 
The experiment has been build at the LNGS and detects
$\nu_e$ via  the electron scattering process. The detector is a
sphere of 300 tons liquid scintillator (100 ton fiducial mass)
viewed by 2200 photomultipliers.
The low threshold has been obtained  after many years of
R$\&$D. 
\par The measurement of neutrinos below 1 MeV  allows to
study the region between the vacuum  and MSW regimes. The best value for
the counting rate is $$ 47 \pm 7_{stat} \pm 12_{syst}~ counts/100 ton 
/day$$
in good agreement with $49\pm 4$ predicted by the solar model 
with the solar neutrino oscillation parameters derived from previous 
experiments 
(the so called Large Mixing Angle solution). 
The rate expected with no
oscillation is $75\pm 4$ counts/100~ton~/day.

\par The aim of the experiment is to measure the Be$^7$
 flux at 5$\%$ level.

\subsubsection{Summary of solar neutrino experimental results:}

The results presented above are summarized in Table~\ref{solarrates} from
which the following conclusions can be drawn:
\begin{table}[htpb!]
\begin{small}
\begin{tabular}{|c|c|c|c|c|c|} \hline
reaction & experiment & results & SSM (*) & data/SSM$\approx$ & notes \\ \hline
Cl$^{37}$$\rightarrow$ Ar$^{37}$  &Homestake~\cite{cleveland}& 2.56$\pm0.16\pm0.16$~SNU
 &7.6$^{+1.3}_{-1.1}$&0.34&  ~~\\
Ga$^{71}$ $\rightarrow$Ge$^{71}$  &Gallium~\cite{gavrin}& 
 67.6 $^{+3.7}_{-3.7}$~SNU&128$^{+9.}_{-7.}$ &0.53 & (1)  \\
$\nu_{x}+e\rightarrow\nu_{x}+e$ &Kamiokande~\cite{fukuda96}&2.8$^{+ 
0.19}_{-0.19}$$^{+0.33}_{-0.33} 10^{6}cm^{-2}sec^{-1}$ & 5.05$^{+1.}_{.8} $ &0.55& (2)\\
$\nu_{x}+e\rightarrow\nu_{x} +e$ &SK~\cite{hosaka06}&  
2.35$^{+0.02.}_{-0.08}10^{6}cm^{-2}sec^{-1}$
&5.05$^{+1.}_{-.8}$&0.47 & (2)\\
$\nu_{x}+e\rightarrow\nu_{x}+e$& SNO~\cite{SNO}& 2.39$^{+0.24}_{-0.23}$$^{+0.12}_{-0.12} 10^{6}cm^{-2}sec^{-1} $ &
5.05$^{+1.}_{-.8}$&0.47&(2) \\
$\nue+d\rightarrow p+p+e^{-}$& SNO~\cite{SNO}& 1.76 
$^{+.06}_{-.05}$$^{+0.09}_{-0.09} 10^{6}cm^{-2}sec{-1} $&5.05$^{+1.}_{-.8} $&0.35& (3)\\
$\nu_{x} +d \rightarrow \nu_{x}+p+n$& SNO~\cite{SNO}
 & 5.09$^{+.44}_{-.43}$$^{.46}_{-0.46} 10^{6}cm^{-2}sec^{-1}$
&5.05$^{+1.}_{-.8}$&1.& (4)\\
$\nu_{x}+e\rightarrow\nu_{x}+e$&Borexino~\cite{bore}&$47^{+7}_{-7}$$^{+12}_{-12}counts/day/100ton$  
&75$^{+4}_{-4}$&0.60&(2)\\ \hline
\end{tabular}
\end{small}
\caption{(*) from reference~\cite{bahcall:2001},
(1) average of the 3 experiments~\cite{gavrin}, (2) mainly $\nue$ elastic scattering,
(3) $\nue$ charged current interactions, (4) neutral current process.}
\label{solarrates}
\end{table}

\begin{itemize}
\item The flux ratio R = measured/SSM predictions is equal to 1 for the NC SNO
measurements. This is a convincing proof of the validity of the solar model predictions.

\item  All experiments  that are sensitive mainly to $\nue$
obtain a ratio R  smaller than 1.
\item The ratio R depends on the threshold of the 
experiment i.e. on the flux composition of
the observed events. The depression is dependent on the neutrino energy.

\end{itemize}

\subsubsection{Determination of the mixing matrix elements:}

\par For sin$\theta_{13}$=0 electron neutrinos are a mixture of $\nu_1$
and $\nu_2$ and so the oscillation can be studied in terms of
$\dmq_{12}$ and $\theta_{12}$. Solar neutrino data identify a
unique solution for the above parameters: the Large Mixing Angle solution
(LMA)~\cite{nakamura}. Solar matter effects largely determine this 
solution. The
matter mixing angle given in Section~\ref{msw} is computed using 
$\epsilon(x)=2\sqrt{2}G_{F}N_{e}(x)E/\dmq_{12}$, where 
$ N_{e}(x)$ is the electron 
density
 at position $x$ from the sun center. In the region identified by the LMA solution, accounting for the non constant solar density, 
the $\nu_e$ survival probability can be written as~\cite{foglifits}
 $$P_{ee}={{1}\over{2}}  +{{1}\over{2}} \cos2\theta_{m12}\cos2\theta_{12}$$
where $\cos2\theta_{m12}$ has been computed with the  electron  density 
at the center of the Sun.
\par For pp neutrinos 
$\cos2\theta_{m12}\simeq \cos2\theta_{12}$ and so 
$P_{ee})=1-{{1}\over{2}} \sin^{2}2\theta_{12}.$
\par For $^8$B neutrino energies  $\epsilon(x)\simeq 1$,
$\cos2\theta_{m12}\simeq -1$ and so $P_{ee}\simeq \sin^{2}2\theta_{12}$.
\par The  SNO results on the flux ratio of CC/NC= $R_{ee}=P_{ee}/1$ then give 
a direct 
measurement of $\sin^2\theta_{12}$. 

Flux differences between day and night (day-night effect), due to MSW effect 
inside the Earth, are expected to be small for the oscillation parameters 
of the 
LMA solution. No evidence for such effect has indeed been 
found by SK~\cite{smy} and SNO~\cite{SNOs}. Distortions of the energy spectra were also not 
observed by 
these 
experiments, as expected. 
 
\begin{figure}[hbtp!]

\begin{center}
\epsfig{figure=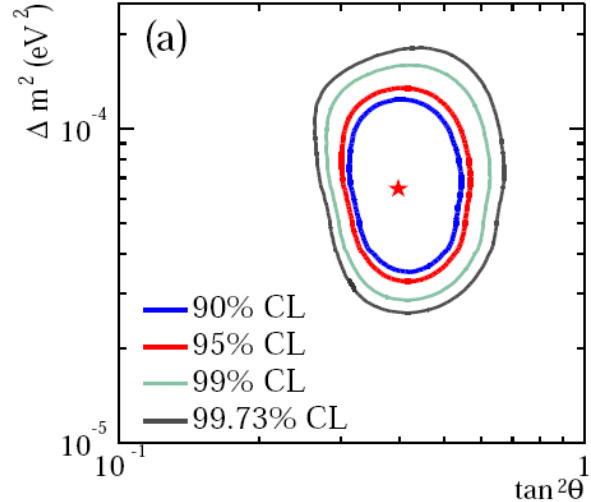,width=8.0cm}
\caption {Results of combined SNO,SK,CL,Ga results in the parameter plane,
from reference ~\cite{saltpa}, \cprD{20042004}, the central values for the parameters  are 
$\dmq_{12} =6.5\times 
10^{-5}$~eV$^2$
and tan$^{2}\theta_{12}=0.4 $.}

\label{solarfinal}
\end{center}
\end{figure}

\par  In conclusion the solar results are  
given in Figure~\ref{solarfinal}.  
\par The correctness of the LMA solution has been confirmed by the KamLAND
reactor neutrino experiment, as will be shown in Section~\ref{kamla}.

\subsection{ Reactor  neutrinos}
\label{reactor}
\par Reactor experiments are designed to detect \nueb via the reaction
 $$\nueb+p\rightarrow e^{+}+n$$
\par At short distances (see Section~\ref{oscil2}) the obtained
limits can be
interpreted in terms of
$\theta_{13}$ (CHOOZ results);
 at  large distances the KamLAND 
experiment results can
be interpreted in the two flavor mixing scheme  in terms  of the 1,2
mixing parameters, the solar ones.

\par A discussion of main characteristics  of experiments with reactor 
neutrinos is given 
in~\cite{bemporad}.
Detectors consist of a tank containing a  liquid scintillator
surrounded by photomultipliers.
The \nueb interactions are detected by a coincidence between the
prompt signal of the $e^{+}$ and a delayed signal from  gamma rays emitted
 in a capture process of the neutron  after its thermalization.
 The neutron receives negligible kinetic energy so the
E($\nueb$) is given by the relation
  $$ T(e^{+})=E(\nueb) +m(p)-m(n)-m(e)=E(\nueb)-1.8 \mathrm{MeV} $$
where T$(e^{+}$) is the kinetic energy of the positron. 
\par
\begin{figure}[h]
\centerline{\psfig{figure=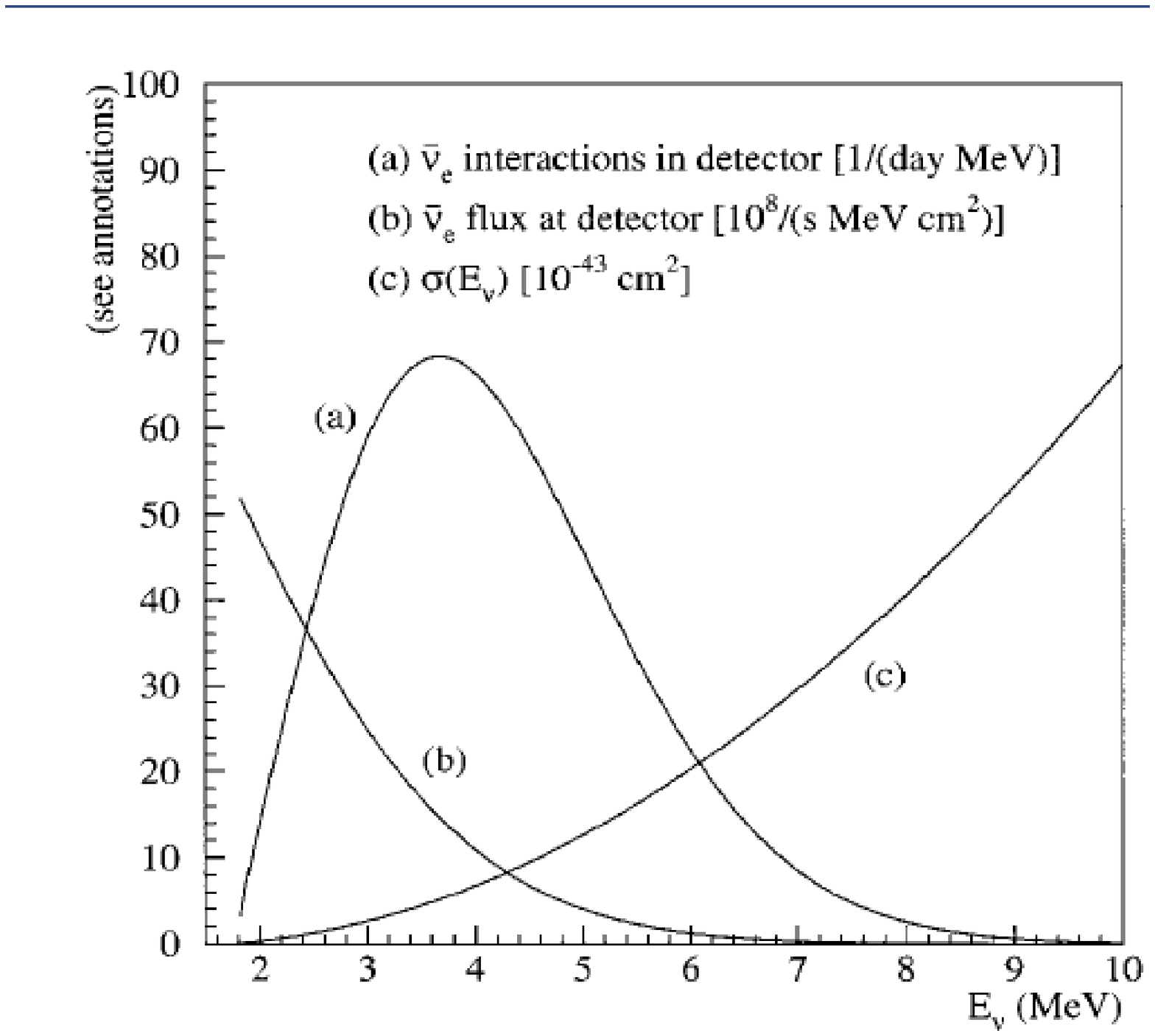,
width=8.0cm}}
\caption {curves a and b are referred to a 12t fiducial mass detector positioned
at 0.8~km from a 12~GW reactor, from reference ~\cite{bemporad}, \cprD{2002}.}
\label{vis}
\end{figure}

 From the above relation we see that the process has a threshold at
1.8~MeV.
The number of events collected depends on the mass of detector, on the flux
of \nueb and on the cross section for the process.
Figure \ref{vis}   shows the
anti-neutrino flux (b) cross section (c) and the interaction rate (a)
for a 12t detector at 0.8~km from a
reactor  with thermal power W=12GW. 
 \par In the last 20 years many experiments on \nueb from reactor  have
been made~\cite{kwon,karno,zacek,bugey}. The ratio 
$L/E$ of these experiments
was such that the minimum  $\dmq$ that could be reached was of
the order of   $10^{-2}~\mathrm{eV}^{2}$.
\par  Two recent experiments CHOOZ~\cite{apollonio}
 and KamLAND~\cite{kamland} have given
relevant results in  the oscillation field. 
\par  Results compatible with 
the CHOOZ ones have been obtained by the Palo Verde 
Experiment~\cite{palo}.

\subsubsection{CHOOZ experiment:}
\par
\begin{figure}[hbtp!]
\begin{center}
\epsfig{figure=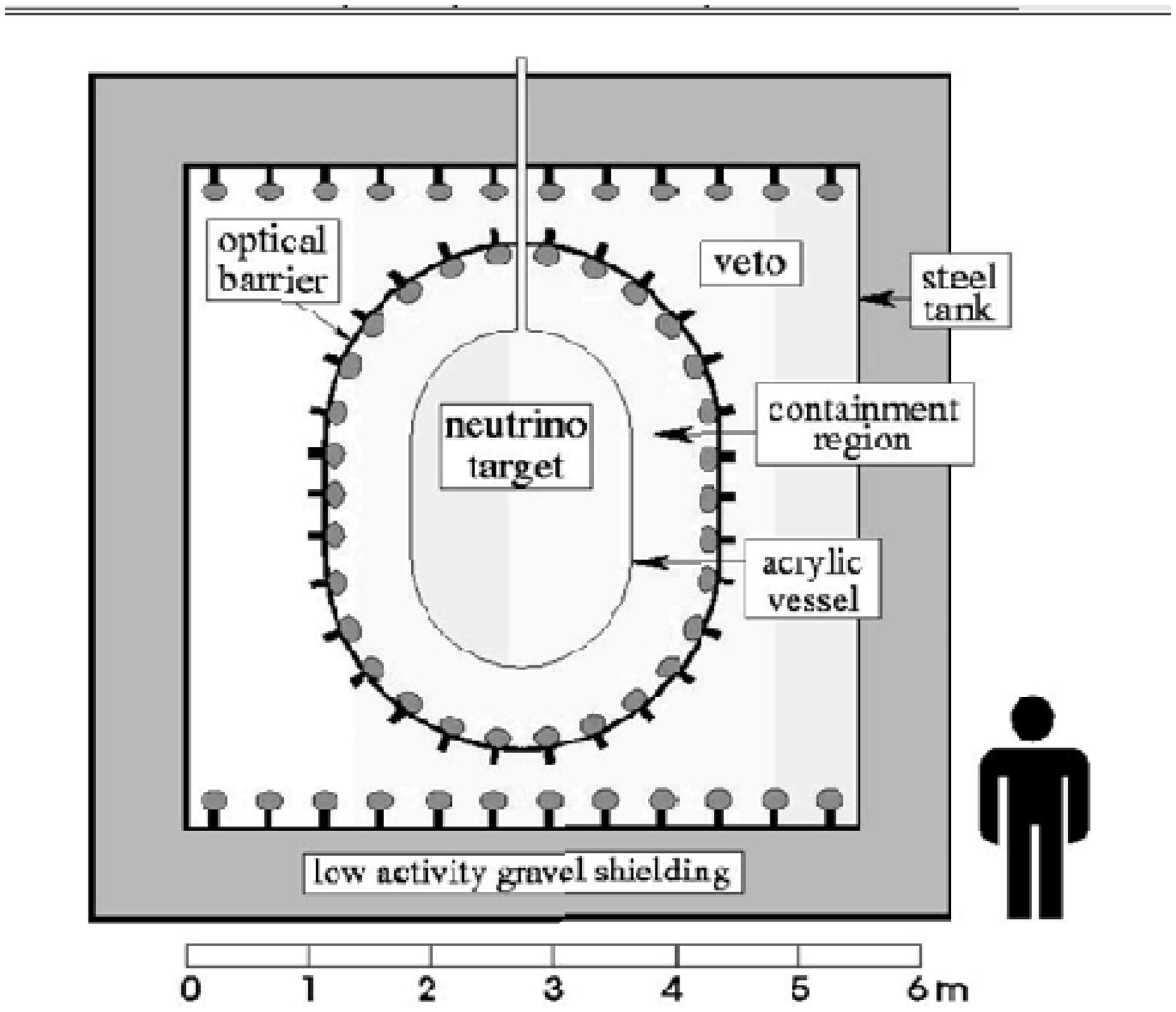,width=8.0cm}
\caption {The CHOOZ detector, from reference ~\cite{bemporad}, \cprD{2002}.}
\label{choos}
\end{center}
\end{figure}

 The experiment was located close to the nuclear power plant of CHOOZ
(north of France), a schematic drawing of the detector is shown in 
Figure~\ref{choos}.
\par The detector  used gadolinium loaded scintillator as neutrino
target . Gadolinium has high thermal neutron capture cross section
and releases about 8 MeV energy in the process.
\par The detector was  located at about 1 km from the neutrino source in an
underground laboratory to reduce the muon flux by about a factor 300
compared to the surface one.  Muons produce  neutrons by spallation
in the
 material surrounding the detector; these neutrons are one of main
sources of background.
 The detector consisted of a central region  
 filled with 5 tons of gadolinium loaded scintillator (0.09$\%$), an
intermediate region (107 tons) filled with undoped scintillator to contain 
the
electromagnetic energy produced by the neutron capture in gadolinium and an
external region still filled with scintillator, used for muon anti-coincidence.
\par Data were taken from March 97 to July 98 . The selection criteria
for $\nueb$ interactions were
\begin{itemize} \item positron energy $\le$ 8 MeV
\item gamma  energy released in the neutron capture$\le$ 12 MeV and $\ge$ 6 
MeV
\item interaction vertex distance from wall $\ge$ 30 cm
\item  distance electron-neutron $\le$ 100 cm
\item  neutron delay $\le$ 100 $\mu$sec
\item  neutron multiplicity =1
\end{itemize}

\begin{figure}[hbtp!]
\begin{center}
\epsfig{figure=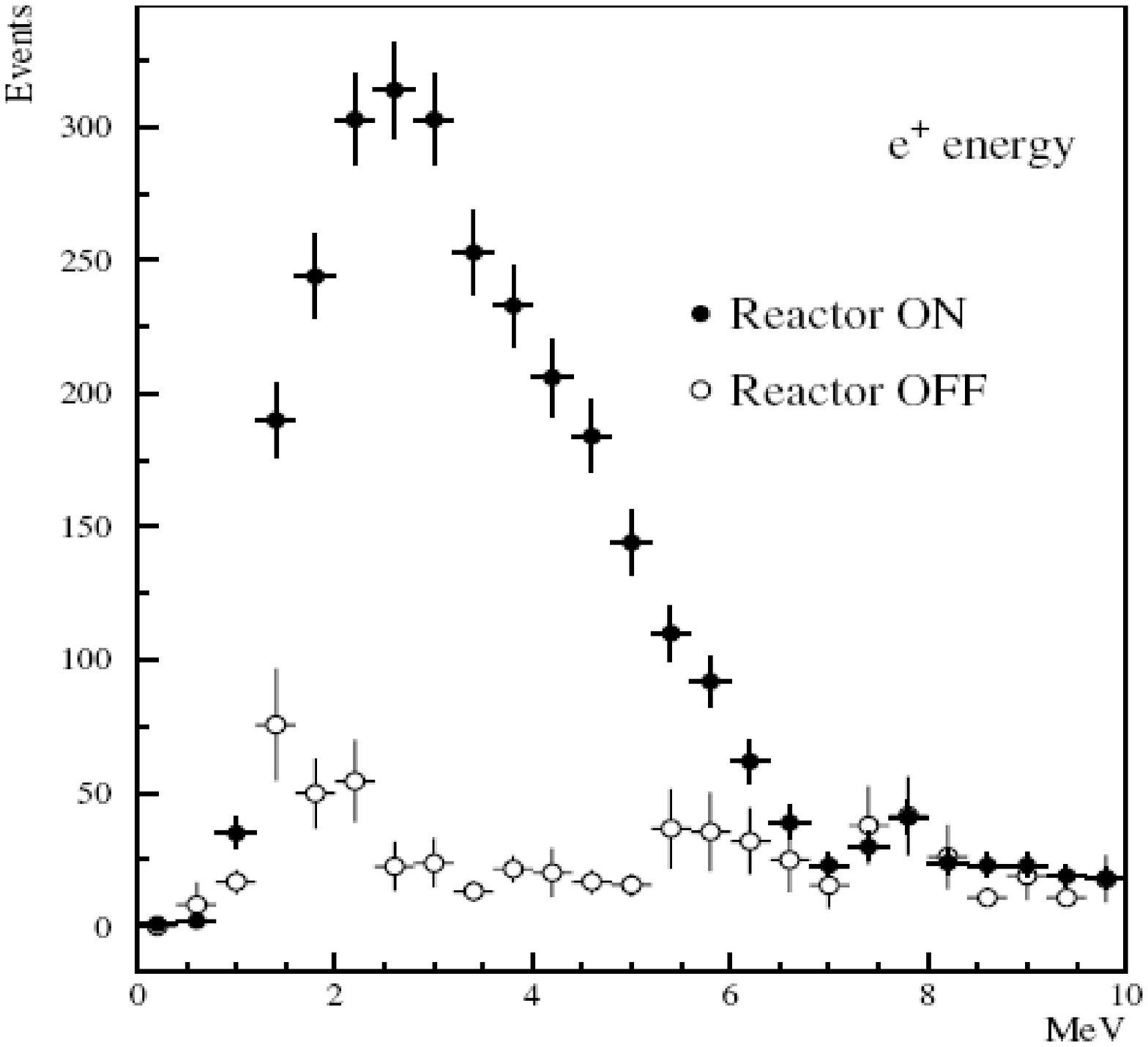,width=10.0cm}
\caption {Positron spectra in CHOOZ with reactor ON and 
OFF, from reference ~\cite{apollonio}, \cprB.}
\label{choospec}
\end{center}
\end{figure}

\par Figure~\ref{choospec} shows the positron energy spectra with reactor on and 
reactor off. 
The positron spectrum after the reactor off  spectrum has been
subtracted is shown in Figure~\ref{SpecChooz}.

\par The analysis of these data has given, for the ratio of the flux to 
the unoscillating expectation, the following result
 $$ R=1.01 \pm 2.8 \% (stat) \pm 2.7 \% (syst).$$
\begin{figure}[hbtp!]
\begin{center}
\epsfig{figure=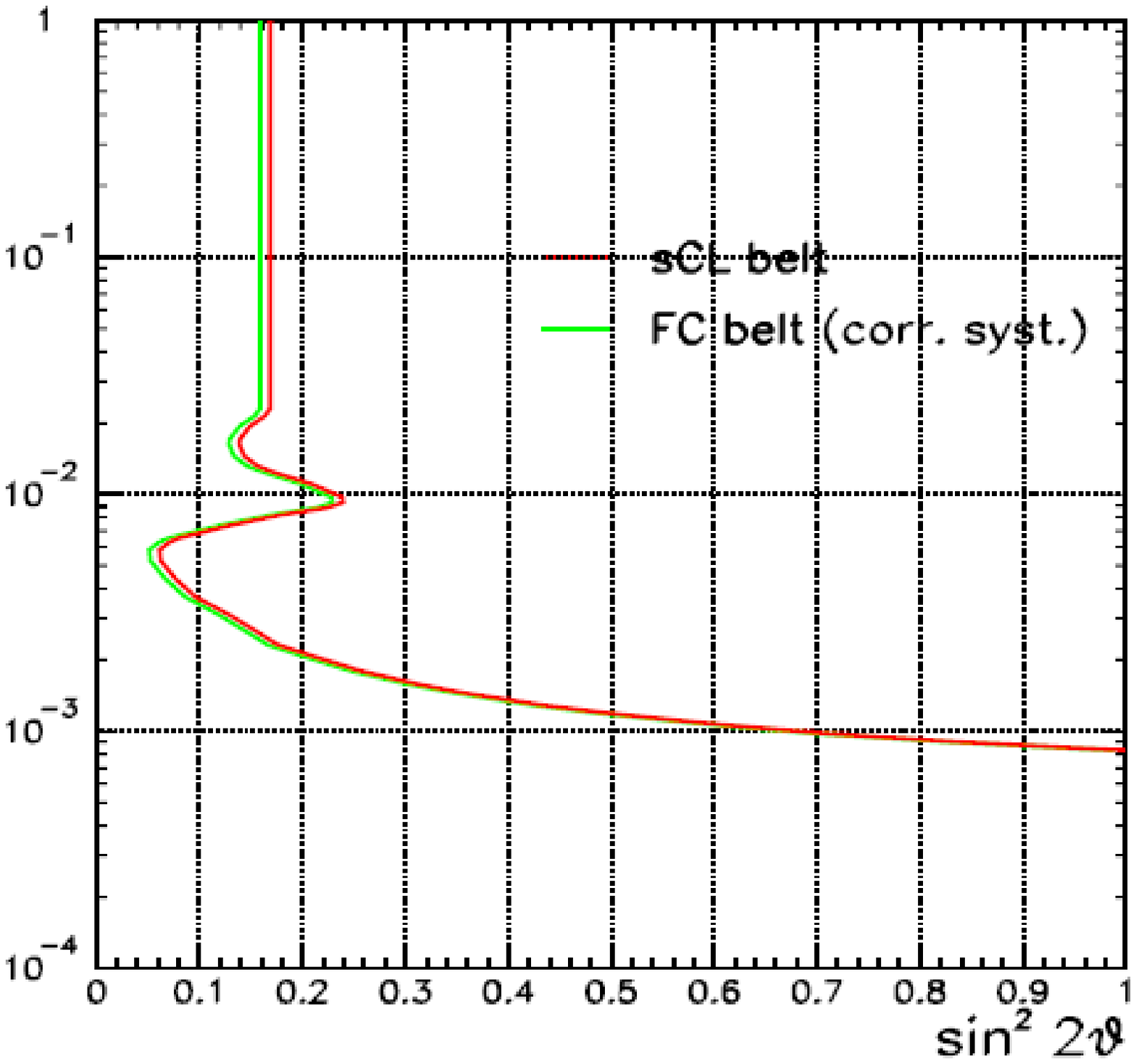,
width=10.cm} \caption {CHOOZ results, from reference ~\cite{apollonio}, \cprB.}
\end{center}
\label{choores}
\end{figure}

\par Figure~\ref{choores}  translates this result into limits 
on the oscillation parameters 
obtained in the two flavor mixing model. 
Oscillations $\nueb \rightarrow \nuex$ are
excluded
for $\dmq  \ge 8\times10^{-4}$~eV$^2$.  Limits on  $\sin^{2}2\theta$
depend on the  assumed  $\dmq$. For the value of $\dmq$ given by the  
atmospheric 
neutrino the limit  $\sin^{2}2\theta \le 0.13 $ is obtained.
This  limit excludes  $ \numu \rightarrow  \nue $ oscillations with this $\dmq$ value and therefore the possibility of interpreting 
the SK atmospheric  muon 
neutrino
deficit in terms of $ \numu \rightarrow  \nue $
oscillations. \vskip 2cm

\subsubsection{Palo Verde experiment:}
\par the Palo Verde experiment was built at the Palo Verde Nuclear 
Generating Station in Arizona. There were  3 identical reactors with a 
thermal power of 11.6 GW.
The detector consisted in 66 acrylic tanks filled with gadolinium loaded
scintillator, with a total mass of 11 ton.
The experiment did run from 1998 to 2000. The final 
result expressed as the ratio R, observed rate over expected one with no 
oscillation, was~\cite{palo}: 
 $$ R=1.01 \pm 2.4 \% (stat) \pm 5.3 \% (syst).$$

\subsubsection{KamLAND  experiment:}
\label{kamla}
\par KamLAND  is situated under 2700~MWE in the Kamioka (Japan)  mine
laboratory in the old site of the Kamiokande experiment.
Data reported here have been taken between March 2002 and January
2004.

\begin{figure}[hbtp!]
\begin{center}
\epsfig{figure=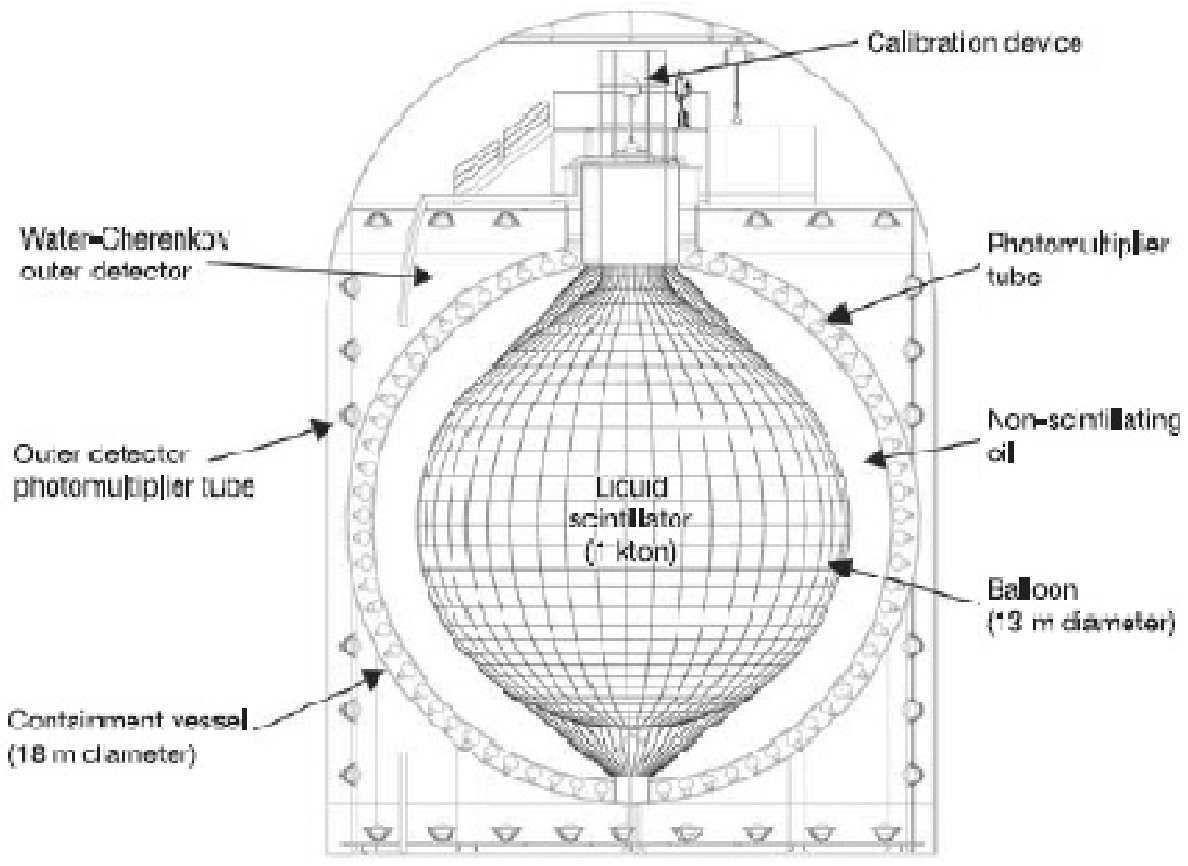,width=12.0cm}
\caption {The KamLAND detector, from reference ~\cite{kamland1}, \cprD{2003}.}
\label{kamdraw}
\end{center}
\end{figure}

53 power reactors surround KamLAND at an average distance of  150 km. 
The detector  consists of 1 kton pure scintillator contained
in a 13~m  diameter balloon suspended in non scintillating
oil. The balloon is viewed by 1879 photomultipliers  (Figure~\ref{kamdraw}).

\par Neutrons  are detected by the capture of neutron on proton (capture
energy=2.2~MeV). The selection criteria were
\begin{itemize} \item fiducial volume with radius $\le$ 5~m
\item gamma energy released in the neutron capture $\le$ 2.6 MeV and $\ge$ 
1.8 MeV 
\item distance from wall $\ge$ 30~cm
\item  distance electron-neutron $\le$ 160 cm
\item  neutron delay $\le$ 660  $\mu$sec
\item  neutron multiplicity =1
\end{itemize}
\par The main sources of background are neutrons from spallation
produced by fast muons and delayed neutrons emitted by He$^8$ and Li$^7$.
The expected non oscillation number of events above 2.6 MeV was
 365 $\pm$ 23 (syst). The number of observed events was 258, with an expected 
background of 17.8 $\pm$ 7.3 events. 
The survival probability has been estimated  to be $$0.658 \pm .044 stat
\pm 0.047 syst.$$
\par
\begin{figure}[hbtp!]
\begin{center}
\includegraphics[width=0.49\columnwidth]{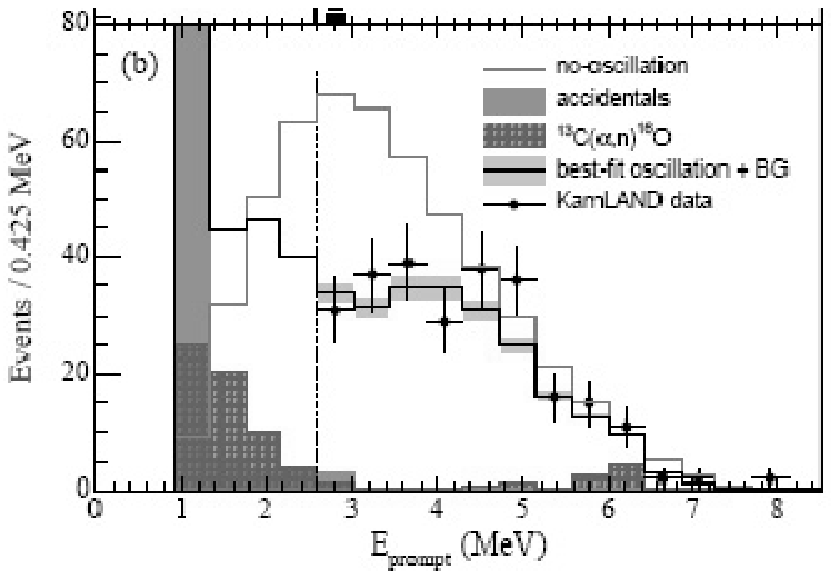}
\includegraphics[width=0.43\columnwidth]{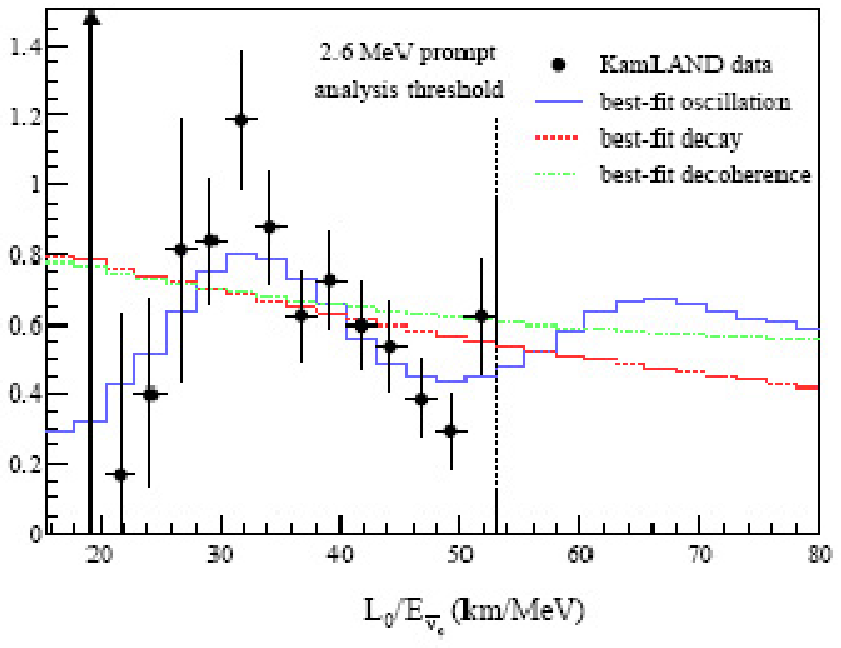}
\caption {left: Energy distribution; right: L/E distribution
 for anti-neutrinos, from reference \cite{kamland}, \cprD{2005}.}
\label{kamle}
\end{center}
\end{figure}

 The total spectrum is shown in Figure~\ref{kamle}left. Above 2.6 MeV 
one 
can see
data and expected spectrum without oscillation. Below 2.6~MeV, subtracting
background, one can estimate 25 $\pm$ 19 events that could be
indication of geological neutrinos. Geological 
neutrinos are generated by the decay of
radioactive elements (uranium, thorium and potassium) inside the earth, they 
are of geological interest.   
Figure \ref{kamle}right gives the $ L/E$ 
distribution
of events above 2.6 MeV. The blue line gives the best fit result for
oscillation.  Alternative models, neutrino decay~\cite{decay} and decoherence
~\cite{deco}, are ruled out. These data 
therefore support the interpretation of the effect as due to
neutrino oscillation.
The neutrino spectrum modulation of the KamLAND Experiment
allows a measurement of $\dmq_{12}$ more precise than the one obtained 
by solar neutrino experiment.
\par  A global two flavor analysis of KamLAND data and solar data~\cite{kamland}
 gives $$\dmq=(7.9^{+0.6}_{-0.5})\times  10^{-5}~\mathrm{eV}^2 $$
$$ \tan^{2}\theta=0.40^{+0.10}_{-0.07}.$$

\begin{figure}[hbtp!]
\begin{center}
\epsfig{figure=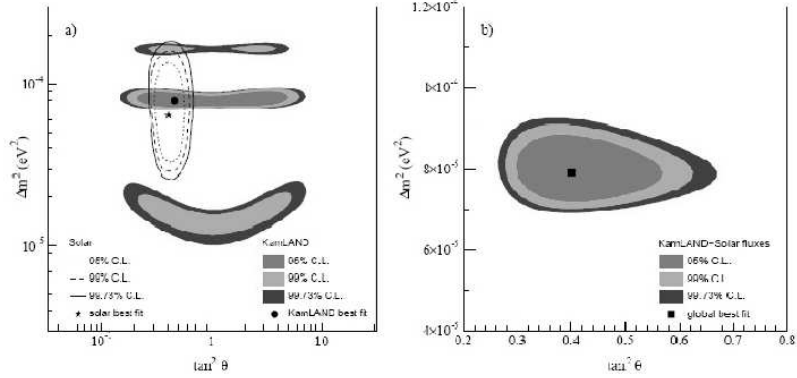
,width=12.0cm} 
\caption {Accepted values for oscillation parameters KamLAND +Solar 
a) dashed region KamLAND allowed region and solar neutrino experiments 
(lines) b) combined results from reference~\cite{kamland}, \cprD{2005}.}
\label{resol}
\end{center}
\end{figure}

 Figure~\ref{resol} presents the final result of the  KamLAND+Solar 
parameters determination a) KamLAND+Solar results, b) combined fit.

\par A complete discussion of the oscillation parameters will be made in
section \ref{mixing}.
\par After the KamLAND result the LMA solution is well established
and the oscillation parameters 1,2  are determined with a good 
accuracy.
\par Future experiments will be mainly devoted to obtain more 
information
on solar model and checks of the LMA solution for oscillations.
\par KamLAND has started a second phase of the experiment in which
elastic scattering  of solar neutrinos will be detected with the 
same aim of Borexino.
The background level will be reduced  at least a factor 100
compared to the present one. If the goal of background rejection will 
be reached the expected rate from
Be$^7$ in the energy window 280-800~KeV 
will be   much larger than in Borexino (1000 ton against the 
100 ton of Borexino).
\par  Proposals for pilot experiments and  R$\&$D 
for a
series of future experiments aiming at the detection of pp, 
CNO and Be$^7$ neutrinos have 
been presented~\cite{heeger}. 

\subsection{ Atmospheric neutrinos:}

\par Atmospheric neutrinos must be  observed in underground  detectors
because of the background due to cosmic rays.

\par For low energy neutrinos the observation of the neutrino interactions
with fully contained reaction products is possible with
reasonable efficiency (fully contained events, FC).
\par When the energy increases, the muon produced in $\numu$ CC 
interactions has a
high probability to escape the detector  (partially contained events,
PC).
\par There is a third category of $\numu$CC events: upward going muons 
produced in the rock. They can stop (stopping muons) or
traverse the detector (through-going muons).
Cosmic rays muons cannot be distiguished from the neutrino produced
ones so this technique cannot be used for muons coming  from  top.
 The typical energy is of the
order of 10 GeV for stopping muons and 100 GeV for traversing ones. The
neutrino energy  will be larger than the observed muon one.

\par To study the neutrino interaction Monte Carlo programs have been
developed 
\cite{Honda:1995hz,ATL97:113,PhysRevD.53.1314,Battistoni:2002ew,Honda:2004yz,Honda:2006qj} 
to predict the ratio 
$\nue$/$\numu$ to be compared with the
experimental observations. The double ratio,
  $$\frac{(\numu/\nue)_{data}}{(\numu/\nue)_{MC}}$$
expected to be 1 in absence of oscillations,  has been determined
by several experiments and   has 
always been found to be 
smaller than 1
\cite{sou,IMB,Kamioatmo,SuperKamat}.
\par The rate of  up going muons can be compared with the MC predictions
and also here  the rates are smaller than expectations 
\cite{Macro,kamio1}. The amount of the effect
depends on the used Monte Carlo generator more than the double
ratio.
\par The final confirmation of the interpretation of the deficit as due to
$\numu$  neutrino oscillation came
 in 1998 \cite{Skat1} when   Super-Kamiokande  demonstrated a
clear difference between upward and downward muon neutrinos, while no
difference was seen in the electron neutrino.
\par The upward  neutrinos traverse the Earth (12000~km), the downward come
from the atmosphere (20 km). We shall discuss in the following some of the 
cited experiments.
\subsubsection{ The Kamiokande and Super-Kamiokande experiments:}
\par The Kamiokande and Super-Kamiokande detectors already discussed in 
the
solar neutrino section have been used in the detection of
atmospheric neutrinos. Now the energy range (see Figure~\ref{honda}) of
studied events is of the order of GeV, so \numu can be detected via
 their CC reactions. The flavor of $\nu$ is determined through the
observation of  the shape of Cherenkov light  emitted by the lepton
produced in the final state.
\begin{figure}[htp]
\begin{center}
\epsfig{figure=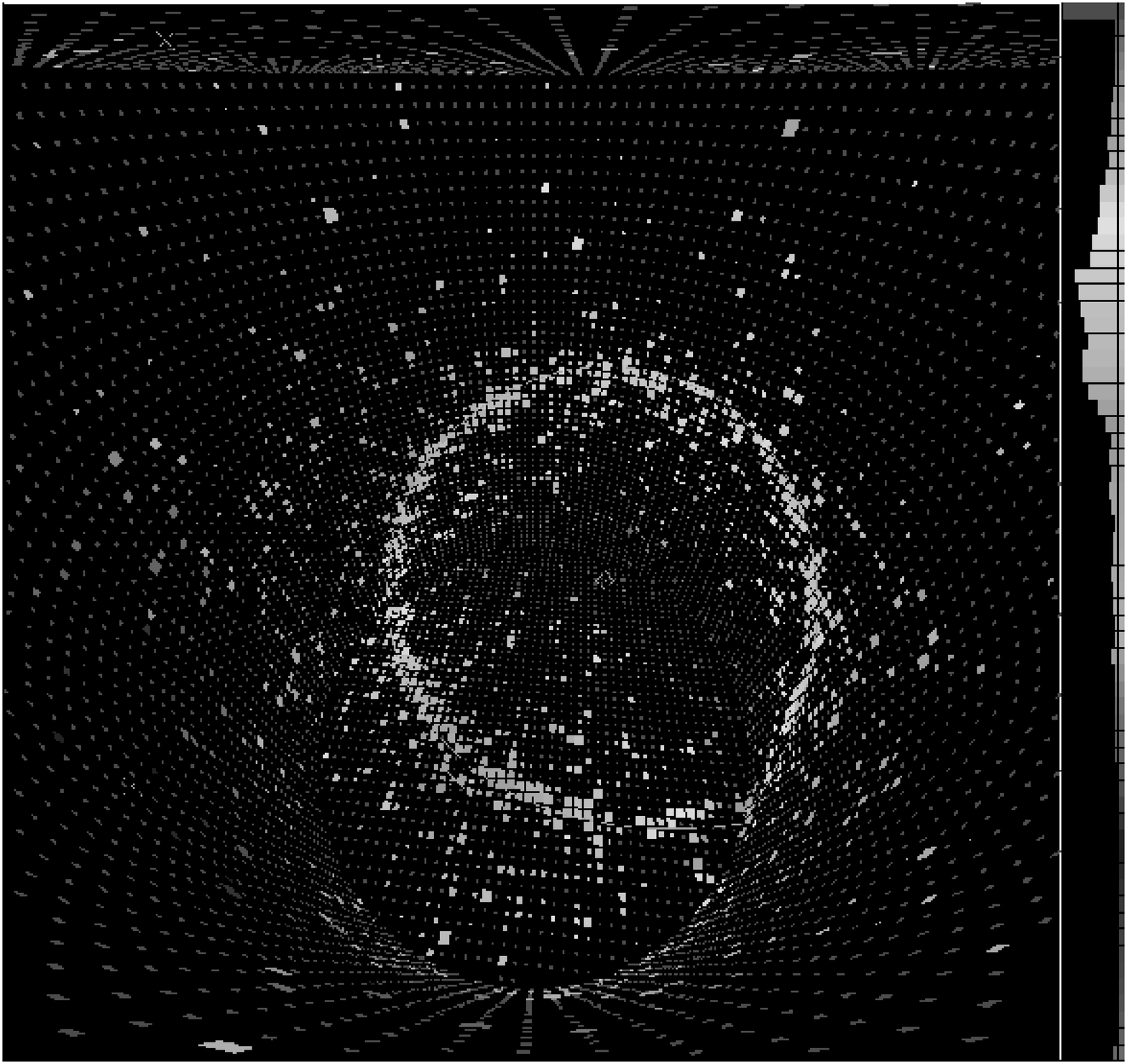,width=6.0cm}
\epsfig{figure=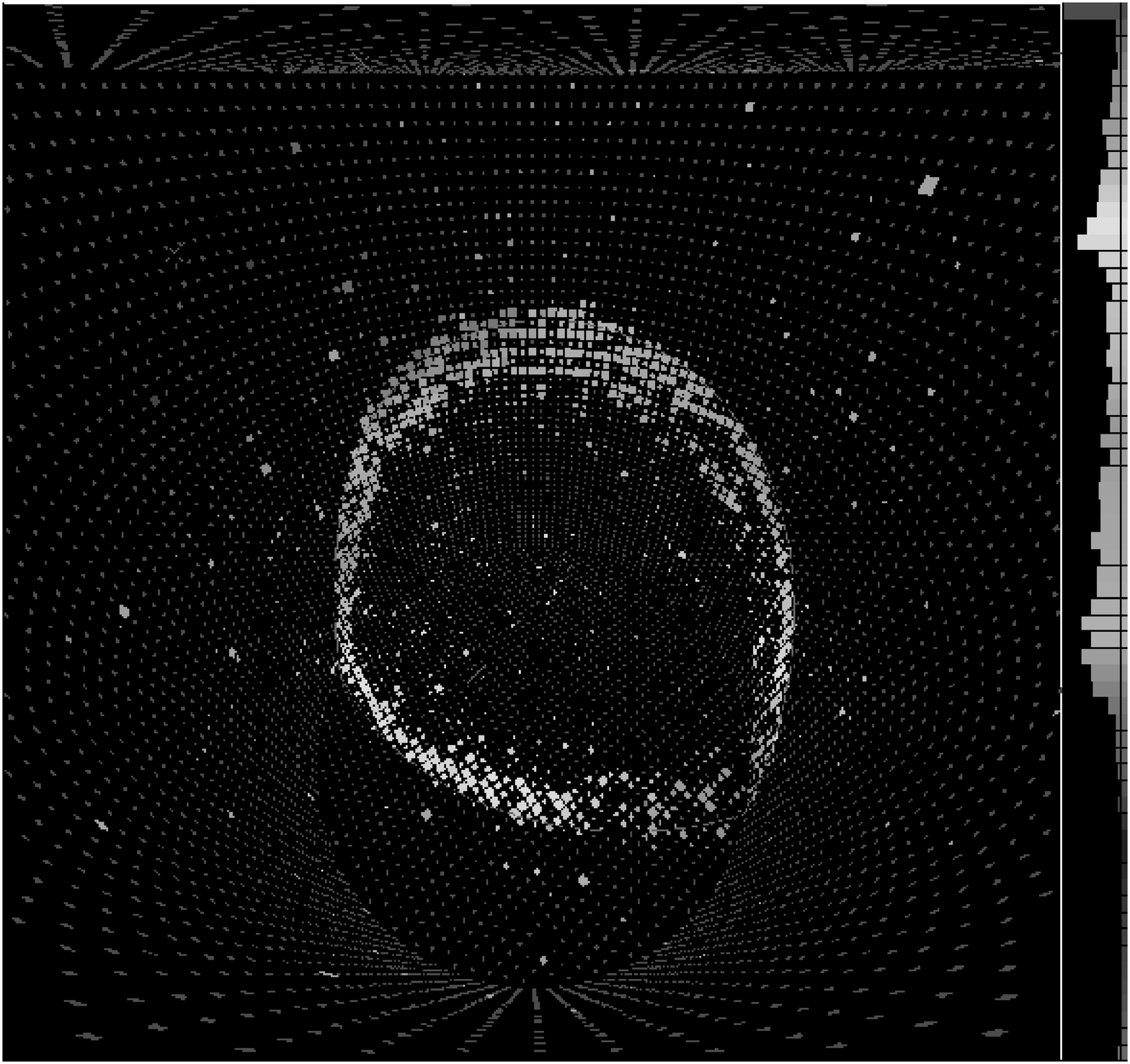,width=6.0cm}
\caption {Examples of an electron and a muon ring in SK detector, from 
http://www.ps.uci.edu/$\sim$tomba/sk/tscan/pictures.html}
\label{rings}
\end{center}
\end{figure}

Muons originate a ring with well defined borders while electrons have blurred
contours (Figure~\ref{rings}).
\par  Super-Kamiokande \cite{Skat1} demonstrated a clear difference 
between 
upward
and downward going muon neutrinos compared with the MC predictions, while no
difference was seen for electron neutrinos.
\par In the analysis  atmospheric neutrino data were  subdivided in

\begin{itemize}
\item Fully contained (FC) events Sub-GeV Evis $\le$ 1.33 GeV
\item Fully contained events Multi-GeV Evis $\ge$ 1.33 GeV
\par FC events were divided in single ring or multiple ring. Single ring
were classified as e-like or $\mu$-like according to the characteristic
of the Cherenkov  cone. Multiring were classified as e-like or
$\mu$like according to the characteristic of the highest energy
cone.

\item Upward going muons. 
\par Muons  traveling up were  divided in muons
stopping in the detector
 (stopping muons) or traversing  (through-going muons).
\end{itemize}
\par
\begin{figure}[htp]
\epsfig{figure=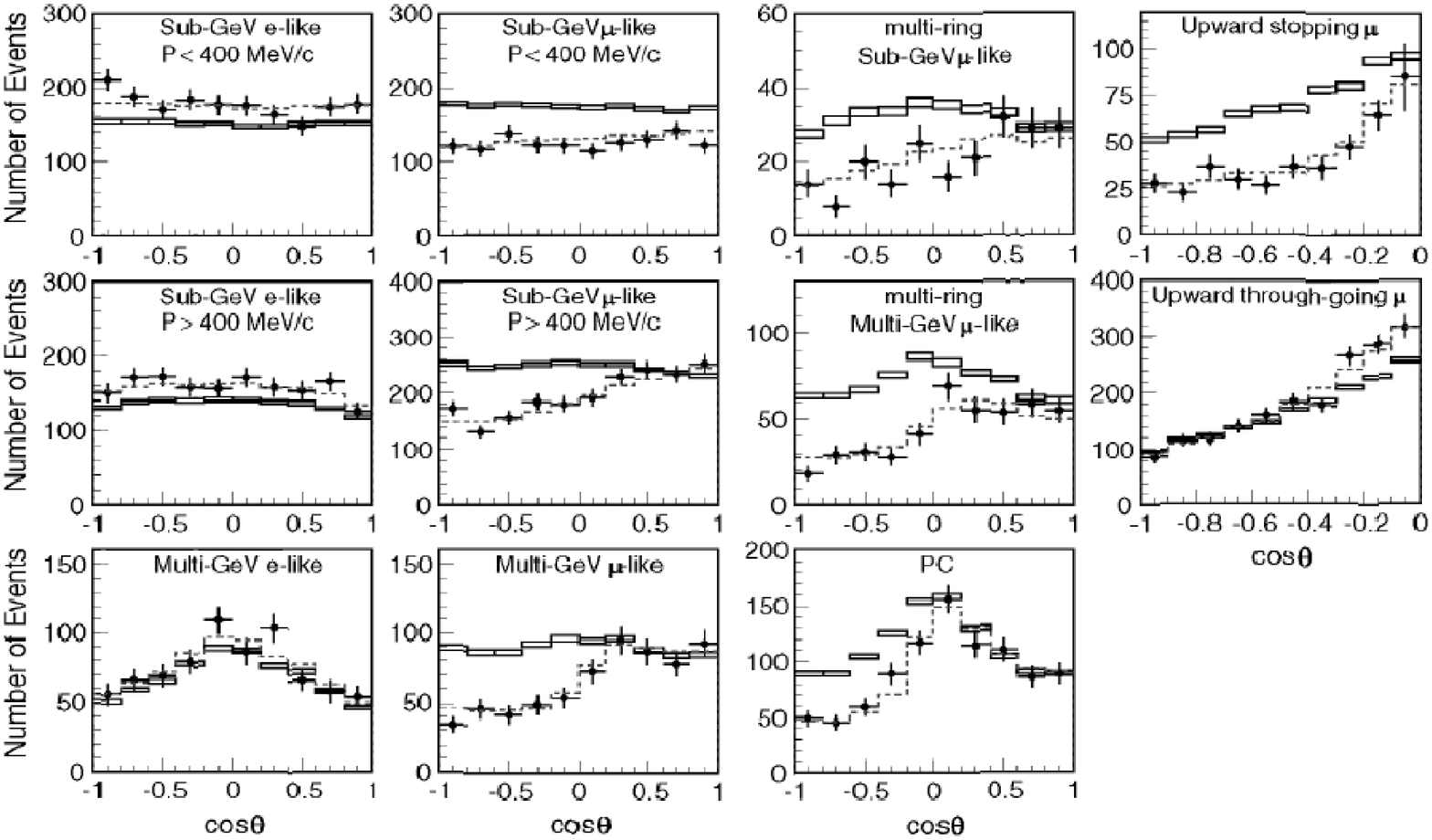,width=18.0cm}
\epsfig{figure=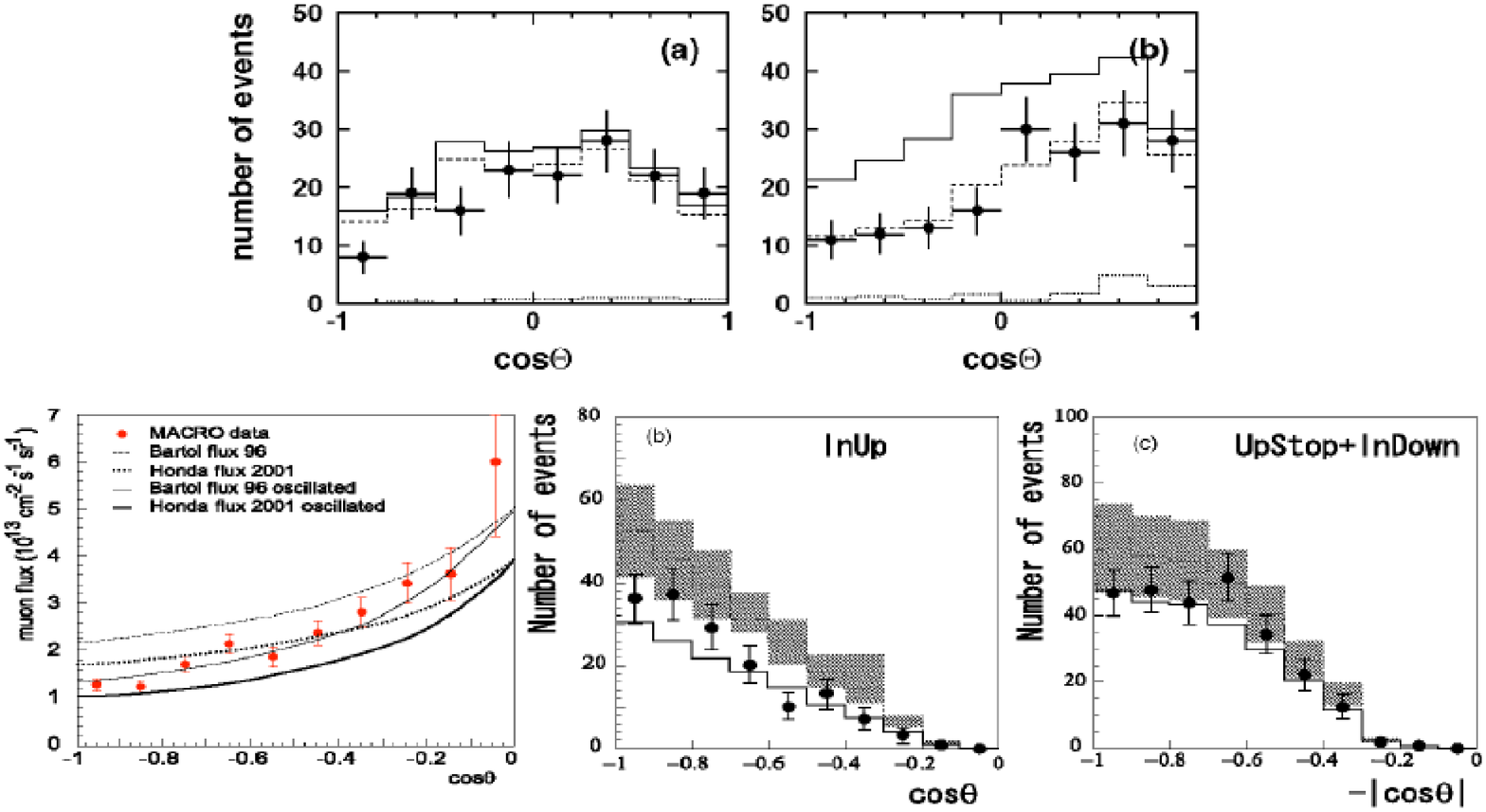,width=18.0cm}
\caption {Angular distributions of atmospheric events, from 
reference ~\cite{kailip}, \cprC{2005}.}
\label{atmspe}
\end{figure}

 The results of Super-Kamiokande, Soudan-2 and MACRO are shown in 
Figure~\ref{atmspe}~\cite{kailip}. 
\par Figure~\ref{atmspe}(top) from reference~\cite {kailip} shows the angular 
distribution for the  SK events categories
defined above.\par We see that
 for electrons the distributions are well
represented by the MC while for muons events coming from bottom, negative
cos$\theta$ events are missing.
\par Figure~\ref{atmspe}(middle) shows  the  angular distribution for the 
Soudan-2
experiment, we still see missing events in the muon distribution (b).
\par Figure~\ref{atmspe} bottom shows the angular distribution of upgoing 
muons in MACRO.

\par
\begin{figure}[htp]
\begin{center}
\epsfig{figure=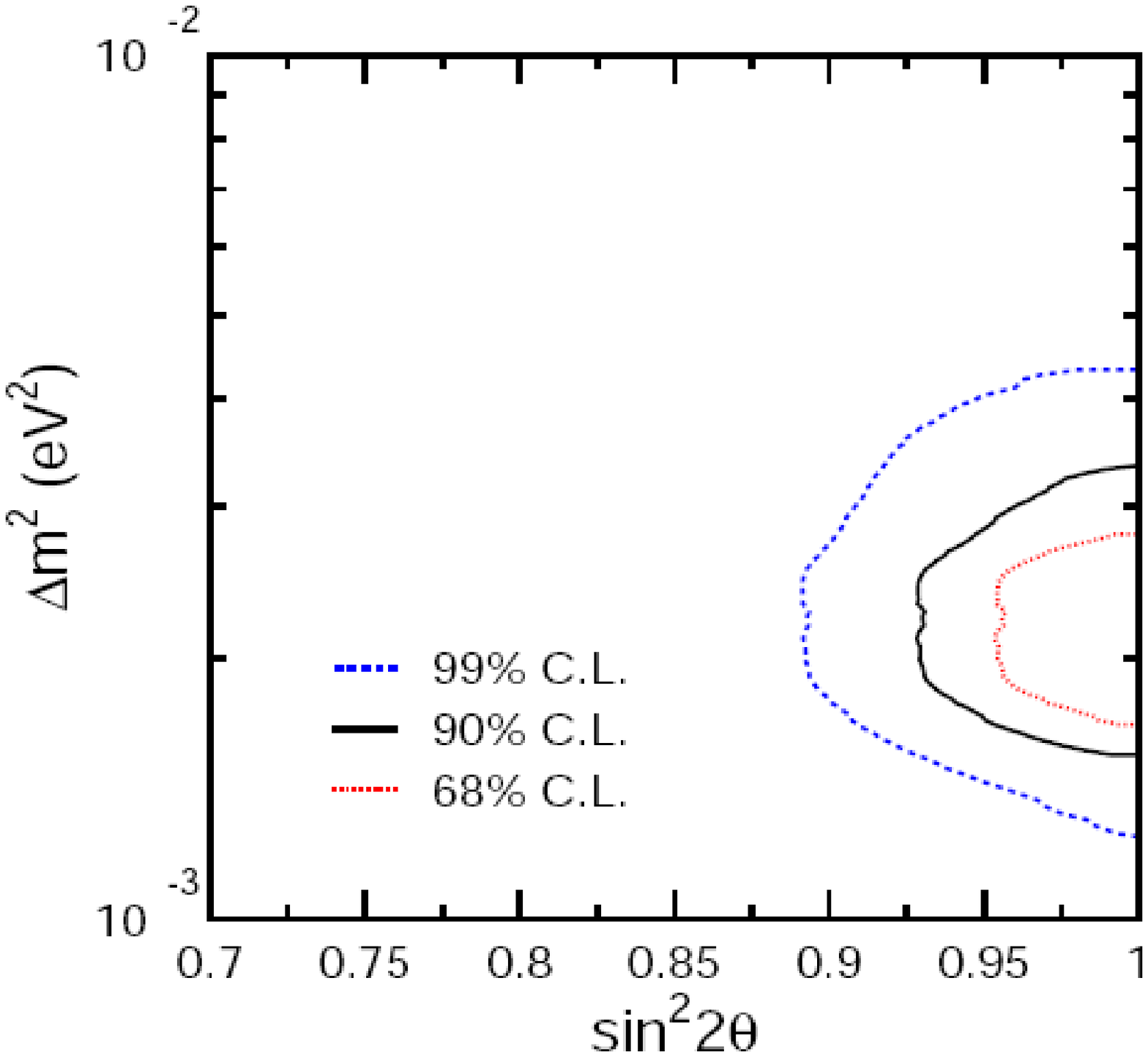,width=12.0cm}
\caption {SK atmospheric fit results, from reference ~\cite{SuperKamat}, \cprD{2005}.}
\label {fitsat}
\end{center}
\end{figure}

 The SK Collaboration~\cite{SuperKamat}, for
the ratio $(\mu/ e)_{data}/( \mu/ e)_{MC}$ that should be 1 in the
absence
of oscillations, quotes for Sub-GeV events  $$ R=0.658 \pm 0.016 \pm
0.035$$
\par and for  for Multi-GeV+PC
 $$ R=0.702 \pm 0.03  \pm.101 $$
\par a two flavor oscillation analysis has been been made 
\cite{SuperKamat}
with  results $$\sin^{2}2\theta\ge 0.92 (90\% CL)$$ 
\par  

$$ 1.5\times 10^{-3} \le \dmq \le 3.4\times 10^{-3}\mathrm{eV}^{2}$$
(see Figure~\ref{fitsat}).

\par Analysis in the three flavor mixing scheme is discussed in \cite{skam3f}.
\begin{figure}[htp]
\begin{center}
\epsfig{figure=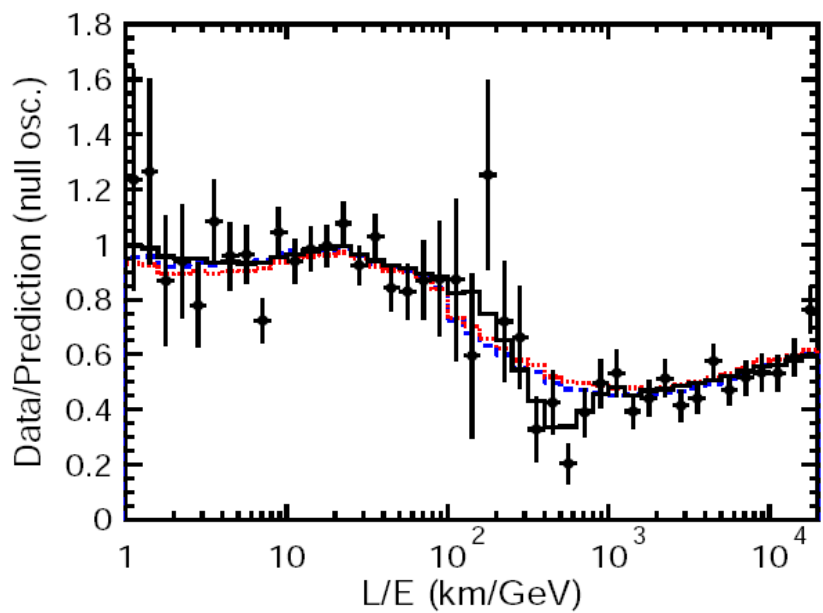,width=12.0cm}
\caption{Ratio of data to MC events without neutrino oscillation (points) as a function of the reconstructed L/E together with the best-fit expectation for two flavor $\numu\rightarrow\nutau$ oscillations(solid line)~\cite{skle}, \cprD{2004}. Also shown are the best-fit expectation for neutrino decay (dashed line) and neutrino 
decoherence (dotted line).}
\label{atmdip}
\end{center}
\end{figure}
\par Selecting well measured events a plot of L/E  has been obtained
and is shown in Figure~\ref{atmdip}~\cite{skle}.
The presence of a dip in the L/E distribution gives strong support
to the oscillation interpretation against other possible explanations. 
Figure~\ref{atmdip} in fact shows that alternative explanations do not 
reproduce the dip.
\par
\begin{figure}[htp]
\begin{center}
\epsfig{figure=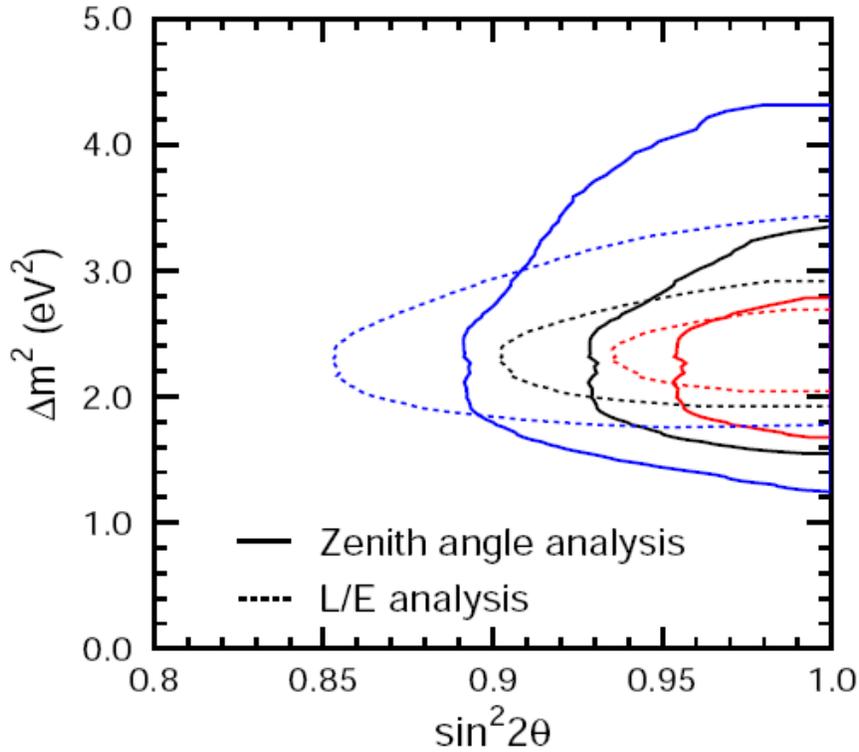,width=12.0cm}
\caption {The 68, 90 and 99 $\%$ confidence level allowed oscillation
parameter regions obtained by the SK L/E and zenith angle
analysis,
from reference ~\cite{SuperKamat}, \cprD{2005}.}
\label{atmcon}
\end{center}
\end{figure}

  Figure~\ref{atmcon} shows the results of the  zenith angle analysis 
and of the
L/E one. The position of the dip allows a better determination of the
$\dmq$ region in the L/E analysis compared with the one 
obtained from analysis of the
zenith angle.  
\vskip 0.5cm
\par We will now briefly describe the other experiment that did  confirm 
the SuperKamiokande results.
\subsubsection{The MACRO experiment:}
\par the MACRO Experiment~\cite{Macro} located in the Gran Sasso 
Laboratory (LNGS) took 
data from 1995 to 2000. It did consist of three independent detectors: 
liquid scintillators counters, limited streamer tubes and nuclear 
track detectors (not used in the oscillation search). The detector did 
reveal upgoing muons coming from interactions in the rock. In the analysis 
the angular distribution and the absolute flux compared with the 
Monte Carlo predictions were used, see Figure~\ref{atmspe}(bottom). The 
 analysis in terms of oscillation did favour maximum mixing
 and $\dmq= 0.0025~eV^{2}$.
\subsubsection{ The Soudan-2  experiment:}
\par Soudan-2~\cite{sou} was  a   770 ton  fiducial mass detector that  
did operate as a time 
projection chamber. 
The active elements of the experiments
were plastic drift tubes. 
The detector was located in 
Minnesota (USA). The experiment did run from 1989 to 2001 with a total 
exposure of 5.90 kton-years. 
An analysis in terms of
oscillation parameters of the L/E distribution gave as result     
 $\dmq=0.0025~eV^{2}$ and $sin^{2}2(\theta)=0.97$.

 \subsubsection{ The MINOS experiment:}

\par The far detector of the MINOS experiment \cite{michae}, described in  
Section~\ref{sec:minosexperiment}, is designed to study 
neutrinos coming
from the neutrino beam NuMI at the Fermilab National Laboratory. 
The experiment can 
also detect atmospheric neutrinos and being a magnetized detector 
it has the advantage to observe  separately $\nu$ and \nubar measuring the
charge of muons in the  magnetic field.
\par The data relative to a period of 18 months (2003-2005) are consistent
with the same oscillation parameters for neutrinos and
anti-neutrinos. In fact MINOS quotes~\cite{minosat}

$$\frac{(\numub/\numu)_{expt}}{(\numub/\numu)_{mc}}~~=.96^{+0.38}_{-0.27}
(stat) \pm 0.15(syst)$$

 \par and 
$$\frac{(up/down)_{exp}}{(up/down)_{mc}}~~=.62^{+0.19}_{-0.14}(stat)
 \pm .02 (syst)$$
\par that still gives an indication of upward going muon disappearance.
These results are statistically limited  and correspond to a
statistics of 4.54 kiloton-year.
From their analysis the hypothesis of no oscillation is excluded
at the $98\%$ of CL.

\subsection{Accelerator neutrinos:}
\vskip 1 cm
\par Neutrino beams (see Section \ref{beams}) have been produced in 
accelerators 
since the 60's. The possibility of doing neutrino experiments at 
accelerators was first proposed by B. Pontecorvo in 1957~\cite{Pontecorvo:twoneut} and M. Schwartz in 1960~\cite{Schwartz}.
Following these suggestions  an experiment was performed at the  
Brookhaven National Laboratory in which the muon neutrino was discovered~\cite{bnl2ne}.
For what concerns oscillation experiments we can divide them in two 
categories, 
short baseline (see Section~\ref{short}) 
and long baseline (see Section~\ref{long}). The 
range of the $\dmq$ that have been detected has pushed toward the second type 
experiments.

\subsubsection{Short baseline experiments:} 
\vskip 0.1cm
\label{short}
\paragraph{Search for  $\numu\rightarrow \nue$}.
\par  A) Bubble chamber experiments
\par Bubble chambers experiments did begin in the '70s. These experiments 
that gave important results in neutrino physics could provide only limits 
in the oscillation parameters space.
\par Experiments were made in  CERN Gargamelle~\cite{bellotti},  CERN 
BEBC (the Hydrogen  bubble chamber)\cite {Erriquez}
 and in the Fermilab 15~ft bubble chamber~\cite{baker}. 
The last
experiment with bubble chambers in CERN was the BEBC
experiment with a low energy neutrino beam to search for $\numu
\rightarrow \nue$ for values of $\dmq\approx$~1~eV$^2$~\cite {angelini}.
\begin{table}[htbp!]
\begin{tabular}{|l|c|c|c|c|}\hline
{\bf Experiment} & Beam Mean Energy&$\dmq$ (eV$^{2}$) & $\sin^{2}2\theta 
\times 
10^{-3}$
\\
 & (Gev) &$(\sin^{2}2\theta=1)$ & (large~\dmq) \\\hline  

Gargamelle CERN \cite{bellotti}& 300& 1.2&10.\\\hline   
BEBC CERN \cite{Erriquez}&300&1.7 &10. \\\hline
15 foot BC Fermilab  \cite{baker} &30&0.6 & 6.\\\hline
BEBC CERN \cite{angelini}&1.5&0.09   & 13. \\\hline
\end{tabular}
\caption{$\numu-\nue$ limits in bubble chamber experiments .}
\label{tab:bubblec}
\end{table}

\vskip 0.1cm
\par  B) Electronic detectors experiments
\vskip 0.1cm

\par Electronic detectors searches were made using general purpose
neutrino detectors~\cite{cdhsdet,charmdet,charm2det,e734det,e776det,nutevdet}
or dedicated detectors\cite{nomaddet,chorusdet}.
\par Several experiments were made to search for  $\numu\rightarrow \nue$
with electronic detectors. A non exhaustive list is given in Table \ref{tab:eltx}.

\begin{table}[htbp!]
\begin{tabular}{|l|c|c|c|}\hline
{\bf Experiment} & Neutrino Mean Energy & $\dmq$ (eV$^{2}$) & 
$\sin^{2}2\theta \times 10^{-3}$ \\
 & (GeV)& $(\sin^{2}2\theta=1)$ & (large~\dmq) \\\hline
CHARM CERN \cite{bergsma}&25.& 0.19& 8\\\hline
E776 BNL \cite{borodvsky}&5. &0.075  & 3 \\\hline
E734 BNL \cite{ahrens} &5.&0.03 & 3.6\\\hline
CHARM2 CERN \cite{vilain}&25.&  8.5   & 5.6 \\\hline
NUTEV FNAL~\cite{avvaku} &140.& 2.6 &1.1\\\hline
NOMAD CERN~\cite{astier}& 25.&    0.4 & 1.4   \\\hline
\end{tabular}
\caption{$\numu-\nue$ limits in accelerator experiments.}
\label{tab:eltx}
\end{table}

\par All these experiments were made with conventional neutrino beams
and gave negative results. The $\nue$ were detected through their 
charged current  
interactions  giving an electron. The $\nue$ contamination of the beam that
had to be subtracted was  one of the main sources of systematic errors.
The other systematic error was the contamination of gamma rays from
$\pi^{0}$ decay.

\vskip 0.1cm
\paragraph{Search for  $\numu \rightarrow \nu_{x}$}.
\vskip 0.1cm

\par Experiments were also  made on disappearance of $\numu$, $\numu
\rightarrow \nu_{x} $. Muons
from  CC $\numu$ interactions were  counted.
In this case two detector systems  at different distances were used
to eliminate
 the uncertainties on the knowledge of neutrino fluxes. For two
detectors
experiments the excluded region closes up at high  $\dmq$ when 
oscillation happens in both detectors.
Results are summarized in Table~\ref{tab:disappeare}.

\begin{table}[htbp!]
\begin{tabular}{|l|c|c|c|c|}\hline
{\bf Experiment} &  Neutrino Mean Energy(GeV)& $\dmq$~min~eV$^{2}$ 
&$\dmq$~max~eV$^{2}$   \\\hline
CDHS  CERN~\cite{cdhs}  &25. &0.23 &100.\\\hline
CHARM CERN~\cite{charm} &25. &0.29 &22.6 \\\hline
FNAL~\cite{baker} &140. &8.  &1250 \\\hline
\end{tabular}
\caption{Limits on the disappearance for $\numu-\nux$}
\label{tab:disappeare}
\end{table}
\paragraph{Search for  $\numu \rightarrow \nutau$}. 
\par The detection in appearance mode of $\numu \rightarrow \nutau $  
is difficult because of the short
lifetime of the $\tau$ whose flight length is $\le$~1mm.
Following  a negative 
result from the emulsion experiment E531 at
Fermilab\cite{e531},
there have been two experiments at the CERN WBB searching for small mixing
angles and relatively large $\dmq$. In these experiments 
the E/L ratio of the
beam is indeed large because the energy has been set to have an 
appreciable $\nu_{\tau}$ cross
section.
 
\par The CHORUS experiment~\cite{chorusdet} was a hybrid emulsion 
electronic 
detector that had 
excellent space resolution at the $\tau$ decay.
\par The NOMAD~\cite{nomaddet} experiment, where the vertex resolution was not 
good enough to
see the tau decay, applied kinematical criteria to search for $\nutau$CC. 
Both experiments
gave a negative result as shown in Table~\ref{tab:tau}.

\begin{table}[htbp!]
\begin{tabular}{|l|c|c|c|}\hline
{\bf Experiment} & Neutrino beam energy (GeV)& 
$\dmq$~eV$^{2}~(\sin^{2}2\theta=1)$ &
$\sin^{2}2\theta(large \dmq)$ \\\hline
NOMAD CERN~\cite{nomado} &25.&0.7 & 3$\times 10^{-4}$\\\hline
CHORUS CERN~\cite{choruso} &25.&0.6 &4.4$\times 10^{-4}$\\\hline
\end{tabular}
\caption{$\numu-\nutau$ limits.}
\label{tab:tau}
\end{table}

\vskip 0.5cm
\subsubsection {Other short baseline experiments:} 
\par LSND, KARMEN and MiniBooNE.
\par There is
 one experiment that has claimed to  have seen oscillations in the region
$\dmq \leq 1$~eV$^{2}$, the LSND~\cite{lsnd}  experiment.
\par The experiment was  run in 
1993-1998 at the LAMPF accelerator in
Los Alamos (USA). The detector  consisted of a tank containing 168 ton of
liquid scintillator equipped on the inside surface with 1220 photomultipliers. 

\par The intense proton beam ($\simeq$~~1mA), at an energy of 798 MeV, 
produces a
large
number of pions mostly $\pi^{+}$  that then decay in $\mup +\numu$. The
 $\mup$ decays
at rest  in $e^{+}+\nue +\numub$. Practically all  $\pi^{-}$  are absorbed 
in the
shielding. 
The $\nueb$ flux coming from the
 $\mum$ decay   at rest,  where $\mum$ are produced in
  the rare $\pi^{-}$  decay in flight, constitutes a small fraction of the
$\numub$ one. 
Consequently the
experiment, through  the  study of the process $\nueb+p \rightarrow e^{+}+ n$,
allows the study of the $\numub \rightarrow \nueb$ oscillation.
\par
\begin{figure}[hbtp!]
\begin{center}
\epsfig{figure=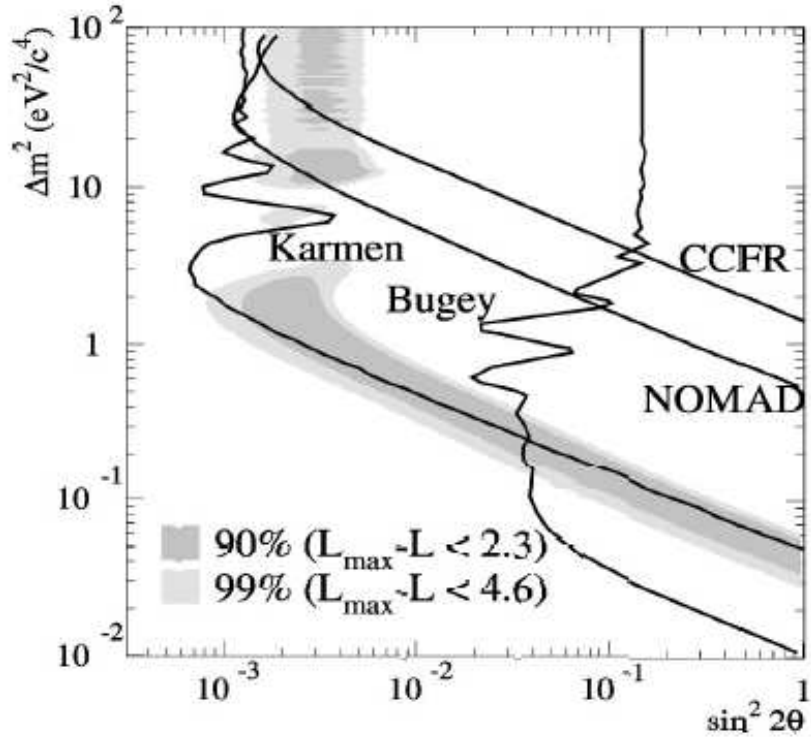,width=12.0cm}
\caption {LSND results ,($\sin^{2}2\theta,\dmq$oscillation parameters fit),
the inner and outer regions correspond to 90$\%$ and  99$\%$ allowed, from reference ~\cite{lsnd}, \cprD{2001}.}
\label{lsnd}
\end{center}
\end{figure}

 The study of the process $\nue+C ~\rightarrow e^{-} +N$, using only
electrons above the Michel endpoint to eliminate the  $\nue$ from  $\mup$ 
decay, did allow  the study of the process $\numu~\rightarrow~\nue$.
LSND found an excess of e$^+$ (e$^-$)~\cite{lsnd} and made a claim for oscillations with 
parameters 
$\dmq=1.2$~eV$^{2}$,
$\sin^{2}2\theta= 0.003 $, as shown in Figure~\ref{lsnd}.
\par A similar experiment, KARMEN~\cite{karmen},
 ran at the ISIS pulsed spallation neutron source in UK in 1997 and 1998 and   
did not give any
positive evidence. It covered a large fraction of the LSND results, as shown in 
Figure~\ref{lsnd}.
  New experiments were  needed. The MiniBooNE experiment, designed at
Fermilab, is now running. The experiment uses the Fermilab booster  (8 GeV
protons)  neutrino beam.  The detector is a spherical tank of inner
radius of 610~cm filled with 800 tons of mineral oil. The 
Cherenkov  and scintillation light is collected by photomultipliers.

\begin{figure}[hbtp!]
\begin{center}
\epsfig{figure=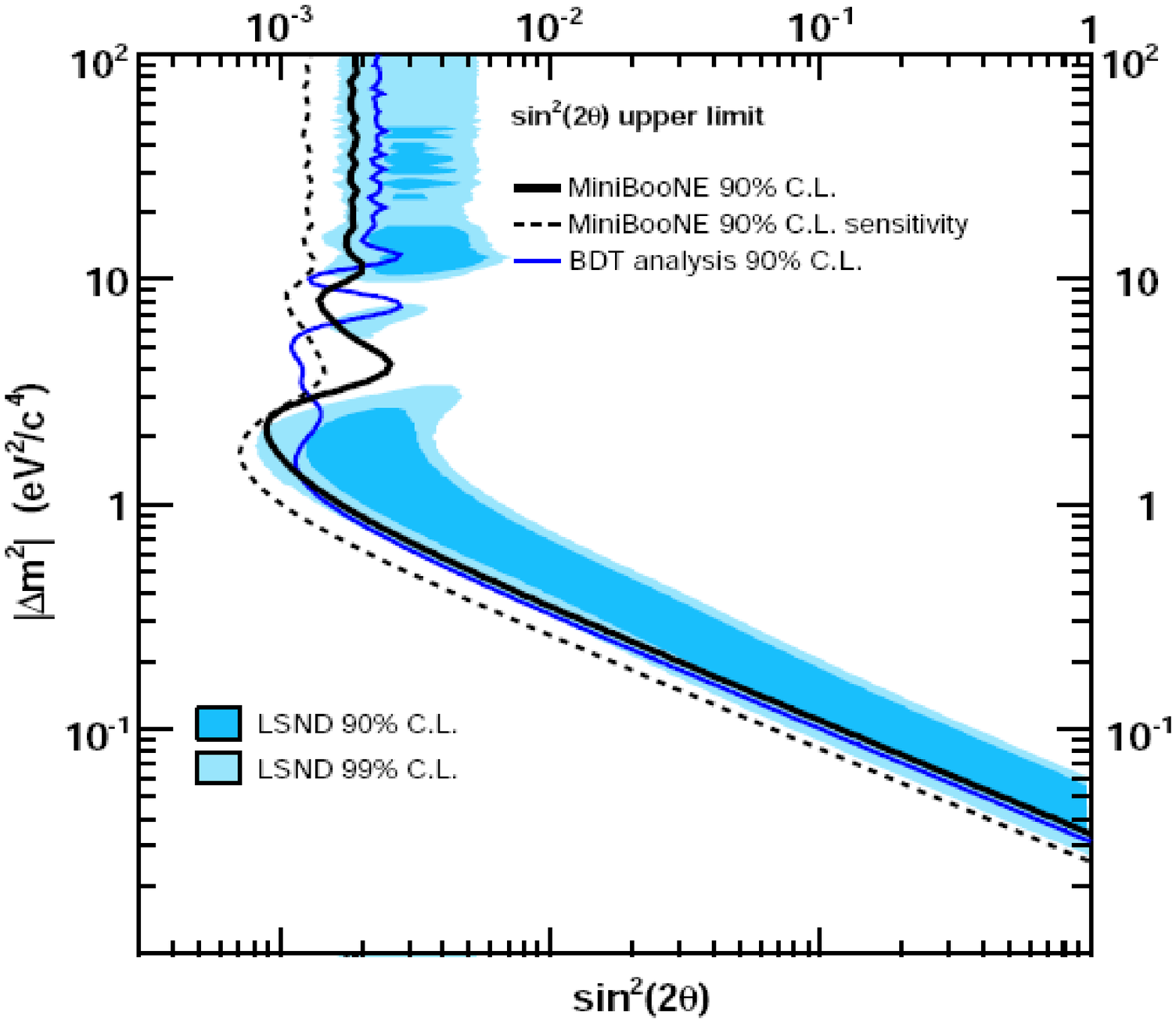,width=12.0cm}
\caption {MiniBooNE 90$\%$ confidence level, shaded area
corresponds to LSND result~,from reference \cite{miniboone}, \cprD{2007}.}
\label{mini}
\end{center}
\end{figure}

The first publication of the experiment
 does not confirm the LSND results~\cite{miniboone} (see Figure~\ref{mini}).
\par Had the LSND claim been  confirmed, then
a major
change in the theory would  have been needed. With only 3 neutrinos
there are two independent $\dmq$ values, 
that we identify with 
the solar and atmospheric ones. The
LSND result, introducing a third $\dmq$ value, would  have   required a fourth, unobserved, sterile neutrino.

\subsubsection{Long baseline accelerator experiments:}
\label{long}
\par Man-made neutrino sources experiments are very important in providing the
final 
confirmation of neutrino oscillations. 
The solar result confirmation was given by KamLAND.
To confirm the atmospheric ones, dedicated long baseline neutrino experiments have been 
conceived, providing access to the same L/E range.
The K2K experiment has been completed and first results
from MINOS have been given, while
OPERA starts to take data. The three experiments are described below.
\par
\paragraph{The K2K experiment:}
\par The experiment~\cite{k2k} used  an accelerator produced  $\numu$ 
neutrino beam of
an average energy of 1 GeV, the neutrino interactions were  measured in 
the
SK detector located at 250 km from the source and in a close detector
located at 300~m from the target.
A total of 10$^{20}$ protons have been delivered to the  target  in the
data taking period 1999-2001.
The SK detector has already  been  presented in Section~\ref{SKC}. 
The close detector consists of 1 kiloton water Cherenkov detector and a
scintillating fiber water target (SCIFI). In the second data
taking period (K2K II) a segmented scintillator tracker (SCIBAR) and a
muon ranger (MRD) were added to it. 
The experiment is a disappearance experiment since the
energy of the beam is below the threshold for $\nutau$ production. 
The oscillations are detected by the measurement of the flux ratio in
the two detectors and by the modulation of the energy distribution
of CC produced events. The energy distribution of events can be
obtained from the SK 1-ring events that are assumed to be quasi
elastic (at the K2K  energies 1 ring mu  events have a high
probability to be quasi elastic). In this approximation
$$ E_{\nu}^{rec}=(M_nE_\mu-m_{\mu}^{2}/2)/(M_n-E_{\mu}+P_{\mu}\cos
\theta_{\mu}).$$
\par
\begin{figure}[hbtp!]
\begin{center}
\epsfig{figure=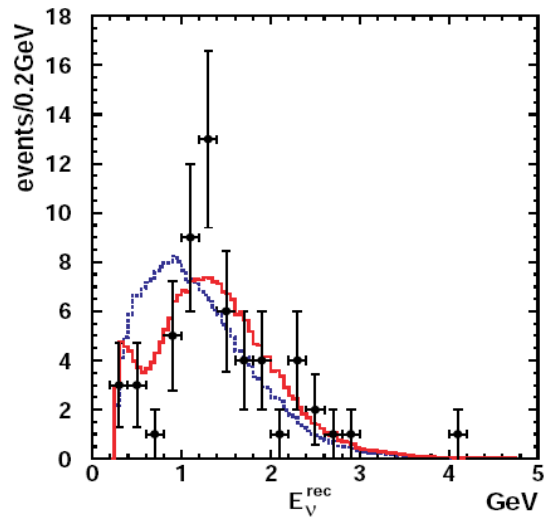,width=8.0cm} 
 \caption {K2K $ E_{\nu}$ distribution for 1 ring $\mu$~events.
Points with error bars represent data, 
solid line the best fit  with oscillations while the dashed 
line shows expectation without oscillations, from  reference~\cite{k2k}, \cprD{2006}.}
\label{k2kmod}
\end{center}
\end{figure}

 The expected number of events in SK in the  absence of oscillation is 158, 
the measured
one is 122. The expected number has been obtained from  the rate of events
measured in the close detector.
The comparison between the SK spectrum and the expected one in absence of 
oscillation is shown in Figure~\ref{k2kmod}.
\par The best fit results~\cite{k2k}  obtained combining the information from 
the spectrum shape and the
normalization
are
$$ \sin^2 2\theta =1~~~~ \dmq= 2.8\times 10^{-3} \mathrm{eV}^2~( 1.8\times 10^{-3}\le 
\dmq \le 3.5\times 10^{-3}~\mathrm{at}~90\%~~CL) $$
the probability of no oscillation hypothesis is 0.0015$\%~(4.3 ~\sigma)$.
\par
\begin{figure}[hbtp!]
\begin{center}
\epsfig{figure=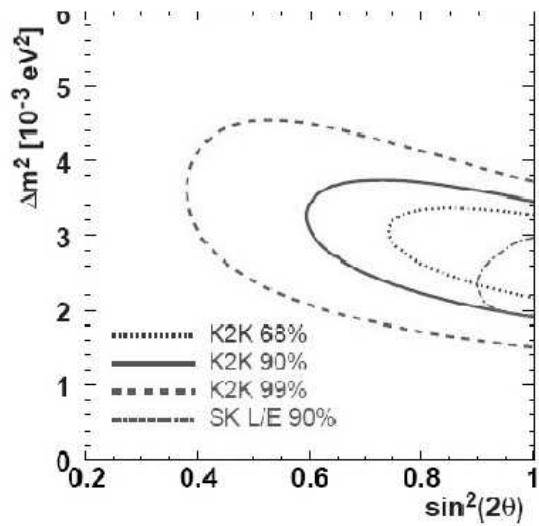,width=8.0cm} 
\caption{Comparison of the K2K results  with the SK atmospheric 
neutrino in the parameter space, dotted, solid, dashed and dashed-dot 
represent the 68, 90, 99 $\%$ CL allowed regions of K2K and 90 $\%$  
CL for SK atmospheric, from reference ~\cite{k2k}, \cprD{2006}.}
\label{k2kfits}
\end{center}
\end{figure}

 Figure~\ref{k2kfits} shows  K2K results compared to Super-Kamiokande 
results obtained with atmospheric neutrinos.
\par
\par A search for $\numu \rightarrow \nue$ has been also 
performed~\cite{yamamo}
and the result for $\dmq=2.8\times 10^{-3} \mathrm{eV}^2$ is $\sin^2 2\theta_{\mu e}$ 
$<$ 0.13 at 90 $\%$ CL. 
This limit will be discussed in 
Section~\ref{mixing}.
together with all the results on  $\sin^2 2\theta_{13}$. 
\par

\paragraph{The MINOS  experiment:}\label{sec:minosexperiment}

The MINOS experiment~\cite{michae} is a $\numu$ disappearance experiment using 
two detectors,
the 
Near Detector (ND) and the
Far Detector (FD).
\par The ND detector (0.98 kton) is located at
103 m underground and at a distance of  1 km  from the source. The FD detector, 
705 m
underground, is located at a distance of 735 km. 
The detectors are magnetized iron
calorimeters made of steel plates of 2.54 cm thickness interleaved with 
plastic scintillator planes segmented into 
strips (4.1 cm wide and 1 cm thick).
\par Data have been collected in the period May 2005-February 2006 and a total
of 1.27$\times 10^{20}$ protons were used
in the target position that gives the ``LE Beam'' (see Figure~\ref{NUMI}) 
the one that maximizes the neutrino flux 
at low energies~\cite{minos}.
\par   215 events with an energy $\le$ 30 GeV have been collected in the FD
to be compared with an expected number of 336 $\pm$ 14.
\par
\begin{figure}[hbtp!]
\begin{center}
\epsfig{figure=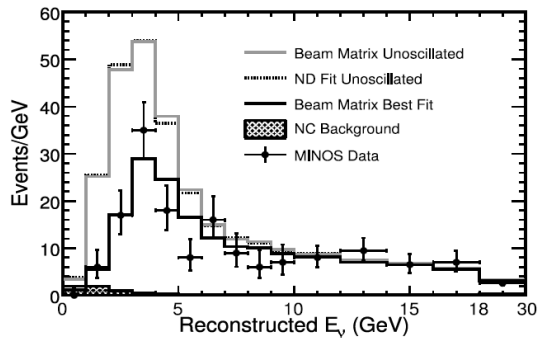,width=8.0cm}
\caption{MINOS comparison of the $E_{\nu}$ spectra with oscillations with 
the no oscillations one, from reference~\cite{minos0}, \cprD{2006}.}
\label{minoscomparison}
\end{center}
\end{figure}

\begin{figure}[hbtp!]
\begin{center}
\epsfig{figure=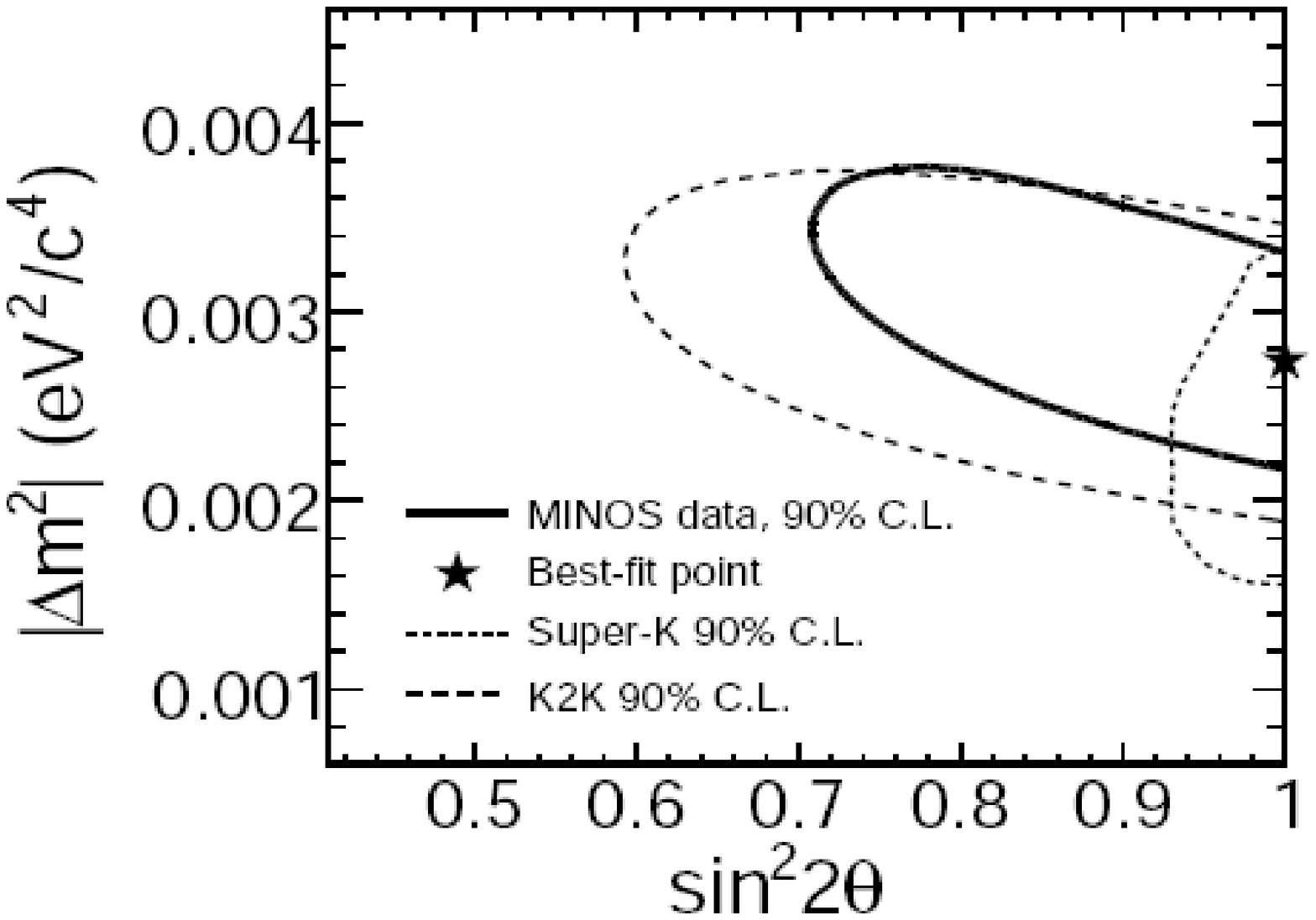,width=8.0cm}
\caption {MINOS confidence limits for the oscillations parameters,
  from reference~\cite{minos}.} 

\label{minosx}
\end{center}
\end{figure}

 The observed reconstructed number of events 
is 
compared (bin by bin) in Figure \ref{minoscomparison} 
to the expected number of events for the oscillation hypothesis. The
results are  $$\dmq_{23} =2.74^{+0.44}_{-0.26}\times 10^{-3}~\mathrm{eV}^2$$ and
$\sin^{2}2\theta_{23}\ge
0.87~at~68\%~CL$~\cite{minos} (figure \ref{minosx}).

\par Preliminary results with  increased statistics (2.5$\times 10^{20}$)
protons  
have been presented 
at 
the TAUP2007 Conference; the updated value for  $\dmq_{23}$ is $2.38^
{+0.20}_{-0.16}\times 10^{-3}$~eV$^2$~\cite{Blake}. 
\vskip 1cm

\paragraph{The OPERA  experiment:}

\par The overall neutrino oscillation picture is still lacking the direct 
observation of a different flavor in a neutrino $\numu$
beam. This is the aim of the OPERA~\cite{opera} experiment that is designed to 
detect $\nutau$ appearance in a $\numu$ beam. The high mass of the $\tau$
lepton requires a high energy neutrino beam. The CNGS (CERN to
Laboratori Nazionali del Gran Sasso) neutrino beam has been optimized to study
these oscillations. The average energy at  LNGS is 17 GeV, the
contamination of $\nue$ or $\nueb$ is smaller than 1 $\%$ and the
$\nutau$ one is completely negligible.
\par
\begin{figure}[hbtp!]
\begin{center}
\epsfig{figure=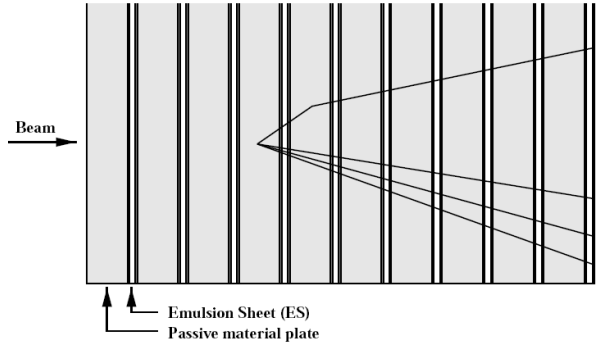,width=8.0cm}
\caption {ECC structure in OPERA, from reference~\cite{opera}}
\label{opera}
\end{center}
\end{figure}

 The detector is made  of two identical super modules, each one consisting of a
target section of 900 ton lead/emulsion modules (using the
Emulsion Cloud chamber technique illustrated in Figure~\ref{opera}), 
of a scintillator 
tracker detector and of a muon spectrometer. The high spatial resolution 
(1 micron)  of the emulsions allows the detection of the $\tau$ flight
path before its decay. Decay lengths are
of the order of 1 mm.
\par  In 5 years of run, with $4.5\times 10^{19}$ p.o.t./year, 30k neutrino
interactions will be detected.
Assuming a $\dmq_{23}$ of $2.5\times 10^{-3}$~eV$^2$ the number of tau detected will be
order of 10
with a background of about 1. It must be noted that at large $\sin^{2}2\theta$
 and $\dmq \ll
L/E$ the number of produced events depends quadratically from $\dmq$, 
so the number of
detected events
will be 14 at $\dmq =3\times 10^{-3}$~eV$^2$ and 6 at  $\dmq =2\times 10^{-3}$~eV$^2$.
\par The experiment will also be  able to give limits on $\numu \rightarrow 
\nue$. A limit of  0.06 on $\sin^{2}2\theta_{13}$ can be reached with
 $\sin^{2}2\theta_{23}$=1 and $\dmq_{23}=2.5\times 10^{-3}$~eV$^2$~\cite{komatsu}.
\par The beam and the detector performances (no emulsion inserted) have
been successfully tested in August 2006~\cite{opera1}. Reconstruction
 of neutrino events  
in the EEC  has been accomplished in 2007. Details on the reconstruction 
of these events is given 
in reference~\cite{ariga}.

\section{ Present knowledge of the parameters of the mixing matrix}
\label{mixing}
\par Flavor and mass eigenstates are connected by the unitary matrix U
that in the general case of (3,3) mixing is defined by 3 angles and
possibly a phase factor  $\delta$ (see Section~\ref{3flav}). 
With 3 neutrino species there
are two independent mass square differences.
\par While the present experiments cannot access  $\delta$, we will now
summarize our present knowledge of the above  parameters.
The small value of $\alpha=\dmq_{solar}/\dmq_{atmospheric}$ and the
 smallness of $\sin^{2}\theta_{13}$ allows in first approximation the
two flavors treatment of neutrino oscillations for atmospheric and solar neutrinos.
\par\underline{$ \dmq_{23}$ and
$\sin^{2}(2\theta_{23})$}
\par
The atmospheric experiments, K2K and MINOS, measure essentially the 
$\numu$ survival probability, which in the limit of 
$\sin\theta_{13}
\simeq 0$ and 
$\sin^{2}(\dmq_{12}L/4E) \ll 10 $, can be expressed as 
(see Section~\ref{twoflav})  
$$ P(\numu \rightarrow 
\numu)=1-\sin^{2}(2\theta_{23})\sin^{2}(\dmq_{23}L/4E) $$
identifying $\theta_{23}$ and $\dmq_{23}$ with $\theta_{atm}$ and 
$\dmq_{atm}$ and with K2K and MINOS parameters.

\par The more recent results for these parameters are
given in Table~\ref{dm2}.
\begin{table}[htbp]
\begin{center}
\begin{tabular}{|c|c|c|}\hline
{\bf experiment} & $\dmq \cdot 10^{-3}~\mathrm{eV}^2$ &
$\sin^{2}2\theta$
\\\hline
ATMO SK~\cite{SuperKamat}&1.5-3.4& $\ge$ 0.92    \\\hline
K2K~\cite{k2k}~&1.5-3.9  & $\ge$ 0.58 \\\hline
MINOS~\cite{minos}&2.48-3.18     &$\ge$ 0.87\\\hline
\end{tabular}
\label{dm2}
\caption{limits on the 23 mixing parameters}
\end{center}
\end{table}
\par \underline{$\dmq_{12}$ and
$\sin^{2}(2\theta_{12})$}
\par The $\nu_e$ solar experiments   are sensitive mainly to these two
quantities   (only to these in the two flavor mixing scheme).
The long distance reactor experiment KamLAND on $\nueb$ is 
also sensitive
(Section~\ref{twoflav}) to  $\sin^{2}(2\theta_{12})$ and  $\dmq_{12}$. In 
this experiment  the 
shape
of the energy distribution allows a precise determination of $\dmq_{12}$ 
while the solar experiments have 
a better sensitivity to $\theta_{12}$. A combined
analysis using  these informations (and assuming CPT invariance)  has
given the following results~\cite{kamland}
 $$ \dmq_{12}~~=7.9^{+0.6}_{-0.5}\times 10^{-5}~\mathrm{eV}^2$$
 $$ \tan^{2}\theta_{12}=0.40^{+0.10}_{-0.07}$$
\par \underline{$\sin^{2}(2\theta_{13})$}
\par Short distance reactor experiments are sensitive to
$\sin^{2}(2\theta_{13})$  (see Section~\ref{reactor}).
The following limits (90$\%$ CL) were obtained:

$$ CHOOZ~\cite{apollonio} \sin^{2}(2\theta_{13})\le 0.13$$
$$ Palo Verde~\cite{palo} \sin^{2}(2\theta_{13})\le 0.17 $$
\par In the three flavor mixing scheme with one $\dmq$ dominance
 we have in   $\numu \rightarrow \nue $ experiments  for 
$\sin^{2}\theta_{23}=0.5$
$$\sin^{2}2\theta_{\mu e}=\sin^{2}2\theta_{13} \sin^{2}\theta_{23}={{1}\over{2}} \sin^{2}2\theta_{13}$$
\par K2K limit is $\sin^{2}(2\theta_{13})\le 0.26$~\cite{yamamo}, while SK on atmospheric neutrinos gives $\sin^{2}(\theta_{13})\le 0.14$~\cite{skam3f}. 

\par \underline{Global fits}
\par Several global fits to neutrino oscillations have been published
(Maltoni~\cite{Maltoni}, Fogli~\cite{foglifits}, 
  Schwetz~\cite {Schwetz}).
\par We give as examples

\par A) The Fogli results
 $$ \sin^{2}\theta_{13}~=0.9^{+2.3}_{-0.9}\times 10^{-2}$$
 $$ \dmq_{12}~~=7.92^{+0.09}_{-0.09}\times 10^{-5}~\mathrm{eV}^2$$
 $$ \sin^{2}\theta_{12}~=0.314^{+0.18}_{-0.15} $$
$$ \dmq_{23}~~=2.4^{+0.21}_{-0.26}\times 10^{-3}~\mathrm{eV}^2$$
 $$ \sin^{2}\theta_{23}~=0.44^{+0.41}_{-0.22} $$
\par B) The  Schwetz  results
 \par ($ \sin^{2}\theta_{13}$~ not fitted~ and assumed to be $\le~~ 
0.025~~$ 
at 2$\sigma~~ level$)
 $$ \dmq_{12}~~=7.9^{+0.3}_{-0.3}\times 10^{-5}~\mathrm{eV}^2$$
 $$ \sin^{2}\theta_{12}~=0.30^{+0.02}_{-0.07} $$
$$ \dmq_{23}~~=2.5^{+0.20}_{-0.25}\times 10^{-3}~\mathrm{eV}^2$$
 $$ \sin^{2}\theta_{23}~=0.50^{+0.08}_{-0.07} $$
\par These two fits have been made using all the available information
and provide compatible results, also in good agreement with the
independent two flavor analysis.

\par The present situation is that we have two values of $\dmq$ but what 
is still
not measured is the sign of $\dmq_{23}$ (i.e. mass hierarchy). 
In the current data there is  not enough information to 
determine the phase of the mixing matrix.
In conclusion the missing measurements are 
\begin{itemize}
\item $\sin^{2}\theta_{13}$ 
\item mass hierarchy
\item phase $\delta$
\end{itemize}
\par To these points will be dedicated the new experiments
that will be described in next two sections.

\section{Next generation of  oscillation experiments}
A relatively large value of
$\theta_{13}$ above $10^{-3}$ would open the possibility of 
studying of CP  violation in the leptonic sector. Therefore future experiments will 
be  mainly
devoted to the measurements of the $\theta_{13}$ parameter. There are two
possibilities for measuring  $\theta_{13}$: accelerator and reactor 
experiments.
Accelerator $\numu\rightarrow \nue$ appearance experiments 
allow the measurement of the three oscillation parameters 
(sign of $\dmq$, $\theta_{13}$, $\delta$). This 
apparent advantage 
introduces ambiguities in the interpretation of the 
results and correlations between the measured parameters.
Reactor experiments, 
being disappearance experiments, cannot display CP  or T 
violations~\cite{cabibbo}  and therefore determine directly the angle $\theta_{13}$. 

\subsection{Reactor experiments}
\par Several experiments have been proposed, some of them are 
already approved
 (at least at a level of R\&D) by funding agencies.
Table ~\ref{futurr}, 
based on  
the presentation of  K.Heeger
in TAUP2007 conference~\cite{heeger},
summarizes these projects. 

\begin{table}
\label{futurr}
\hspace{-1.5cm}
\begin{small}\begin{center}
\begin{tabular}{|lc|c|c|c|c|c|c|}\hline
location & power& dist Near/Far  &depth  
&target mass &limit & time  \\
 & (GW)& (m) &(MWE)
& (kton) &(10$^{-2}$)& (year *) \\\hline
ANGRA (1)~\cite{Angra}, Brasil        &4.6  & 300/1500 &  250/2000  &500  
&0.5.  &\\
DAYA BAY (2),China~\cite{xinheng}        &11.6  &360(500)/1750&260/910  & 
40 
&1.  &3\\
Double CHOOZ (3) ,Fr~\cite{double}&6.7   &1050/1067    & 60/300  &10.2&3.  
&5\\
KASKA (4), Japan~\cite{kaska}         & 24.  &350/1600      & 90/260  & 
6.& 2 
&\\
RENO (5) Korea~\cite{reno}          &17.3  &150/1500      &230/675  &20 & 
2&3\\\hline
\end{tabular}\end{center}
\end{small}
\caption{proposed reactor  $\theta_{13}$ neutrino Experiments, (*) time 
needed to reach limit in years after completition of construction.
(1) Angra proposed and R$\&$D, (2) DAYABAY construction starts in 2007,
(3) Double CHOOZ construction  under way,
 (4) Proposal, (5)Proposal}
\label{atmos}
\end{table}
\hspace{0cm}
\par Main points to increase the sensitivity of future experiments will be

\begin{itemize}
\item higher reactor power, for the reduction of statistical errors
\item   at least two  detectors configuration, for the reduction of 
reactor systematic
errors
\item sufficient over burden and active shielding for reduction of background
\item improved  calibrations and monitoring
\end{itemize}

Within the approved experiments the best 
sensitivity is claimed by DAYA BAY \cite{xinheng} that will give an 
improvement of a 
factor  $\sim$10 over present limits. Similar results will be obtained 
on the same time scale by
the Double Chooz experiment \cite{double}.
\subsection{Accelerator experiments}
\label{futuracc}
\par Accelerator experiments will be focused  on  the measurement of 
$\theta_{13}$ 
through the detection of the sub--leading   $\numu \rightarrow 
\nue$ oscillation. 
This is an  appearance experiment,  that can
give information on all the oscillation   parameters. The probability 
can be written, in  
the lowest order approximation
in the form of equation~\ref{eq:approx}.(section \ref{appro})

\par For experiments made at the first oscillation maximum for atmospheric 
neutrinos parameters,  if MSW 
effects are negligible, 
the leading term is the one in the first line of the above quoted formula: 
\par $P= sin^{2}2\theta_{13}
sin^{2}\theta_{23}sin^{2}(\dmq_{23}L/4E)$
\par $P={{1}\over{2}} sin^{2}2\theta_{13}$ 
\par The last step assumes $sin^{2}(\theta_{23})=0.5$ and $sin^{2}(\dmq_{23}L/4E)\simeq 1$.
\par Searching for leptonic CP  violation  one will look for different
 appearance
probabilities  for neutrino and anti-neutrino due to the change of
 the sin$\delta$ term.
Using neutrino and anti-neutrino beams we can measure the asymmetry
of the appearance probability:
$$  Asym=\frac{P(\numu -\nue)-P(\numub -\nueb)}{P(\numu -\nue)+P(\numub
-\nueb)}$$
that is given in vacuum by
$$ Asym=\dmq_{12}L/(4E_\nu)\cdot 
sin2\theta_{12}/sin2\theta_{13}\cdot sin\delta$$
\par The  MSW effect changes sign for 
neutrino and anti-neutrino so, when it cannot be neglected, the effects of 
$\delta$ and MSW must be 
disentangled. A further complication comes in
because the value of A as given in section \ref{appro} will change sign according to  
the 
sign of  
$\dmq_{23}$.
\par  In general the measurement of oscillation probabilities
will not give unique solutions for the oscillation parameters,
correlations and degeneracies will be found. The correlation $\delta$ vs
 $sin^{2}2\theta_{13}$ is shown in Figure~\ref{novasens}, where  the 
degeneracies are also
shown. Furthermore the sign of
$\dmq_{23}$ and the interchange $ (\theta_{23},\pi/2-\theta_{23})$
can lead to an eight-fold  degeneracy in the determination of oscillation
parameters. No single experiment will be able to solve these
degeneracies
and  proposals to solve the problem have been made
\cite{Barger}, \cite{Burguet}, \cite{Migliozzi}.
 \subsubsection{T2K Experiment}
\par
\begin{figure}[hbtp!]
\begin{center}
\epsfig{figure=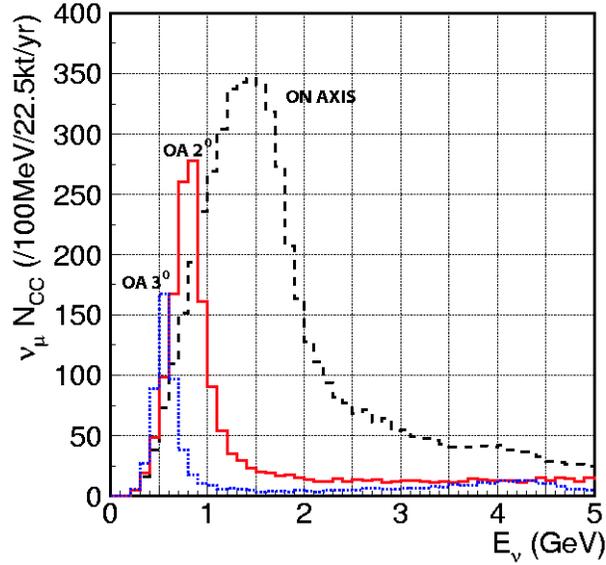,width=8.0cm}
\caption {T2K beam neutrino energy spectrum for different off-axis 
angles, from reference
\cite{ito}}
\label{t2kbeam}
\end{center}
\end{figure}

 The T2K~\cite{ito}, \cite{t2k}, experiment is under 
construction and the first data 
will be collected
in 2009. It adopts the same principle of the K2K experiment: it is a 
two detectors
experiment, with a  far detector (SK-3) at 295 km from the 50 GeV
accelerator at JPARC complex in Japan, and a close detector that will be 
at
a distance of 280 meter. The neutrino beam will be an off axis beam
at an energy of 0.6 GeV. The neutrino momentum distribution is 
shown in Figure~\ref{t2kbeam} for various off axis angles.
The reduced average energy has the advantage of
reducing the
number of
$\pi^0$ produced, of
gamma
rays from $\pi^0$ decay and consequently the
background to the detection of electrons from $\nue$ interactions.

The aim of what is called phase I (JPARC proton beam power 0.75 MW) are
\begin{itemize}
\item In appearance mode a sensitivity (for $\delta=0$) down to 0.008 on  
$sin^{2} 2\theta_{13}$. 
\par The correlation between $\delta$ and $sin^{2}2\theta_{13}$ sensitivity 
is shown in Table~\ref{delta}.
\item In disappearance mode 
$$ \sigma(\dmq_{23})=10^{-4}eV^2~~~ \sigma( sin^{2}\theta_{23})~=~0.01$$

\item And a 
 search for $\numu \rightarrow \nutau$ by measurement of neutral
current events.
\end{itemize}

\begin{table}
\begin{center}
\begin{tabular}{|c|c|} \hline
$\delta$ & $\sin^{2}2 \theta_{13}$ \\ \hline
0 & 8 $\times 10^{-3}$ \\ \hline
-$\pi$/2 & $3 \times 10^{-3}$ \\ \hline
$\pi$/2 & $2 \times 10^{-2}$ \\ \hline
$\pi$ & $8 \times 10^{-3}$ \\ \hline
\end{tabular}
\caption {$\delta$ vs $sin^{2}2 \theta_{13}$ sensitivity for T2K, from 
reference ~\cite{t2k}.}
\label{delta}
\end{center}
\end{table}

\par These numbers have been computed for a 5 years run with $5\times 10^{21}$
protons.
\par Given the low neutrino energy, matter effects will be small. 
\subsubsection {NO$\nu$A Experiment}
\label{nova}
\par
\begin{figure}[hbtp!]
\begin{center}
\epsfig{figure=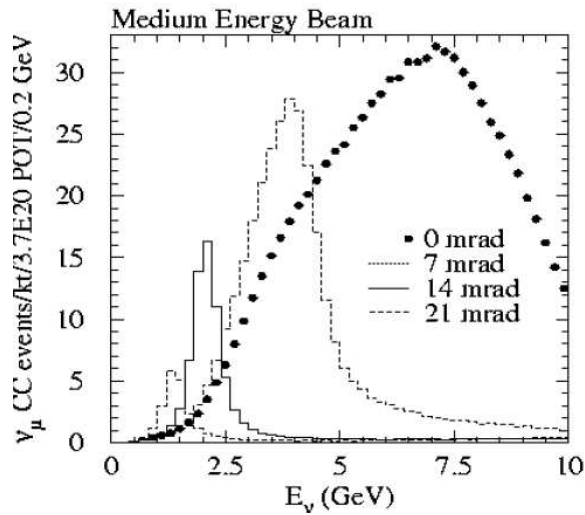,width=8.0cm}
\caption {Rates in NO$\nu$A beam, from reference ~\cite{novaprop}.}
\label{novafig}
\end{center}
\end{figure}

 The NO$\nu$A
experiment has been proposed at Fermilab~\cite{novaprop} and is now in an R$\&$D
phase on the way for approval.
It will be a two detector experiment, with a 810 km baseline, from NUMI 
beam at
Fermilab, at 2.5 degrees off axis. The beam will be at an average
momentum of 2.3 GeV. The momentum distribution of interacting neutrinos for 
 various off  axis angles is shown in Figure~\ref{novafig}.

The far detector will be made of  planes of PVC structures
containing liquid scintillator, the close detector will have the same
structure followed by a muon catcher. The experiment is on the way of approval.
\par Initially the experiment will run with a proton beam power of  
0.3 MW, then of 0.7 MW, finally of 1.2 MW.
\par The main aim of the experiment will be the detection of $\numu \rightarrow
\nue$ so the  
detector will be optimized to separate  electron events.
\par
\begin{figure}[hbtp!]
\begin{center}
\epsfig{figure=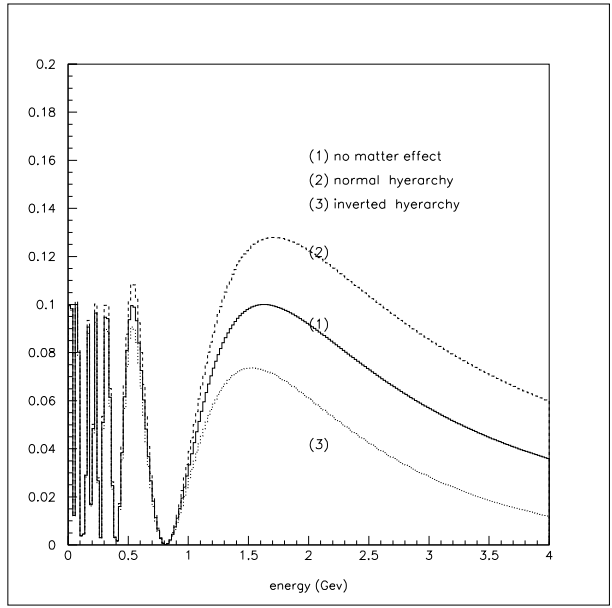,width=8.0cm} \caption
{Matter effects in NO$\nu$A experiment.}
\label{novamatter}
\end{center}
\end{figure}

  The experiment will be sensitive  
to the mass hierarchy (see Figure~\ref{novamatter}) through matter effects.
In fact  at the first maximum of the oscillation probability
we can write (see equation~\ref{eq:approx}):
$$ P_{mat}(\numu \rightarrow \nue)=(1+2A)P_{vac}(\numu \rightarrow
\nue)$$
Introducing $E_{R}=\dmq_{23}/2\sqrt{2}G_{F}N_{e}=E\dmq_{23}/|B|$,
with $G_F$ Fermi constant and $N_{e}$ electron number density,
the above expression can be rewritten as
$$ P_{mat}(\numu \rightarrow \nue)=(1\pm2E/E_R)P_{vac}(\numu \rightarrow
\nue).$$ 
The sign in front of the $E_R$ depending term is + for neutrinos and - for 
anti-neutrinos.
$E_{R}$ will be positive or negative 
according to the sign of $\dmq_{23}$.
\par In the case of NO$\nu$A, for the normal hierarchy, matter effects increase
by about 30$\%$ the
oscillation probability or decrease it by the same amount 
for the inverted one, in the neutrino
case.  The opposite is true for anti-neutrinos.

\par As an example Figure~\ref{novamatter} shows $ P(\numu \rightarrow 
\nue)$ computed for
L=800 km,
$\dmq_{23}$=0.0025~eV$^2$, sin$^{2}2\theta_{13}$=0.1 and sin$^{2}2\theta_{23}$=1.

\begin{figure}[hbtp!]
\begin{center}
\epsfig{figure=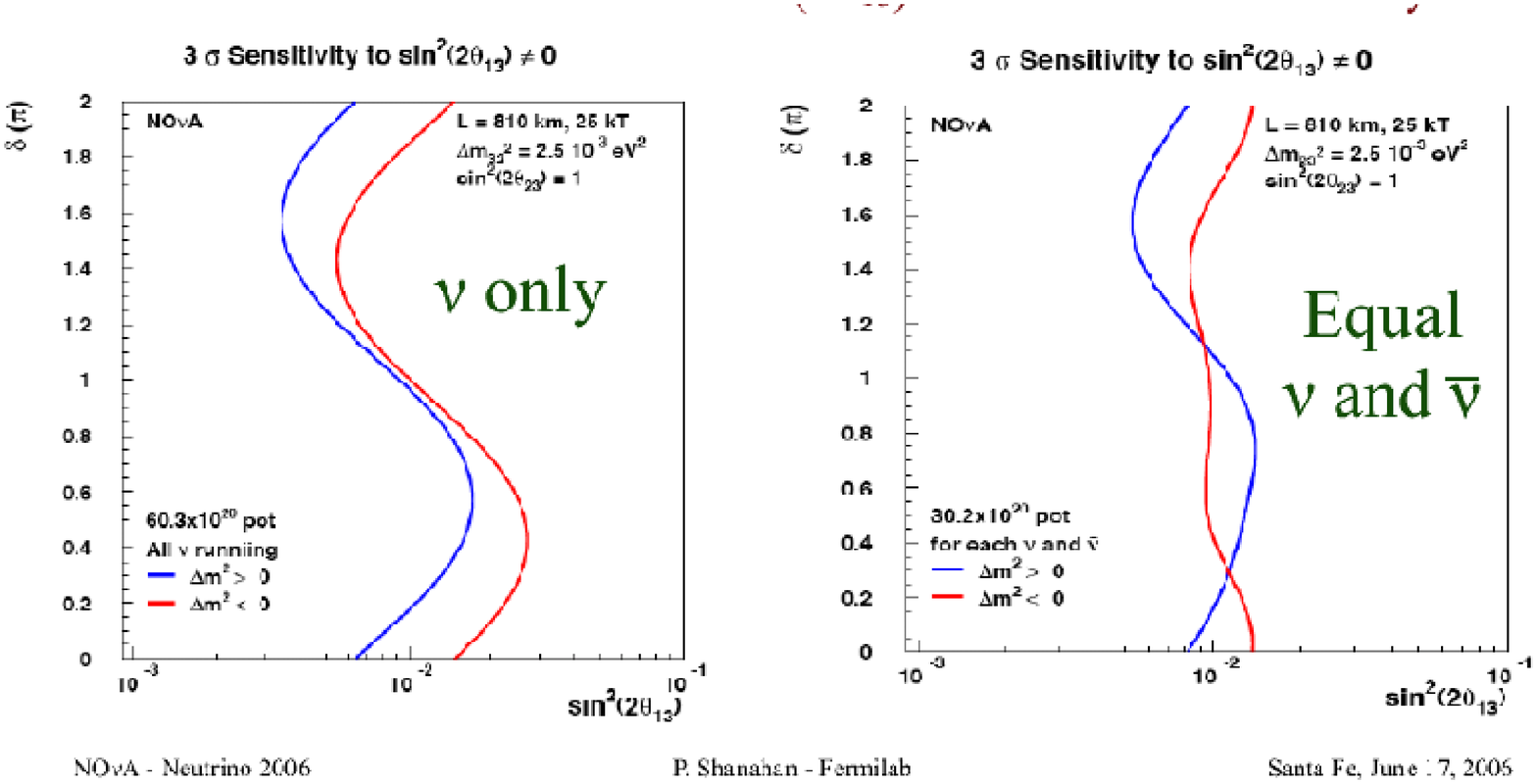,width=12.0cm}
\caption {NO$\nu$A sensitivities, $\delta$ versus $sin^{2}(2\theta_{13}$). 
Left panel neutrino only, right panel neutrinos and  anti-neutrinos. Lines 
represent $\dmq_{23}$ positive and negative values, from reference 
~\cite{shana}.}
\label{novasens}
\end{center}
\end{figure}

The probability of oscillation will depend on all the still unknown
parameters.
The discovery limit for sin$^{2}2\theta_{13}$ at $\delta$ =0 will be
8$\times 10^{-3}$     or 1.5$\times 10^{-2}$ for normal or inverted  hierarchy. 
The limit
will depend on the value of $\delta$ as shown in Figure~\ref{novasens}a. 
Because the
anti-neutrinos have an opposite dependence of $\delta$ on
sin$^{2}\theta_{13}$, running neutrinos and anti-neutrinos the correlation
will be largely reduced(Figure~\ref{novasens}b).

\section{Long term plans for  oscillation experiments}
\par After 2010 the proposed reactor experiments will have improved our
knowledge of $\sin^{2}2\theta_{13}$ by about a factor 10 compared to the
present limit. Being disappearance experiments they will not give
 informations on the other missing parameters:
mass hierarchy  (sign of $\dmq_{23}$) and value of $\delta$.
These informations will be given by the measurement of P($\numu \rightarrow
\nue$) at an L/E corresponding to the value of $\dmq_{23}$ given by the
atmospheric neutrinos. First informations will be given by T2K and NO$\nu$A,
for which improvements have been proposed.
\subsection{Improvements of T2K and NO$\nu$A}
\par \underline{T2K experiment}
\par The improvements will consist in
\begin{itemize}
\item
increase of JPARC proton beam power from 0.75 MW to 4MW
\item new far detector Hyper-Kamiokande (HK) with a mass of 0.5 megaton
\item run with anti-neutrino   beam. 
\end{itemize}
\par Another possible development proposed is the construction of 
T2KK~\cite{t2kk},
a detector  in Korea, located at the second oscillation maximum.
T2KK will improve the sensitivity on $\delta$, and given the
longer distance, matter effects will become considerable with a possibility of
determining the mass hierarchy.
\par \underline{NO$\nu$A experiment}
The upgrade would consists in
\par a) final proton beam power 1.2 MW
\par b) a second detector at a different distance possibly using novel
technologies (Liquid Argon detector).
\par At the maximum proton power NO$\nu$A will be able to explore the full phase
space for $\delta$ provided sin$^{2}\theta_{13} \ge 10^{-2}$. 
If this was the case, in combination with the upgraded T2K
experiment, a resolution of 2$\sigma$ would be reached in the
determination of mass hierarchy.

\subsection{SPL beam to Fr\'ejus}
\par
\begin{figure}[htb]
\begin{center}
\epsfig{figure=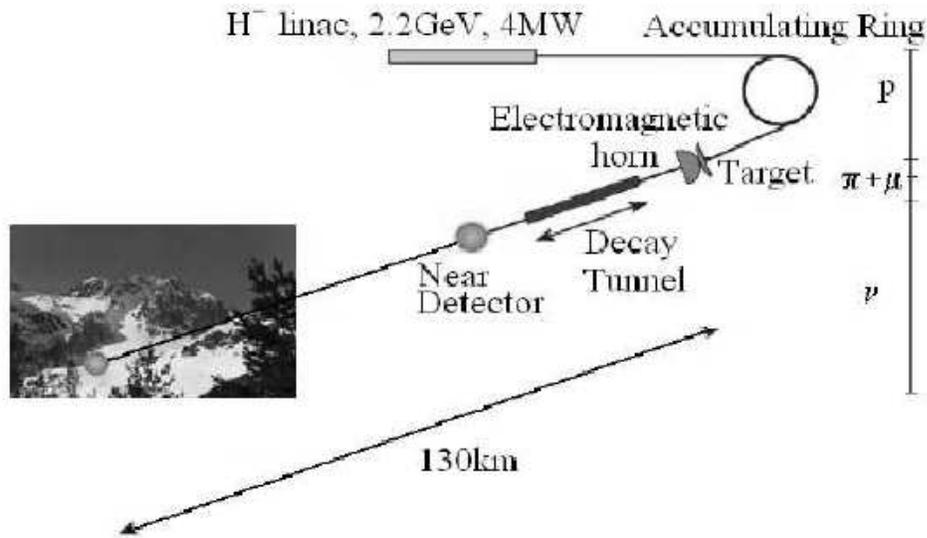,width=14.0cm}
\caption {SPL setup, from reference~\cite{Campagne}. }
\label{SPL}
\end{center}
\end{figure}

  Still in the line of using conventional beams, a proposal has been
presented for the Superconducting Linac beam at the Fr\'ejus tunnel.
The European project~\cite{SPL} foresees (see Figure~\ref{SPL}):
\begin{itemize}
\item  a super conducting proton Linac with a power of  4 MW, and an energy 
up to 5 GeV, at CERN;
\item a neutrino beam energy of about 300 MeV, 
optimized to give maximum sensitivity
on the far detector located in the Fr\'ejus tunnel (that is at a distance
of 130~km from CERN);
\item a far detector (MENPHYS~\cite{memphys}) of 500 kton water Cherenkov
at a  depth 4800~MWE;
\item  a close detector in the CERN site. 
\end{itemize}
\par 
Competing proposals for the water Cherenkov detector 
can be found  in UNO~\cite{UNO} and 
Hyper-Kamiokande proposals~\cite{t2k}. 
\par In ten years of running a sensitivity of 0.001 at 90$\%$  CL  for 
$\sin^{2}2\theta_{13}$ can be obtained~\cite{memphys}.
\subsection{Atmospheric neutrinos}
\par A large amount  of information could be obtained from an underground
large magnetized detector of atmospheric neutrinos.
 A calorimeter (ICAL) of this type  as been proposed by the
Indian Neutrino Observatory collaboration
(INO) \cite{INO}. Comparison of results obtainable in Iron calorimeters   
and in large water detector can be found in
\cite {Choubey}
\subsection{New ideas}
\par One of the limiting factors in the measurement of P($\numu\rightarrow
\nue$) is the $\nue$ contamination in conventional
neutrino beams.
Novel ideas for high purity beams
 overcoming this problem have been proposed, these are the Beta Beams 
and the Neutrino Factories.
\subsubsection{Beta Beams:}

\begin{figure}[htb!]
\begin{center}
\epsfig{figure=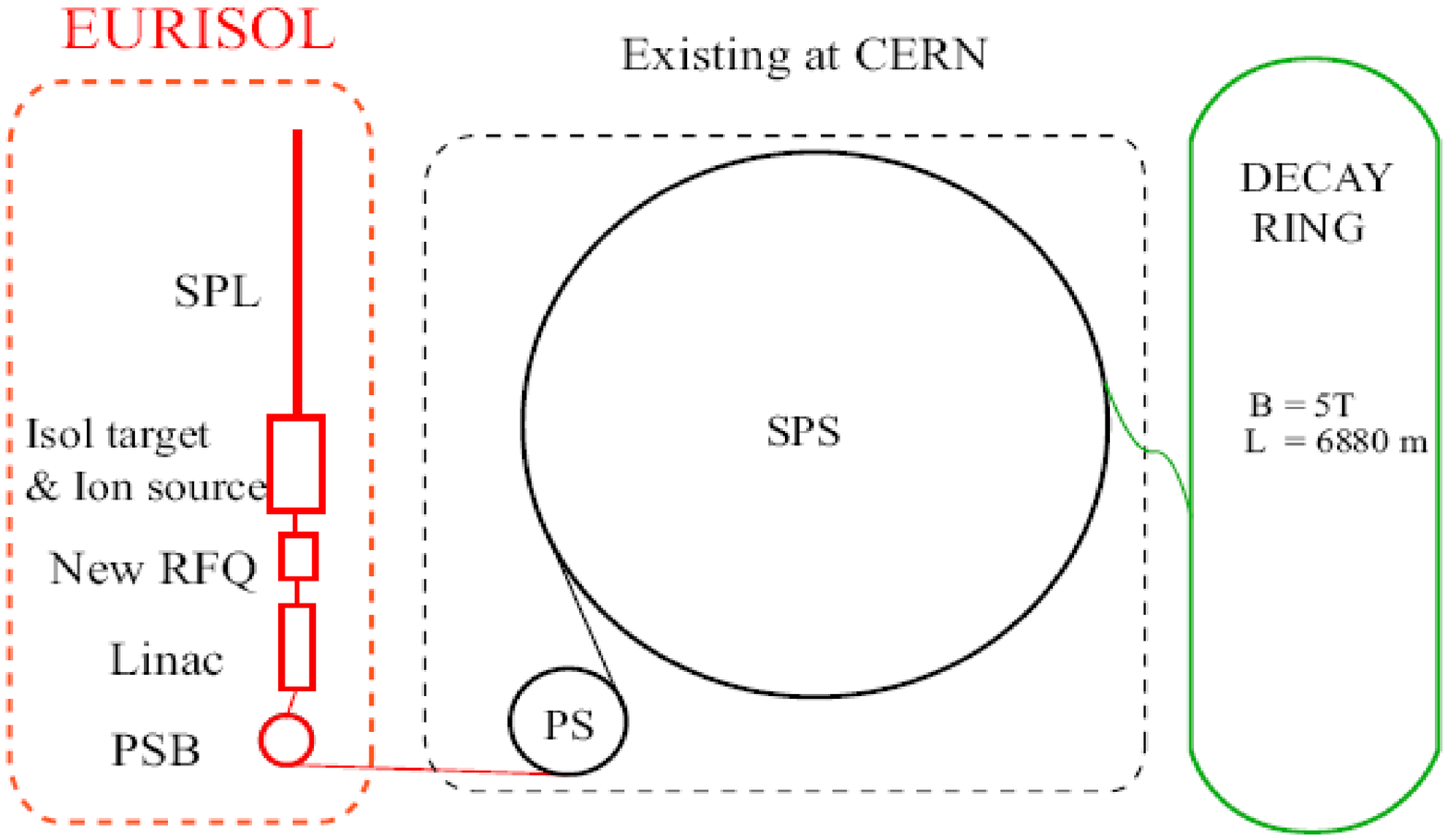,width=12.0cm}
\caption{Beta Beam layout, from reference ~\cite{Mezzetto:2005yf}, \cprC{2005}. }
\label{betabeam}
\end{center}
\end{figure}

\par The Beta Beam idea, introduced by Piero 
Zucchelli~\cite{zucchelli}, is that
  $\beta^{+}$ (or $\beta^{-}$)  decays, from
accelerated radioactive nuclei, produce  
pure forward $\nue$ (or 
$\nueb$) beams.
Radioactive nuclei producing respectively $\nueb$ and $\nue$ are for example 
$He^{6}$ and $Ne^{18}$:
 \par $ He^{6} \rightarrow Li^{6} +e^{-} +\nueb$
\par $Ne^{18} \rightarrow F^{18} +e^{+} +\nue$.
\par According to the type of used radioactive nucleus 
 $\nue$ or $\nueb$ beams will be produced and it will be possible to
study $\nue\rightarrow\numu$ or $\nueb\rightarrow\numub$ channels.

The characteristics of produced beams will be:
\begin{itemize}
\item pure beams with just one flavor
\item very intense beam completely known, their energy  will be determined
 by the beta decay energy and Lorentz factor $\gamma$ 
\item flux normalization from the number of the radioactive ions circulating in
the ring
\item divergence of the beam given by  $\gamma$.
\end{itemize}

\par A conceptual design of a Beta Beam has been proposed at CERN.
 A possible layout is shown in Figure~\ref{betabeam}.
The neutrino beam will be sent to the Underground laboratory at 
the Fr\'ejus tunnel   where a 
megaton Water Cherenkov detector will be deployed, the same proposed for 
the SPL project.
Measurement of  $\sin^{2}2\theta_{13}$ down to 0.0004 
(at $\delta$ =0)
will be possible for 10 years running
 using appearance 
and disappearance channels~\cite{mezzetto}. 
\par The physics reach of the CERN  Beta Beam + SPL
combination is described in \cite{spl+beta}. This combination offers
the possibility of comparing two beams with the same detector thus reducing 
the detector related systematic effects. It  will be 
possible to study
\begin{itemize}
\item CP violation: comparison of  $\numu$ and $\numub$ with SPL and
 $\nue$ and $\nueb$ with Beta Beam
\item T violation: comparison of $\nue \rightarrow \numu$ (Beta Beam) and 
$\numu \rightarrow \nue$ (SPL).
\end{itemize}  
\subsubsection{Neutrino factories:}
\par The principle of the Neutrino Factory \cite{geer} is to produce 
intense neutrino
 beams from the decay of muon stored in a  ring  with long straight 
sections.
\par Several projects  are under study in Europe, USA and 
Japan.
The results that can be obtained in a Neutrino Factory are described 
in
\cite{nf2}. It will be possible
 to reach very small values of sin$^{2}\theta_{13}$ $\simeq 
10^{-4}$ 
 not reachable with other experiments~\cite{huber03}.
\par The proposed energies are of the order of 30-50 GeV implying 
distances of the order of thousand kilometers and 
thus requiring massive detectors.

\begin{figure}[htb]
\centerline{\psfig{figure=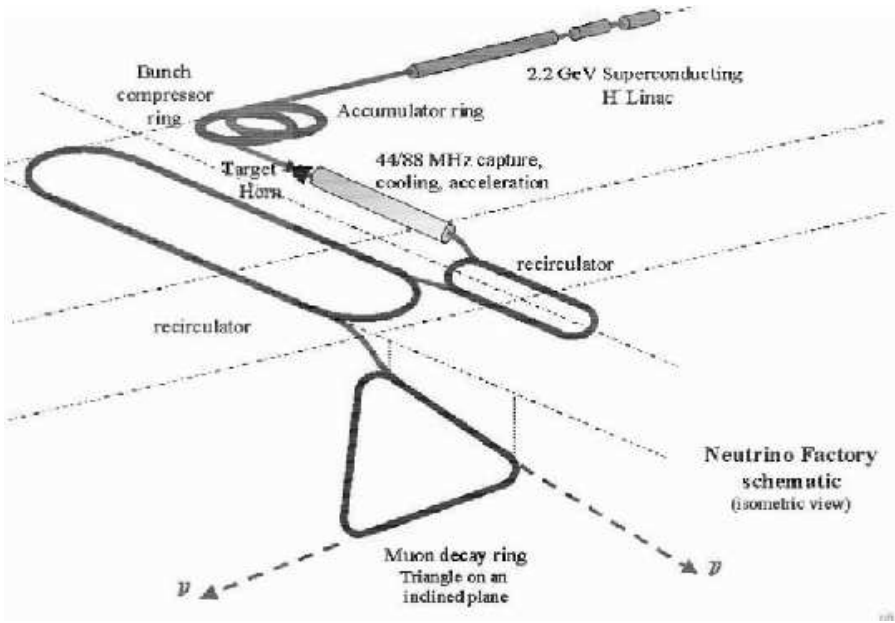,width=12.0cm}}
\caption{Possible layout of a Neutrino Factory, from reference 
\cite{Gruber:610249}. }
\end{figure}
\label{nufac}
\par A Neutrino Factory project will include (figure \ref{nufac})

\begin{itemize}
\item
ion source
\item
proton accelerator
\item
pion to muon decay line with beam cooling
\item
muon accelerator
\item
muon storage in a decay ring
\item  neutrino detectors
\end{itemize}
\par In Neutrino factories it   will be possible to study  many
channels.
A channel  (golden channel\cite{golden}) that will be studied 
will consists in the detection of  ``wrong sign'' 
muons.
If we store $\mup$ they will decay and produce $\numub$ and $\nue$.    
 In  the detector $\numub$ will produce $\mup$.
  $\numu$  from  oscillated $\nue$    will give $\mum$, which 
have ``wrong sign'' with respect to the primary component.
Detection of the sign of muons can be achieved using 
massive magnetized detectors.   
\par The removal of ambiguities and degeneracies has been studied 
by several authors \cite{Barger,Burguet,Migliozzi}.
It has been shown for example  \cite{huber03} that
 running  on  $ \numu$ and $\numub$ and making experiments with
  different
baselines it will be possible to remove completely ambiguities and 
degeneracies and that a sensitivity of 
$\simeq  10^{-4}$ will be reached on 
sin$^{2}2\theta_{13}$.  
\subsection{Comments  on future projects} 
\par Beta Beams and Neutrino Factories are long term projects. They 
will require large funding and intensive R$\&$D. 
On the other hand, conventional beams will give safe results on CP
 violation only if
sin$^{2}2\theta_{13} \ge 10^{-2}$. Below $10^{-3}$ the only viable solution 
would be a Neutrino Factory. For a complete discussion of future plans  see reference 
\cite{Blondel} and references in therein.

\section{Conclusions}
\par In recent years the evidence for neutrino oscillations has become
clear. After 40 years of indications,  that did start with R. Davis   
observation of the solar neutrinos 
deficit, now we know that neutrinos have mass (although small) and that the mass 
eigenstates are not the flavor ones. A large amount of experimental data has 
been collected and many elements of the mixing matrix have been 
determined. 
To have a complete  description of the mixing matrix 
the term sin$2\theta_{13}$ needs a better determination, now only upper limits are 
known, and the phase $\delta$, now completely unknown, must be measured. If  
sin$2\theta_{13}$ 
is not too small the way will be open to studies of CP violation in
the weak interaction sector, the CP violation term
$\delta$ will be accessible.  
\par When the running, approved or on the way of approval,  experiments will give 
their results the
future of oscillation experiments will be made more clear and experiments based 
on novel ideas, Beta Beams and Neutrino Factories, could  become 
necessary.

\section{Acknowledgments}
We warmly thank Paolo Lipari for encouragement, criticism and
discussions. We also gratefully aknowledge the critical reading 
of P.F~Loverre, M.~Mezzetto and L.~Zanello and the support of C.~Mariani 
in the editing of this paper.

\section*{References}
\bibliography{NuOsc}
\end{document}